\RequirePackage{fix-cm}
\documentclass[twocolumn]{svjour3}          
\smartqed  

\usepackage{amsmath,mathrsfs}
\usepackage{amstext}
\usepackage{color}
\usepackage{amssymb,amsbsy}
\usepackage{amsfonts}
\usepackage{graphicx}
\usepackage{epstopdf}
\usepackage{url}
\usepackage[us]{datetime}
\usepackage{cite}
\usepackage[thmmarks, amsmath]{ntheorem}
\usepackage[shortcuts,acronym]{glossaries}
\usepackage{mathtools}
\usepackage{color,colortbl}
\usepackage[table]{xcolor}
\usepackage{accents}
\usepackage{mathtools}
\usepackage{multirow}
\usepackage{bm}
\usepackage{fancyhdr}

\usepackage[OT2,OT1]{fontenc}
\newcommand\cyr
{\renewcommand\rmdefault{wncyr}
\renewcommand\sfdefault{wncyss}
\renewcommand\encodingdefault{OT2}
\normalfont
\selectfont}
\DeclareTextFontCommand{\textcyr}{\cyr}

\definecolor{Gray}{gray}{0.9}
\definecolor{LightCyan}{rgb}{0.88,1,1}
\definecolor{LightMagenta}{rgb}{1,0.88,1}
\definecolor{LightOrange}{rgb}{1,0.99,0.88}
\definecolor{LightGreen}{rgb}{0.9,1,0.8}
\definecolor{DarkOrange}{rgb}{0.98,0.91,0.71}
\definecolor{DarkGreen}{rgb}{0.67,0.88,0.69}

\allowdisplaybreaks[1]
\graphicspath{{../}, {../Fig12/}, {../Fig13/}, {../Fig14/}, {../Fig15/}, {../Fig16/}, {../Fig17/}, {../Fig18/}, {../Fig19/}, {../Fig20/}, {../Fig21/}, {../Fig25/}, {../Table01/}, {../Table02/}, {../Table03/}, {../Table04/}}

\begin{document}

\title{Nonholonomic dynamics and control of road vehicles: \\
moving toward automation}
\titlerunning{Nonholonomic dynamics and control of road vehicles: moving toward automation}        

\author{\mbox{Wubing B. Qin \and Yiming Zhang \and D\'enes Tak\'acs \and G\'abor St\'ep\'an \and G\'abor Orosz}}
\authorrunning{W. B. Qin \and Y. Zhang \and D. Tak\'acs \and G. Orosz \and G. St\'ep\'an} 

\institute{Wubing B. Qin \at
		    Department of Mechanical Engineering\\
    	    University of Michigan, 
    	    Ann Arbor, MI 48109, USA\\
            \email{wubing@umich.edu}
        \and
           Yiming Zhang \at
            Department of Aerospace Engineering\\
    	    University of Michigan, 
    	    Ann Arbor, MI 48109, USA\\
            \email{zhyiming@umich.edu}
        \and
            D\'enes Tak\'acs \at
			MTA-BME Research Group on Dynamics of Machines and Vehicles\\
			Budapest University of Technology and Economics\\
			Budapest, H-1111, Hungary\\
			\email{takacs@mm.bme.hu}
		\and
		    G\'abor St\'ep\'an  \at
		    Department of Applied Mechanics\\
			Budapest University of Technology and Economics\\
			Budapest, H-1111, Hungary\\
			\email{stepan@mm.bme.hu}
		\and
		    G\'abor Orosz \at
		    Department of Mechanical Engineering\\
            Department of Civil and Environmental Engineering\\
    	    University of Michigan, 
    	    Ann Arbor, MI 48109, USA\\
            \email{orosz@umich.edu}
}
\date{Received: date / Accepted: date}

\maketitle

\begin{abstract}
Nonholonomic models of automobiles are developed by utilizing tools of analytical mechanics, in particular the Appellian approach that allows one to describe the vehicle dynamics with minimum number of time-dependent state variables. The models are categorized based on how they represent the wheel-ground contact, whether they incorporate the longitudinal dynamics, and whether they consider the steering dynamics. It is demonstrated that the developed models can be used to design low-complexity controllers that enable automated vehicles to execute a large variety of maneuvers with high precision.
\keywords{Vehicle dynamics and control \and Appellian approach \and Nonholonomic system}
\end{abstract}

\section{Introduction}

\begin{figure*}[!t]
\begin{center}
\includegraphics[scale=1]{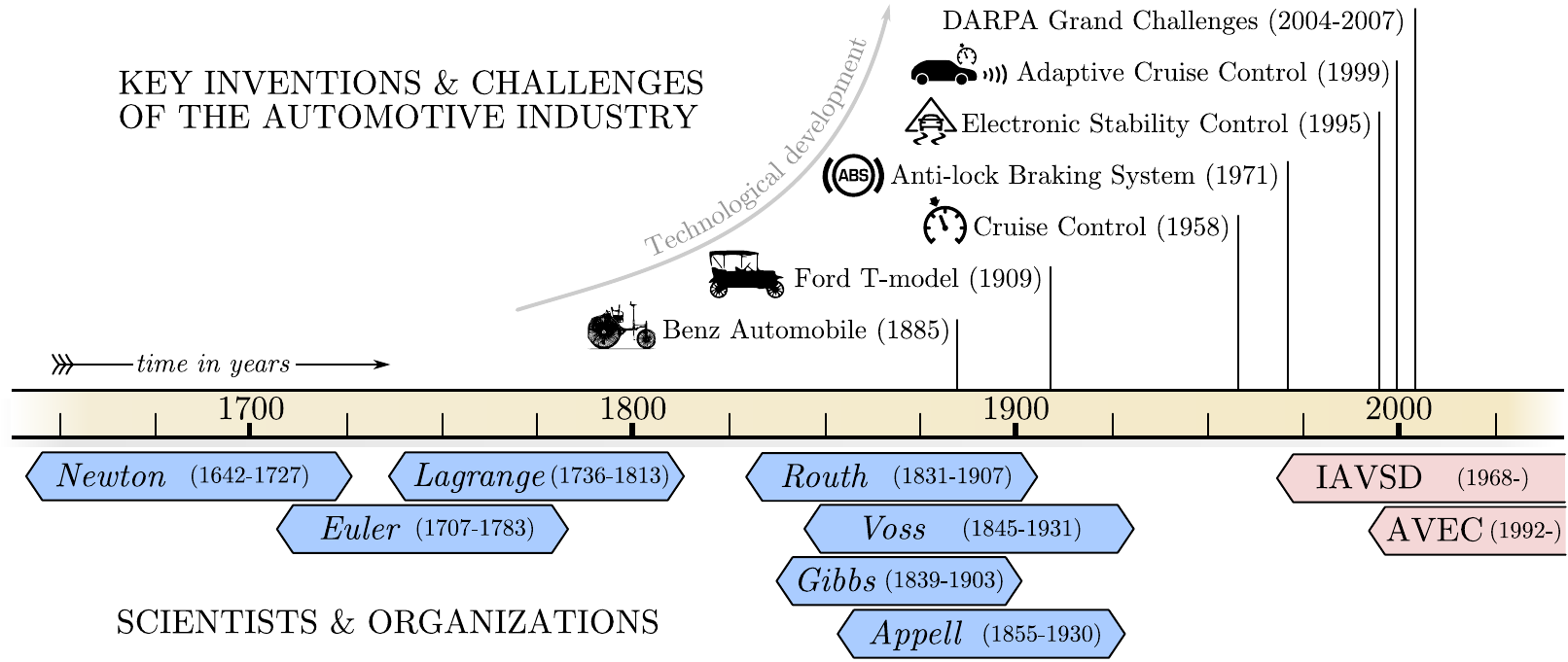}
\end{center}
\caption{Timeline of the life of dynamicists and the development of vehicle dynamics, control, and automation. \label{fig:timeline}}
\end{figure*}

During the last century, we have seen an unprecedented evolution of road transportation. This started with Benz's invention of the \textit{horseless carriage} or \textit{automobile}  at the end of the nineteenth century, which was turned into mass production by Ford during the early twentieth century. It was not until the second half of the twentieth century when engineers started to describe the motion of road vehicles and the subject \textit{vehicle dynamics} was born, as evidenced by the establishment of the International Association of Vehicle Systems Dynamics (IAVSD) and the corresponding journals and symposia. During the last few decades road vehicles transformed from mechanical to electro-mechnical systems by taking advantage of the products of the semiconductor industry. This essentially led to the birth of the subject \textit{vehicle control} which was evidenced by the foundation of the organization Advanced Vehicle Control (AVEC) and the corresponding series of symposia. During the first two decades of the current century, starting with the DARPA Grand Challenges, the notion of \textit{automated vehicle} or \textit{self-driving vehicle} was established, and this is expected to dominate the research and development of vehicle dynamics and control during the next few decades. The timeline of these events is summarized at the top of Fig.~\ref{fig:timeline}.

The field of vehicle dynamics, control and automation has been driven by three main factors: the demand for speed, the demand for maneuverability, and the need for safety. These led to many key inventions during the second half of the last century, including cruise control, anti-lock braking system (ABS), electronic stability control (ESC), adaptive cruise control (ACC), and lane keeping systems; see Fig.~\ref{fig:timeline}. These systems relieve the human driver from some of the driving tasks, while still keeping him/her in the loop. Recent efforts, on the other hand, have been mainly dedicated to handing over all driving tasks to automated systems whose capabilities supersede those of the human drivers. This ambitious goal demands for vehicle models that can describe a large variety of maneuvers and for low-complexity controllers which enable the vehicle to execute those maneuvers. These controllers need to achieve high level of maneuverability without large computational efforts, as the latter would result in time delays and would compromise safety, especially for high speed \cite{LiuBioBis2006,XuPenTan_2021,BerAveHeTakOro2022,LuSteLuTak2022}.

To move towards higher levels of automation, many efforts are put forward both in academia and in the industry.
These include the establishment of research centers and test facilities at universities \cite{universities}, large investments made by traditional automakers, and the creation of many start-up companies which primarily focus on automated driving.
Researchers approach automation from various perspectives including safety, efficiency, security, legal, and even ethical considerations.
Vehicle dynamics and control plays a fundamental role in enhancing these performance measures, and thus, will have a key role in achieving self-driving functionalities.

In order to understand vehicle dynamics, we need to go back a little more in history. Fig.~\ref{fig:timeline} also displays a list of eminent scientists who predated the automotive revolution by centuries. As a matter of fact, most of the current modeling approaches are based on the works of Isaac Newton \cite{Newton_1687}, which were reformulated later by Leonhard Euler \cite{Euler_1736}. This method, which maintains knowledge of reaction forces between bodies, still constitutes the base of vehicle dynamics softwares currently used in academia and industry; we refer to this as the \textit{Newtonian approach}. A method that eliminates the reaction forces arising from geometric constraints is due to Joseph-Louis Lagrange (born Giuseppe Luigi Lagrangia) \cite{Lagrange_1788}. This can also be used for vehicle systems and we refer to this as the \textit{Lagrangian approach}. Kinematic constraints, which can describe the dynamics of rolling wheels, were incorporated in the Lagrangian approach by Edward Routh \cite{Routh_1884} and Aurel Voss \cite{Voss_1885}. Nevertheless, the method developed by Paul Appell \cite{Appell_1900} and independently by Josiah Gibbs \cite{Gibbs_1879} was necessary to eliminate the reaction forces arising from the rolling constraints. We refer to this as the \textit{Appellian approach}, though we remark that similar methods were developed later by Petr Voronets \cite{Voronets1901}, Georg Hamel \cite{Hamel1938} and Thomas Kane \cite{Kane1961}. Various analytic and geometric approaches to kinematic constraints and the so-called nonholonomic systems were later developed and summarized in \cite{Gantmacher_1970,NeiFuf1972,Kane_1985,Baruh_1999,Papastavridis_2002,Greenwood_2003,Koon_Marsden_1997,OstAng1998,Bloch_2003,deSapio_2017}.

The Appellian approach has not yet been utilized widely in the field of vehicle dynamics. Kinematic constraints are often imposed on (potentially complex) vehicle models to form constrained optimization-based control problems, which are computationally expensive. In contrast, the Appellian approach eliminates kinematic constraints by selecting the minimum number of dynamic variables that can be used to describe the motion of the vehicle. It generates a system of first order differential equations, which is ready-made for control design without the need of imposing any additional constraint.
Such simplicity can have tremendous benefit for motion planning and control of automated vehicles, which has not yet been exploited so far, except in case of steering control of kinematic models \cite{Murray_TAC_1993,DeLuca_Planning_1998}.

The first major contribution of this paper is the introduction of the Appellian approach into the field of vehicle dynamics. For the first time, nonholonomic dynamic models are derived, which incorporate the essential longitudinal and lateral dynamics of automobiles, while taking into account the kinematic constraints of rolling. We assume rigid wheels and skates to model the wheel-ground contact, and derive the constraining forces at the contact points which are needed to realize the kinematic constraints.
This approach enables us to reveal the backbone dynamics of automobiles \cite{Antali_2020}, and to describe a large variety of maneuvers without significant increase in model complexity.
In particular, the models remain differentially flat
\cite{MarMurRou2003,AgrParCosRosAmePan2021,MurRatSlu1995,FliLevMarRou1995,Lev2009}, enabling the development of low-complexity planners and controllers with low computational cost.

We remark that in the literature, tire models \cite{Pacejka_2002} are utilized typically to calculate the wheel-ground contact forces. These involve many empirical parameters and the corresponding vehicle models are of higher complexity, making it difficult to identify the backbone dynamics \cite{MenNovFliMou2014,SeoLeeBaeHorCho2022}. We remark that the Appellian approach can also be used when incorporating tire models of different complexity \cite{TakSte2013,OhAveOro2021,BerAveHeTakOro2022}. The arising high-complexity vehicle models may be used to test the controllers designed using the nonholonomic models. We also remark that one may incorporate the elasticity of the suspension while assuming rigid wheel-ground contact \cite{TakSteHog_2008,LimMas2018,CazCorBig_2021}.

To investigate the developed nonholonomic models, we study the path-following control problem by utilizing these models. In their original form, the models describe the absolute position and orientation of the vehicle in an Earth-fixed frame. For path following, however, it is beneficial to use the relative position and orientation with respect to the given path. This can enable the design of analytical path-following controllers for any given path, rather than limiting to straight or circular paths. Therefore, the second major contribution of this paper is to derive a nonlinear transformation analytically, which can be used to transform the vehicle dynamics from absolute coordinates to relative coordinates. This transformation can be applied to any vehicle model (even to those with tires), and here we apply it to the developed nonholonomic vehicle models.

We use the transformed vehicle models to design path-following controllers. In the literature, many sophisticated, high-complexity controllers are available for path-following \cite{Borrelli_IJVAS_2005,Falcone_VSD_2008,BerQuiUnoCai2020,Lietal2020,Rossetter_2006,Talvala_Gerdes_2011,Choi_Hedrick_2015_ECC,Andersen_AIM_2016,XuPenTan_2021,Bae2020,Ave2021}. However, these rarely meet all the expectations of the automotive industry simultaneously, like low computational cost, high maneuverability, increased comfort, and enhanced safety. In particular, controllers based on models with tires require a great effort in parameter identification, whereas the uncertainties and model errors make them only capable of executing maneuvers in restricted scenarios. Consequently, automated vehicles, which are capable of following straight paths and circles, often perform poorly when the curvature changes abruptly or when disturbances occur. This can lead to reduced maneuverability, discomfort, and safety hazards.

The third major contribution of this paper is to create a novel low-complexity nonlinear path-following controller based on the backbone dynamics of nonholonomic models derived by the Appellian approach. We construct a controller by integrating  a nonlinear feedforward controller revealed by the transformed dynamics and a nonlinear feedback controller that is able to handle both small and large errors in lateral deviation and relative yaw angle. We investigate the stability of the controller analytically, and demonstrate via numerical simulations that it enables vehicles to follow different paths with high precision. Such controllers can enable automated vehicles to
execute a large variety of maneuvers without compromising comfort or safety, which will play an essential role as we move toward higher levels of automation.

We start the rest of the paper by discussing constraints in mechanical systems as well as the Newtonian, Lagrangian, and Appellian modeling approaches in Section~\ref{sec:review_model_tech}. The latter one is utilized in the subsequent sections to develop the models for automobiles. Readers who are familiar with these modeling approaches may decide to skip this section. In Section~\ref{sec:bicycle_model}, we describe our modeling assumptions used for the single track models developed in the paper. We categorize the models based on how the wheel-ground contact is modeled (rigid wheels vs skates), whether the longitudinal speed is restricted or the vehicle is driven by forces/torques, and whether the steering angle is assigned or a steering torque is applied. The models with skates are described in detail in Section~\ref{sec:model_skate}, while the models with rigid wheels are discussed in Section~\ref{sec:model_wheels}. In both cases the models are given at the beginning of the sections followed by the detailed derivations in subsections. These derivations may be skipped by the reader depending on his/her interests. In Section~\ref{sec:discussion} we discuss the roles of singularities, the calculation of nonholonomic constraining forces, and  present an analytic method to reformulate the models using path coordinates. These equations are used in control design when the vehicle is intended to follow a given path in Section~\ref{sec:control}. We conclude the paper in Section~\ref{sec:conclusion} where we also point out some future research directions.

\section{Analytical Mechanics for Nonholonomic Systems \label{sec:review_model_tech}}

In this section, we briefly review the concepts involved in nonholonomic systems and the related modeling techniques. The reader may refer to
\cite{Gantmacher_1970,NeiFuf1972,Kane_1985,Baruh_1999,Papastavridis_2002,Greenwood_2003,Koon_Marsden_1997,OstAng1998,Bloch_2003,deSapio_2017} for more details on dynamics of nonholonomic systems. We start with defining constraints and degrees of freedom, and then review the Newtonian, the Lagrangian and the Appellian approaches. For the sake of simplicity, the derivations are carried out for systems of particles and we provide the necessary formulas to allow the reader to generalize the calculations for rigid bodies.

\subsection{Constraints and degrees of freedom\label{sec:constraints}}

\begin{figure}[!t]
\begin{center}
\setlength{\unitlength}{0.012500in}%
\includegraphics[scale=1]{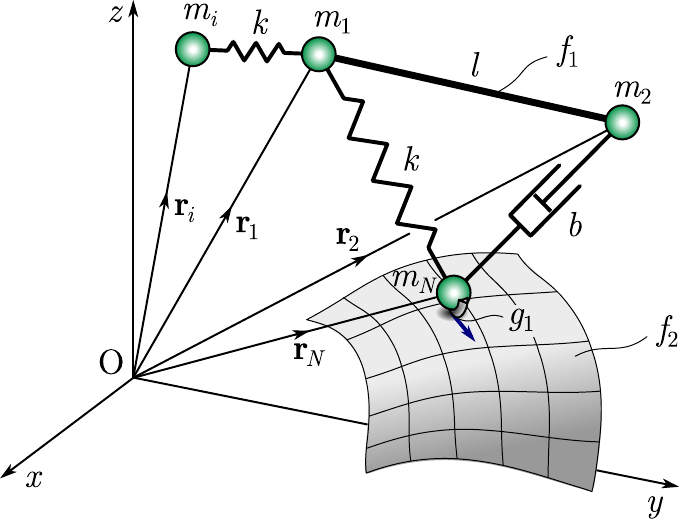}
\end{center}
\caption{Dynamical system of $N$ particles with constraints. A rod maintains the distance $l$ between particles $m_1$ and $m_2$, which is described by the constraint $f_1$. The particle $m_N$ is constrained to a surface while its velocity is directed by the blade of a skate sliding on the surface; the corresponding constraining equations are given by $f_2$  and $g_1$, respectively. \label{fig:Nparticles}}
\end{figure}

Let us consider a system of $N$ particles of mass $m_i$, $i=1,\ldots,N$ as shown in Fig.~\ref{fig:Nparticles}. Without constraints this system has $3N$ degrees of freedom, that is, it requires $3N$ scalar coordinates to unambiguously describe the system. The corresponding Newtonian equations of motion can be formulated as $3N$ second order ordinary differential equations, or equivalently, $6N$ first order ordinary differential equations. In particular, one may use the position vectors $\mathbf{r}_i$, $i=1,\ldots,N$ of the particles to describe their motion uniquely.

Now consider that the system is subject to $g$ geometric (also called holonomic) constraints of the form
\begin{equation}\label{eqn:geom_const}
f_\alpha(\mathbf{r}_i, t) = 0\ , \quad \alpha=1,\ldots,g\ ,
\end{equation}
see the examples $f_1$ and $f_2$ in Fig.~\ref{fig:Nparticles}.
Here we use a simplified notation so that $\mathbf{r}_i$ stands for $\mathbf{r}_1,\ldots,\mathbf{r}_N$ representing that each constraint may depend on the position vectors of all particles as well as on the time $t$ explicitly. This notation is implemented in the rest of this section in order to keep the complexity of formulas manageable. For example, $w_\beta(\mathbf{r}_j, t)$ means that $l_\beta$ may depend on $\mathbf{r}_1,\ldots,\mathbf{r}_N$.
Assume that apart from the geometric constraints, we also have $h$ kinematic (also called nonholonomic) constraints of the form
\begin{equation}\label{eqn:kinem_const}
g_\beta(\mathbf{r}_i,\mathbf{\dot{r}}_i, t) = 0\ , \quad \beta=1,\ldots,h\ ,
\end{equation}
where the dot represents derivative with respect to time $t$; see the example $g_1$ in Fig.~\ref{fig:Nparticles}.

We show that the geometric constraints can be eliminated using generalized coordinates, while the kinematic constrains can be handled using pseudo velocities, although the equations \eqref{eqn:kinem_const} shall be kept as part of the equations of motion. Consequently, one can define the degrees of freedom of the system as ${3N-g-h/2}$  corresponding to the ${6N-2g-h}$ first order ordinary differential equations that govern the motion of the system; see also \cite{Hu_2018}.

In order to simplify the matter, we only consider kinematic constraints that are affine functions of velocities $\mathbf{\dot{r}}_i$:
\begin{equation}\label{eqn:kinem_const_lin}
\sum_{i=1}^N \mathbf{w}_{\beta i}(\mathbf{r}_j,t) \cdot \mathbf{\dot{r}}_i + w_{\beta}(\mathbf{r}_j,t) = 0\ , \quad \beta=1,\ldots,h\ ,
\end{equation}
where $\cdot$ denotes the dot product of vectors. In what follows, all the sliding and/or rolling constraints can be written in the form (\ref{eqn:kinem_const_lin}) which seems to be quite generic in classical multi-body systems.

\begin{figure}[!ht]
\begin{center}
\setlength{\unitlength}{0.012500in}%
\includegraphics[scale=1]{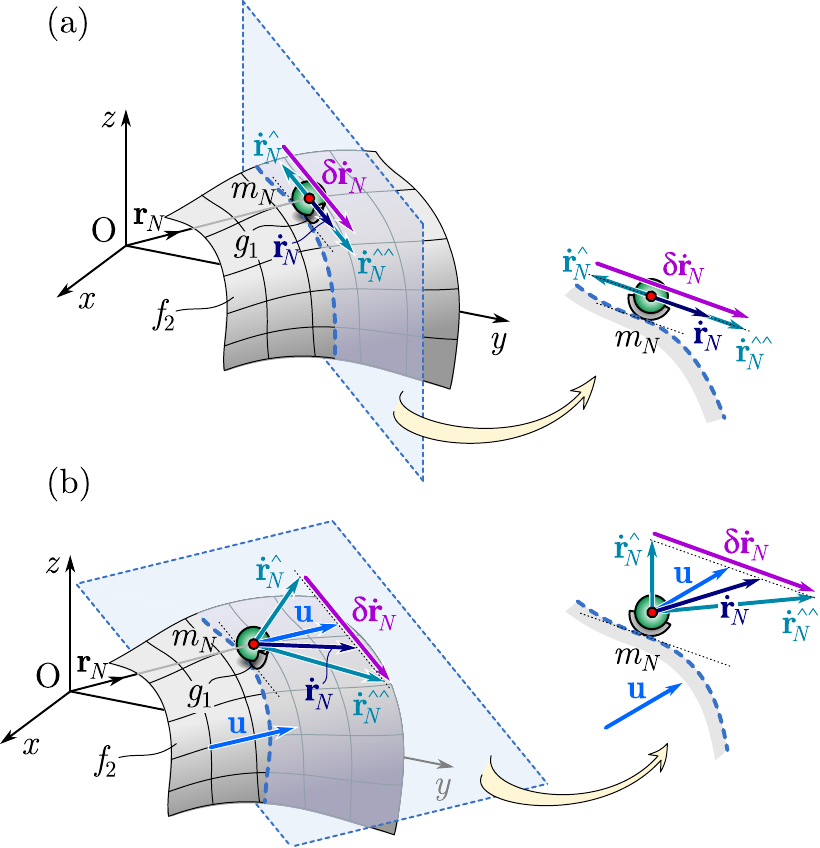}
\end{center}
\caption{Actual, admissible and virtual velocities. (a) Skating on a stationary surface. (b) Skating on a non-stationary surface that translates with velocity $\mathbf{u}$. In both cases the virtual velocity is tangential to the surface. \label{fig:virtualvelocity}}
\end{figure}

Before trying to eliminate the constraints, we list some definitions. As already indicated above, the actual velocity of particle $i$ is denoted by $\mathbf{\dot{r}}_i$; this will be the solution of the equations of motion once they are constructed. The set of velocities that satisfy the constraints above (but may not satisfy the equations of motion) are called admissible velocities and they are denoted by ${\{ \mathbf{{\dot{r}}}^{\wedge}_i, \mathbf{{\dot{r}}}^{\wedge\wedge}_i, \mathbf{{\dot{r}}}^{\wedge\wedge\wedge}_i, \ldots \}}$. One of these velocities is, in fact, the actual velocity. Finally, the virtual velocity is defined as the difference of two admissible velocities, e.g.,
${\delta \mathbf{\dot{r}}_i = \mathbf{{\dot{r}}}^{\wedge}_i - \mathbf{{\dot{r}}}^{\wedge\wedge}_i}$; see illustrations in Fig.~\ref{fig:virtualvelocity}. Notice that despite $\delta$ often refers to variations, the virtual velocities are not infinitesimal quantities.

Using the definitions above, one may reformulate \eqref{eqn:kinem_const_lin} as follows. Since the admissible velocities satisfy the constraints \eqref{eqn:kinem_const_lin}, the velocities $\mathbf{{\dot{r}}}^{\wedge}_i$  and $\mathbf{{\dot{r}}}^{\wedge\wedge}_i$ may be substituted there instead of $\mathbf{\dot{r}}_i$. Then taking the difference of the two leads to
\begin{equation}\label{eqn:kinem_const_lin_virt}
\sum_{i=1}^N \mathbf{w}_{\beta i}(\mathbf{r}_j,t) \cdot \delta \mathbf{\dot{r}}_i = 0\ , \quad \beta=1,\ldots,h\ ,
\end{equation}
that are linear in the virtual velocities $\delta \mathbf{\dot{r}}_i$.

\subsection{Newton's 2nd Law and Jourdain's Principle}

The power of the force $\mathbf{F}_i$ acting on particle $i$ is defined as $P_i = \mathbf{F}_i \cdot  \mathbf{\dot{r}}_i$. Based on the definitions above, the virtual power can also be defined as $\delta P_i = \mathbf{F}_i \cdot \delta \mathbf{\dot{r}}_i$. Then, the ideal constraints are defined by requiring the virtual power of the corresponding constraining forces $\mathbf{K}_i$ to be zero:
\begin{equation}\label{eqn:ideal_constraint}
\sum_{i=1}^N \mathbf{K}_i \cdot \delta \mathbf{\dot{r}}_i = 0\ .
\end{equation}
Here we only consider ideal constraints, and the forces that do not satisfy  \eqref{eqn:ideal_constraint} are called active forces and denoted by $\mathbf{F}_i$.

Thus, Newton's 2nd Law can be written in the form
\begin{equation}\label{eqn:Newton2}
m_i  \mathbf{\ddot{r}}_i = \mathbf{F}_i + \mathbf{K}_i\ , \quad i=1,\ldots,N\ ,
\end{equation}
where the active and the constraining forces are separated. By multiplying this equation with the corresponding virtual velocity $\delta \mathbf{\dot{r}}_i$, and summing them for all particles, we obtain Jourdain's principle
\begin{equation}\label{eqn:Jourdain}
\sum_{i=1}^N  \left( m_i  \mathbf{\ddot{r}}_i - \mathbf{F}_i \right) \cdot \delta \mathbf{\dot{r}}_i =0\ ,
\end{equation}
where \eqref{eqn:ideal_constraint} is utilized. Similar result is obtained by applying D'Alembert's principle with virtual displacements instead of virtual velocities, which is widely used in statics \cite{deSapio_2017}. However, since virtual displacements are infinitesimal quantities, it is challenging to use them in dynamical problems \cite{Antman_1992}.

\subsection{Lagrange Equations of the 2nd Kind}\label{sec:Lagrange}

The Lagrangian approach can be used to eliminate the $g$ geometric constraints \eqref{eqn:geom_const} and the corresponding geometric constraining forces from the governing equations. In order to do this, ${3N-g}$ so-called generalized coordinates ${q_k, k=1,\ldots,3N-g}$ have to be selected intuitively. While the chosen definitions
\begin{equation}\label{eqn:general_definition}
q_k:=H_k(\mathbf{r}_i,t)\,,\quad k=1,\ldots,3N-g\ ,
\end{equation}
are optional, it can be considered appropriate if these generalized coordinates, together with the geometric constraints \eqref{eqn:geom_const}, provide an unambiguous description of the dynamical system. In mathematical terms, this means that the system of $3N$ nonlinear algebraic equations
\begin{equation}\label{eqn:unambiguous_description}
\begin{split}
H_k(\mathbf{r}_i,t)&=q_k\,,\quad  k =1,\ldots,3N-g\ ,
\\
f_\alpha(\mathbf{r}_i, t)&= 0\,,\ \quad  \alpha =1,\ldots,g\ ,
\end{split}
\end{equation}
(cf.~\eqref{eqn:geom_const} and \eqref{eqn:general_definition}) has a unique solution for the $3N$ unknown coordinates of the position vectors $\mathbf{r}_i\,,\ i=1,\ldots,N$. Consequently, if the generalized coordinate selection is appropriate, a unique solution
\begin{equation}\label{eqn:unique_solution}
\mathbf{r}_i(q_k,t)\,, \quad i=1,\ldots,N\ ,
\end{equation}
exists, where again, we use the abbreviated notation that $q_k$ represents ${q_1,\ldots,q_{3N-g}}$. We remark that the explicit time dependence of \eqref{eqn:general_definition}, \eqref{eqn:unambiguous_description} and \eqref{eqn:unique_solution} originates in the fact that, in general, the constraints \eqref{eqn:geom_const}, \eqref{eqn:kinem_const} and \eqref{eqn:kinem_const_lin} can be time dependent.

Taking the time derivative of \eqref{eqn:unique_solution} allows us to express the velocities $\mathbf{\dot{r}}_i$ as an affine function of the generalized velocities $\dot{q}_k$:
\begin{equation}\label{eqn:velocity_transform}
\mathbf{\dot{r}}_i = \sum_{k=1}^{3N-g} \frac{\partial \mathbf{r}_i (q_{\ell},t)} {\partial q_k} \dot{q}_k + \frac{\partial \mathbf{r}_i (q_{\ell},t)} {\partial t}\ , \quad i=1,\ldots,N\ .
\end{equation}
Similarly to \eqref{eqn:kinem_const_lin} and \eqref{eqn:kinem_const_lin_virt}, this can be rewritten for the virtual velocities ${\delta \mathbf{\dot{r}}_i = \mathbf{\dot{r}}^{\wedge}_i - \mathbf{\dot{r}}^{\wedge\wedge}_i}$ by means of the virtual generalized velocities ${\delta \dot{q}_k = \dot{q}^{\wedge}_k - \dot{q}^{\wedge\wedge}_k}$ as
\begin{equation}\label{eqn:velocity_transform_virt}
\delta \mathbf{\dot{r}}_i = \sum_{k=1}^{3N-g} \frac{\partial \mathbf{r}_i (q_{\ell},t)} {\partial q_k} \delta \dot{q}_k\ , \quad i=1,\ldots,N\,.
\end{equation}

Substituting \eqref{eqn:velocity_transform} into \eqref{eqn:kinem_const_lin} yields the kinematic constraints in the form
\begin{equation}\label{eqn:kinem_const_gen}
\sum_{k=1}^{3N-g} A_{\beta k}(q_{\ell},t)\, \dot{q}_k + A_{\beta}(q_{\ell},t) = 0\ , \quad \beta=1,\ldots,h\ ,
\end{equation}
that are expressed with respect to the generalized velocities as
\begin{equation}\label{eqn:kinem_const_gen_coeff}
\begin{split}
A_{\beta k}(q_{\ell},t) &= \sum_{i=1}^N \mathbf{w}_{\beta i}\big(\mathbf{r}_j(q_{\ell},t),t\big) \cdot \frac{\partial \mathbf{r}_i (q_{\ell},t)} {\partial q_k}\ ,
\\
 A_{\beta}(q_{\ell},t) &= \sum_{i=1}^N \mathbf{w}_{\beta i}\big(\mathbf{r}_j(q_{\ell},t),t\big) \cdot \frac{\partial \mathbf{r}_i (q_{\ell},t)} {\partial t}  \\
&+ w_{\beta}\big(\mathbf{r}_j(q_{\ell},t),t\big)\ .
\end{split}
\end{equation}
Substituting \eqref{eqn:velocity_transform_virt} into \eqref{eqn:kinem_const_lin_virt} yields the kinematic constraints in the form
\begin{equation}\label{eqn:kinem_const_gen_virt}
\sum_{k=1}^{3N-g} A_{\beta k}(q_{\ell},t)\, \delta \dot{q}_k = 0\ , \quad \beta=1,\ldots,h\ ,
\end{equation}
expressed with the virtual generalized velocities.

In order to derive the equations of motion in terms of the generalized coordinates $q_k$, we substitute \eqref{eqn:velocity_transform_virt} into Jourdain's principle \eqref{eqn:Jourdain}; this leads to
\begin{equation}\label{eqn:Jourdain_gen}
\sum_{k=1}^{3N-g} \left( \sum_{i=1}^N  m_i  \mathbf{\ddot{r}}_i \cdot \frac{\partial \mathbf{r}_i}{\partial q_k} - \sum_{i=1}^N  \mathbf{F}_i \cdot \frac{\partial \mathbf{r}_i}{\partial q_k} \right) \delta \dot{q}_k =0\ .
\end{equation}
Applying the relationship $\partial \mathbf{\dot{r}}_i / \partial \dot{q}_k = \partial \mathbf{r}_i / \partial q_k$ (cf.~\eqref{eqn:velocity_transform}) and the chain rule, one can  reformulate \eqref{eqn:Jourdain_gen} as
\begin{equation}\label{eqn:Jourdain_gen2}
\sum_{k=1}^{3N-g} \left( \frac{\textrm{d}}{\textrm{d}t}\frac{\partial T}{\partial \dot{q}_k}-\frac{\partial T}{\partial q_k} - Q_k \right) \delta \dot{q}_k =0\ ,
\end{equation}
where
\begin{equation}\label{eqn:kinetic_energy}
\begin{split}
T &= \frac{1}{2}  \sum_{i=1}^N  m_i\, \mathbf{\dot{r}}_i^2
\\
  &= \frac{1}{2} \sum_{j,\,k=1}^{3N-g} \left(\sum_{i=1}^N  m_i \frac{\partial  \mathbf{r}_i}{\partial q_j} \cdot \frac{\partial  \mathbf{r}_i}{\partial q_k} \right) \dot{q}_j \dot{q}_k
\\
  &+ \sum_{k=1}^{3N-g} \left(\sum_{i=1}^N  m_i \frac{\partial  \mathbf{r}_i}{\partial q_k} \cdot \frac{\partial \mathbf{r}_i}{\partial t} \right) \dot{q}_k
  + \frac{1}{2} \sum_{i=1}^N  m_i \left(\frac{\partial \mathbf{r}_i}{\partial t} \right)^2\ ,
\end{split}
\end{equation}
is the kinetic energy of the system, while
the generalized forces are introduced by the definition:
\begin{equation}\label{eqn:generalized_force}
Q_k := \sum_{i=1}^N  \mathbf{F}_i \cdot \frac{\partial \mathbf{r}_i}{\partial q_k}\ , \quad k=1,\ldots,3N-g\ .
\end{equation}

If the virtual generalized velocities $\delta \dot{q}_k$ were independent, one could equate the expression in the bracket in \eqref{eqn:Jourdain_gen2} to zero. Note, however that these quantities cannot be chosen independently since they must satisfy the kinematic constraints \eqref{eqn:kinem_const_gen_virt}. In order to resolve this issue, we introduce $h$ Lagrange multipliers ${\lambda_{\beta}, \beta = 1,\ldots, h}$, one for each kinematic constraint equation in \eqref{eqn:kinem_const_gen_virt}, summarize them and add the sum to \eqref{eqn:Jourdain_gen2}. This yields
\begin{equation}\label{eqn:Jourdain_gen3}
\sum_{k=1}^{3N-g} \left( \frac{\textrm{d}}{\textrm{d}t}\frac{\partial T}{\partial \dot{q}_k}-\frac{\partial T}{\partial q_k} - Q_k - \sum_{\beta=1}^{h}\lambda_{\beta}A_{\beta k} \right) \delta \dot{q}_k =0\ .
\end{equation}
Now, equating the expressions in the parentheses to zero and recalling the kinematic constraints \eqref{eqn:kinem_const_gen_virt}, we obtain the Lagrange equations of 2nd kind generalized for nonholonomic systems in the form:
\begin{equation}\label{eq:RouthVoss}
\begin{split}
&\frac{\textrm{d}}{\textrm{d}t}\frac{\partial T}{\partial \dot{q}_k}-\frac{\partial T}{\partial q_k} = Q_k+\sum_{\beta=1}^{h}\lambda_{\beta}A_{\beta k}\ ,\quad  k=1,\ldots,
3N-g\,,
\\
&\sum_{k=1}^{3N-g} A_{\beta k}\dot{q}_k+A_{\beta} = 0\ , \quad  \beta=1,\ldots,h\ .
\end{split}
\end{equation}
This is a set of ${3N-g+h}$ algebraic differential equations that needs to be solved for the ${3N-g+h}$ unknown time histories of the generalized coordinates ${q}_k(t)$, ${k=1,\ldots,3N-g}$ and the magnitudes $\lambda_{\beta}(t)$, $\beta = 1,\ldots, h$ of the generalized constraining forces that ensure the kinematic constraints to be satisfied.

Since the Lagrange multipliers ${\lambda_{\beta},\ \beta = 1,\ldots, h}$ appear linearly in \eqref{eq:RouthVoss}, they can be eliminated by algebraic manipulations, to obtain ${3N-g-h}$ second order ordinary differential equations that are still augmented with $h$ first order ordinary differential equations (the kinematic constraints). This is equivalent to having ${2(3N-g-h)+h = 6N-2g-h}$ first order ordinary differential equations, that is, ${3N -g- h/2}$ degrees of freedom. In this interpretation, each geometric (holonomic) constraint reduces the number of degrees of freedom by one, while each kinematic (nonholonomic) constraint reduces the number of degrees of freedom by one half.

The generalized forces \eqref{eqn:generalized_force} may be calculated by noticing that the virtual power of the active forces is the same as that of the generalized forces, that is, using \eqref{eqn:velocity_transform_virt} yields
\begin{equation}\label{eqn:virtual_power}
\delta P = \sum_{i=1}^N  \mathbf{F}_i \cdot \delta \mathbf{\dot{r}}_i
=  \sum_{k=1}^{3N-g}  \sum_{i=1}^N  \mathbf{F}_i \cdot \frac{\partial \mathbf{r}_i}{\partial q_k} \delta \dot{q}_k = \sum_{k=1}^{3N-g} Q_k  \delta \dot{q}_k\ .
\end{equation}

\begin{figure}[!t]
\begin{center}
\includegraphics[scale=1]{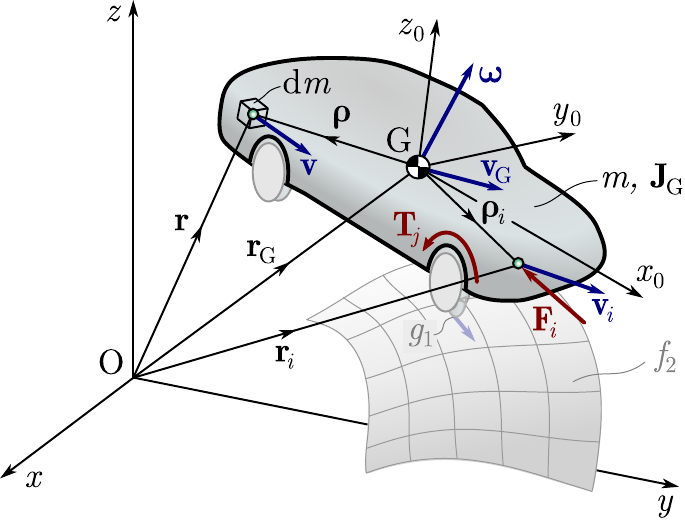}
\end{center}
\caption{Derivation of the kinetic energy and the virtual power of active forces acting on rigid bodies.  \label{fig:rigidbody1}}
\end{figure}

Finally, as mentioned above, the theory also works for rigid bodies. In that case, for each rigid body in the system, the kinetic energy \eqref{eqn:kinetic_energy} has to be calculated. To do this, one should sum (integrate) the kinetic energy of each particle of the rigid body, that is,
\begin{equation}
T = \frac{1}{2} \int_{(m)} \mathbf{v}^2\, \mathrm{d}m\,,
\end{equation}
where $\mathbf{v}\equiv\dot{\mathbf{r}}$ refers to the velocity of a particle of the rigid body (see Fig.~\ref{fig:rigidbody1}), which can be calculated as
\begin{equation}
\mathbf{v} =  \mathbf{v}_{\rm G} + \boldsymbol{\omega}\times  \boldsymbol{\rho} \,,
\end{equation}
where $\mathbf{v}_{\rm G}$ is the velocity of the center of mass G (for which $\int_{(m)} \boldsymbol{\rho}\, \mathrm{d}m = \mathbf{0}$), $\boldsymbol{\omega}$ is the angular  velocity vector of the body. The position vector $\boldsymbol{\rho}$ points from the center of gravity G to the particle, and $\times$ denotes the cross product of vectors. Thus,
\begin{equation}\label{eq:kineticRB}
\begin{split}
T =& \frac{1}{2} \int_{(m)} (\mathbf{v}_{\rm G} + \boldsymbol{\omega}\times  \boldsymbol{\rho})^2\, \mathrm{d}m\,
\\
=& \frac{1}{2} \int_{(m)} \mathbf{v}^2_{\rm G} \,\mathrm{d}m
+  \int_{(m)} \mathbf{v}_{\rm G} \cdot (\boldsymbol{\omega}\times  \boldsymbol{\rho}) \,\mathrm{d}m
\\
&+ \frac{1}{2} \int_{(m)} \underbrace{(\boldsymbol{\omega}\times  \boldsymbol{\rho})\cdot(\boldsymbol{\omega}\times  \boldsymbol{\rho})}_{=\boldsymbol{\omega}\cdot(\boldsymbol{\rho}\times(\boldsymbol{\omega}\times\boldsymbol{\rho}))} \,\mathrm{d}m
\\
=& \frac{1}{2} \underbrace{\int_{(m)} 1 \,\mathrm{d}m}_{=m} \mathbf{v}^2_{\rm G}
+  \mathbf{v}_{\rm G} \cdot \bigl(\boldsymbol{\omega}\times  \underbrace{\int_{(m)} \boldsymbol{\rho}\,\mathrm{d}m}_{=\mathbf{0}} \bigr)
\\
&+ \frac{1}{2} \boldsymbol{\omega}\cdot \int_{(m)} \underbrace{\boldsymbol{\rho}\times(\boldsymbol{\omega}\times\boldsymbol{\rho})}_{=\boldsymbol{\rho}^2 \boldsymbol{\omega}-(\boldsymbol{\omega}\cdot\boldsymbol{\rho})\boldsymbol{\rho} } \,\mathrm{d}m
\\
=&  \frac{1}{2} m \mathbf{v}^2_{\rm G}
+ \frac{1}{2} \boldsymbol{\omega}\cdot \underbrace{\int_{(m)}( \boldsymbol{\rho}^2\mathbf{I} -\boldsymbol{\rho}\otimes\boldsymbol{\rho} )\,\mathrm{d}m}_{=\mathbf{J}_{\rm G}} \, \boldsymbol{\omega}\,,
\end{split}
\end{equation}
where $m$ is the mass of the body, $\otimes$ is the diadic product, and $\mathbf{J}_{\rm G}$ is the mass moment of inertia tensor about the center of mass G. Hence, the kinetic energy becomes
\begin{equation}\label{eqn:kinetic_energy_ridig_body}
T = \frac{1}{2} m \mathbf{v}_{\rm G}^2 + \frac{1}{2} \boldsymbol{\omega} \cdot \mathbf{J}_{\rm G} \boldsymbol{\omega}\ .
\end{equation}

Moreover, the virtual power \eqref{eqn:virtual_power} of the active force system acting on each rigid body can be calculated via the summation of the virtual powers of each active force $\mathbf{F}_i$ that acts on the $i$-th particle of the rigid body. In addition, the torques $\mathbf{T}_j$ acting on rigid bodies also have to be considered, namely:
\begin{equation}\label{eqn:virtual_power_rigid_body_org}
\begin{split}
\delta P =& \sum_i \mathbf{F}_i \cdot \delta \mathbf{v}_{i}+ \sum_j \mathbf{T}_j \cdot \delta \boldsymbol{\omega}\\
=&  \sum_i \mathbf{F}_i \cdot \delta (\mathbf{v}_{\rm G}+\boldsymbol{\omega}\times \boldsymbol{\rho}_i) + \sum_j \mathbf{T}_j\cdot \delta \boldsymbol{\omega} \\
=& \underbrace{\bigl( \sum_i \mathbf{F}_i \bigr)}_{=\mathbf{F}} \cdot \delta \mathbf{v}_{\rm G} + \underbrace{\bigl(\sum_i \boldsymbol{\rho}_i \times \mathbf{F}_i + \sum_j \mathbf{T}_j\bigr)}_{=\mathbf{M}_{\rm G}} \cdot \delta\boldsymbol{\omega}\,.
\end{split}
\end{equation}
So, the virtual power can be calculated
\begin{equation}\label{eqn:virtual_power_rigid_body}
\delta P = \mathbf{F} \cdot \delta \mathbf{v}_{\rm G} + \mathbf{M}_{\rm G} \cdot \delta \boldsymbol{\omega}\ ,
\end{equation}
where $\mathbf{F}$ is the resultant force, while $\mathbf{M}_{\rm G}$ is the resultant torque about the center of mass G.

\subsection{Appell Equations}\label{sec:Appell}

The Appellian approach allows one to eliminate the kinematic (nonholonomic) constraints by selecting
intuitively ${n=3N-g-h}$ so-called pseudo velocities $\sigma_j$, ${j = 1,\ldots,n}$. These have to be  defined appropriately as the linear combinations of the generalized velocities $\dot{q}_k$, ${k = 1,\ldots,3N-g}$, such that
\begin{equation}\label{eqn:pseudo_velocities}
\sigma_j := \sum_{k=1}^{3N-g} B_{jk}(q_{\ell},t)\, \dot{q}_k  \ , \quad  j=1,\ldots,n\ .
\end{equation}

Similarly to the requirements for the selection of the generalized coordinates in \eqref{eqn:general_definition}, the otherwise optional functions $B_{jk}$ must be selected in a way that the defined pseudo velocities $\sigma_j$ provide an unambiguous description of the system dynamics. Accordingly,
the definitions of the pseudo velocities \eqref{eqn:pseudo_velocities} together with the kinematic constraints \eqref{eqn:kinem_const_gen} constitute a ${(3N-g)}$-dimensional system of linear algebraic equations with respect to the generalized velocities. This can be written in the form
\begin{equation}\label{eqn:pseudo_and_kinconstraints}
    \underbrace{\begin{bmatrix}
      A_{11} & \dots & A_{1(3N-g)}
      \\
      \vdots & \ddots & \vdots
      \\
      A_{h1} & \dots & A_{h(3N-g)}
      \\
      B_{11} & \dots & B_{1(3N-g)}
      \\
      \vdots & \ddots & \vdots
      \\
      B_{n1} & \dots & B_{n(3N-g)}
    \end{bmatrix}}_{=:\mathbf{\mathbf{C}}}
    \begin{bmatrix}
      \dot{q}_1
      \\
      \vdots
      \\
      \dot{q}_{3N-g}
    \end{bmatrix}
    =
    \begin{bmatrix}
      - A_{1}
      \\
      \vdots
      \\
      - A_{h}
      \\
      \sigma_1
      \\
      \vdots
      \\
      \sigma_{n}
    \end{bmatrix}\ ,
\end{equation}
which must have a unique solution for the generalized velocities $\dot q_k$.
Consequently, the pseudo velocities have to be defined appropriately, that is, the coefficients $B_{jk}$ in \eqref{eqn:pseudo_velocities} must be selected in a way that the coefficient matrix $\mathbf{C}$ in \eqref{eqn:pseudo_and_kinconstraints} is not singular in the configuration space of the generalized coordinates $q_\ell$ at any time:
\begin{equation}\label{eqn:nonsingular}
{\rm det}\bigl(\mathbf{C}(q_\ell ,t)\bigr) \neq 0\ .
\end{equation}

In general, if $\mathbf{C}$ is not singular then the generalized velocities can be expressed as a unique function of the pseudo velocities, generalized coordinates and time:
\begin{equation}\label{eqn:intro:dqk}
\dot{q}_k = \sum_{j=1}^{n} f_{kj}(q_{\ell},t)\,\sigma_j + f_k(q_{\ell},t)\ , \quad k=1,\ldots,3N-g\ ,
\end{equation}
similarly to the generalized coordinates expressed as unique functions of the system position vectors in \eqref{eqn:unique_solution}.

For the single track vehicle models presented in this paper, we will discuss different choices of the pseudo velocities and the corresponding possible singular points of the configuration spaces in Section~\ref{sec:singularities}.

Substituting \eqref{eqn:intro:dqk} into \eqref{eqn:velocity_transform}, we can express the velocities with the pseudo velocities as
\begin{equation}\label{eqn:velocity_transform2}
\mathbf{\dot{r}}_i = \sum_{j=1}^{n} \mathbf{d}_{ij} (q_{\ell},t)\, \sigma_j + \mathbf{d}_i (q_{\ell},t)\ ,
\quad i=1,\ldots,N\, ,
\end{equation}
where
\begin{equation}
\begin{split}
\mathbf{d}_{ij} (q_{\ell},t)
&= \sum_{k=1}^{3N-g} \frac{\partial \mathbf{r}_i (q_{\ell},t)} {\partial q_k} f_{kj}(q_{\ell},t)\ ,
\\
\mathbf{d}_i (q_{\ell},t)
&= \sum_{k=1}^{3N-g} \frac{\partial \mathbf{r}_i (q_{\ell},t)} {\partial q_k} f_k(q_{\ell},t) + \frac{\partial \mathbf{r}_i (q_{\ell},t)} {\partial t}\ .
\end{split}
\end{equation}
In a similar way as \eqref{eqn:velocity_transform_virt} is derived from \eqref{eqn:velocity_transform}, one can obtain
\begin{equation}\label{eqn:velocity_transform2_virt}
\delta \mathbf{\dot{r}}_i = \sum_{j=1}^{n} \mathbf{d}_{ij} (q_{\ell},t)\, \delta \sigma_j \ ,
\quad i=1,\ldots,N\, ,
\end{equation}
from \eqref{eqn:velocity_transform2}. Also, differentiating \eqref{eqn:velocity_transform2} yields the acceleration
\begin{equation}\label{eqn:acceleration_transform}
\mathbf{\ddot{r}}_i = \sum_{j=1}^{n} \mathbf{d}_{ij} (q_{\ell},t)\, \dot{\sigma}_j + \ldots\ ,
\quad i=1,\ldots,N\, ,
\end{equation}
where $\ldots$ represent terms that contain only generalized coordinates $q_{\ell}$, pseudo velocities $\sigma_j$, and time $t$, but do not contain pseudo accelerations $\dot{\sigma}_j$. Observe that \eqref{eqn:acceleration_transform} results in
\begin{equation}\label{eqn:acceleration_transform2}
\frac{\partial \mathbf{\ddot{r}}_i}{\partial \dot{\sigma}_j} = \mathbf{d}_{ij} (q_{\ell},t)\ .
\end{equation}

Recall Jourdain's principle in \eqref{eqn:Jourdain}, where the substitution of \eqref{eqn:velocity_transform2_virt} leads to
\begin{equation}\label{eqn:Jourdain_pse}
\sum_{j=1}^{n} \left( \sum_{i=1}^N  m_i  \mathbf{\ddot{r}}_i \cdot \mathbf{d}_{ij}  - \sum_{i=1}^N  \mathbf{F}_i \cdot \mathbf{d}_{ij}  \right) \delta \sigma_j =0\ .
\end{equation}
Using \eqref{eqn:acceleration_transform2} and the chain rule, one can  reformulate \eqref{eqn:Jourdain_pse} as
\begin{equation}\label{eqn:Jourdain_pse2}
\sum_{j=1}^{n} \left( \frac{\partial S}{\partial \dot{\sigma}_j} - \Pi_j \right) \delta \sigma_j =0\ ,
\end{equation}
where
\begin{equation}\label{eqn:aceeleration_energy}
S = \frac{1}{2}  \sum_{i=1}^N  m_i\, \mathbf{\ddot{r}}_i^2\ ,
\end{equation}
is the so-called acceleration energy (or Gibbs function) of the system, and the pseudo force $\Pi_j$ is defined by
\begin{equation}\label{eqn:pseudo_force}
\Pi_j = \sum_{i=1}^N  \mathbf{F}_i \cdot \mathbf{d}_{ij}\ ,\quad j=1,\ldots,n\ .
\end{equation}

Since the virtual pseudo velocities $\delta \sigma_j$ in \eqref{eqn:Jourdain_pse2} are not constrained, the parentheses can be equated to zero. These, together with \eqref{eqn:intro:dqk}, constitute the Appell equations:
\begin{equation}\label{eqn:Appell}
\begin{split}
&\frac{\partial S}{\partial \dot{\sigma}_j} = \Pi_j\ , \,\,\,\qquad\qquad j=1,\ldots,n \ ,
\\
&\dot{q}_k = \sum\limits^{n}_{j=1}f_{kj}\sigma_j+f_k \ , \quad k=1,\ldots,3N-g\ ,
\end{split}
\end{equation}
which is a system of ${n+3N-g = 6N-2g-h}$ first order ordinary differential equations for the ${n=3N-g-h}$ pseudo velocities $\sigma_j$ and the ${3N-g}$ generalized coordinates $q_k$ corresponding to the ${3N-g-h/2}$ degrees of freedom of the system.

Similarly to the derivation of \eqref{eqn:virtual_power}, the pseudo forces \eqref{eqn:pseudo_force} can be calculated through the virtual power, that is,
\begin{equation}\label{eqn:virtual_power2}
\delta P = \sum_{i=1}^N  \mathbf{F}_i \cdot \delta \mathbf{\dot{r}}_i
=  \sum_{k=1}^{3N-g} Q_k\delta \dot q_k = \sum_{j=1}^{n} \Pi_j  \delta \sigma_j\ .
\end{equation}

\begin{figure}[!t]
\begin{center}
\includegraphics[scale=1]{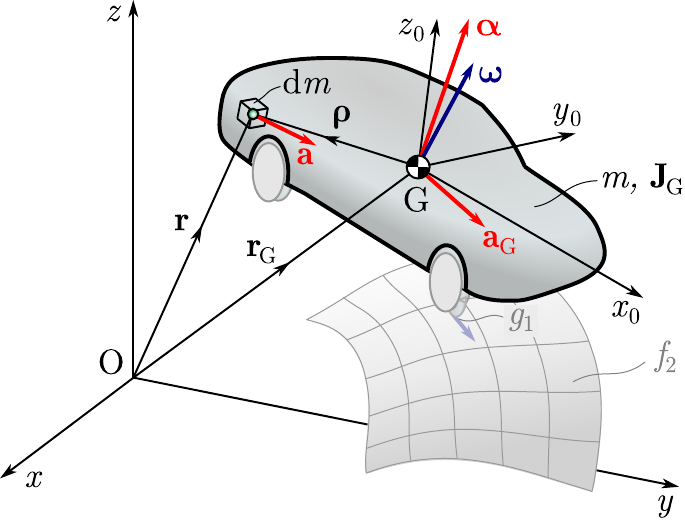}
\end{center}
\caption{Derivation of the acceleration energy for rigid bodies.  \label{fig:rigidbody2}}
\end{figure}

When the theory is applied for multi-body systems, the acceleration energy of each rigid body has to be calculated as
\begin{equation}
S = \frac{1}{2} \int_{(m)} \mathbf{a}^2\, \mathrm{d}m\,,
\end{equation}
where $\mathbf{a}\equiv\ddot{\mathbf{r}}$ refers to the acceleration of a particle of the rigid body; see Fig.~\ref{fig:rigidbody2}. Based on the rigid body kinematics, the acceleration of any particle can be calculated as
\begin{equation}
\mathbf{a} =  \mathbf{a}_{\rm G} +\boldsymbol{\alpha} \times \boldsymbol{\rho} + \boldsymbol{\omega}\times (\boldsymbol{\omega}\times  \boldsymbol{\rho}) \,,
\end{equation}
where $\mathbf{a}_{\rm G}$ is the acceleration of the center of mass G, $\boldsymbol{\alpha}$ is the angular acceleration vector of the body. Thus,
\begin{equation}
\begin{split}
S =& \frac{1}{2} \int_{(m)} \left(\mathbf{a}_{\rm G} +\boldsymbol{\alpha} \times \boldsymbol{\rho} + \boldsymbol{\omega}\times (\boldsymbol{\omega}\times  \boldsymbol{\rho})\right)^2\, \mathrm{d}m\,
\\
=& \frac{1}{2} \int_{(m)} \mathbf{a}^2_{\rm G} \, \mathrm{d}m + \frac{1}{2} \int_{(m)}\underbrace{(\boldsymbol{\alpha} \times \boldsymbol{\rho})\cdot(\boldsymbol{\alpha} \times \boldsymbol{\rho})}_{=\boldsymbol{\alpha}\cdot (\boldsymbol{\rho}\times(\boldsymbol{\alpha} \times \boldsymbol{\rho}))} \, \mathrm{d}m
\\
&+ \int_{(m)} \mathbf{a}_{\rm G}\cdot(\boldsymbol{\alpha} \times \boldsymbol{\rho}) \, \mathrm{d}m
\\
& +  \int_{(m)} \mathbf{a}_{\rm G} \cdot (\boldsymbol{\omega}\times (\boldsymbol{\omega}\times  \boldsymbol{\rho})) \, \mathrm{d}m
\\
& + \int_{(m)} \underbrace{(\boldsymbol{\alpha} \times \boldsymbol{\rho}) \cdot (\boldsymbol{\omega}\times (\boldsymbol{\omega}\times  \boldsymbol{\rho}))}_{=\boldsymbol{\alpha} \cdot (\boldsymbol{\rho} \times (\boldsymbol{\omega}\times (\boldsymbol{\omega}\times  \boldsymbol{\rho})))}  \, \mathrm{d}m
\\
& + \frac{1}{2} \int_{(m)} \underbrace{(\boldsymbol{\omega}\times (\boldsymbol{\omega}\times  \boldsymbol{\rho}))^2}_{\text{does not depend on } \dot{\sigma}_j} \, \mathrm{d}m
\\
=& \frac{1}{2} \underbrace{\int_{(m)} 1 \,\mathrm{d}m}_{=m} \mathbf{a}^2_{\rm G}
+ \frac{1}{2} \boldsymbol{\alpha} \cdot \underbrace{\int_{(m)} (\boldsymbol{\rho}^2 \mathbf{I} - \boldsymbol{\rho}\otimes\boldsymbol{\rho})\, \mathrm{d}m}_{=\mathbf{J}_{\rm G}} \,\boldsymbol{\alpha}
\\
&+ \mathbf{a}_{\rm G}\cdot\biggl(\boldsymbol{\alpha} \times \underbrace{\int_{(m)} \boldsymbol{\rho} \, \mathrm{d}m}_{=\mathbf{0}} \biggr)
\\
&+ \mathbf{a}_{\rm G}\cdot\biggl(\boldsymbol{\omega} \times \bigl(\boldsymbol{\omega}\times \underbrace{\int_{(m)} \boldsymbol{\rho} \, \mathrm{d}m}_{=\mathbf{0}} \bigr)\biggr)
\\
&+\boldsymbol{\alpha} \cdot \biggl( \boldsymbol{\omega} \times \underbrace{\int_{(m)} (\boldsymbol{\rho}^2 \mathbf{I} - \boldsymbol{\rho}\otimes\boldsymbol{\rho})\, \mathrm{d}m}_{=\mathbf{J}_{\rm G}}\, \boldsymbol{\omega}\biggr) + \ldots\,,
\end{split}
\end{equation}
where the same steps are used to extract the mass moment of inertia $\mathbf{J}_{\rm G}$ as in \eqref{eq:kineticRB}, and there is no need to calculate the additional terms referred to by $\ldots$ since they do not contain accelerations, and consequently, their derivatives are always zero with respect to the pseudo accelerations $\dot\sigma_j$ in the Appell equations \eqref{eqn:Appell}.

Finally, the acceleration energy of a rigid body can be calculated as
\begin{equation}
S = \frac{1}{2} m \mathbf{a}_{\rm G}^{2} + \frac{1}{2}\boldsymbol{\alpha} \cdot  \mathbf{J}_{\rm G}\boldsymbol{\alpha} + \boldsymbol{\alpha} \cdot (\boldsymbol{\omega} \times \mathbf{H}_{\rm G}) + \ldots
\end{equation}
where  $\mathbf{H}_{\rm G}=\mathbf{J}_\mathrm{G}\boldsymbol{\omega}$ is the angular momentum vector about the center of mass G, that is, the last term is a scalar triple product of the angular acceleration, the angular velocity, and the angular momentum vectors.

To calculate the right hand side of the Appell equations in case of multi-body system, one can calculate the virtual power of the active forces acting on the rigid body using \eqref{eqn:virtual_power_rigid_body} and identify the pseudo forces via \eqref{eqn:virtual_power2}.

\section{Single Track Models \label{sec:bicycle_model}}

In this section we describe the fundamental abstractions that are used to model the dynamics of automobiles. First, we introduce the so-called single track or bicycle model. Then we discuss different abstraction levels of the wheel, namely, rigid wheel and skate. Finally, we categorize the different models developed in this paper based on the wheel models and constraints considered. Note that the dynamics of real bicycles are in fact quite different and substantially more intricate as discussed, for example, in \cite{MeiPapRuiSch_2007,LimMas2018}.

\begin{figure}[!b]
\begin{center}
\setlength{\unitlength}{0.012500in}%
\includegraphics[scale=1]{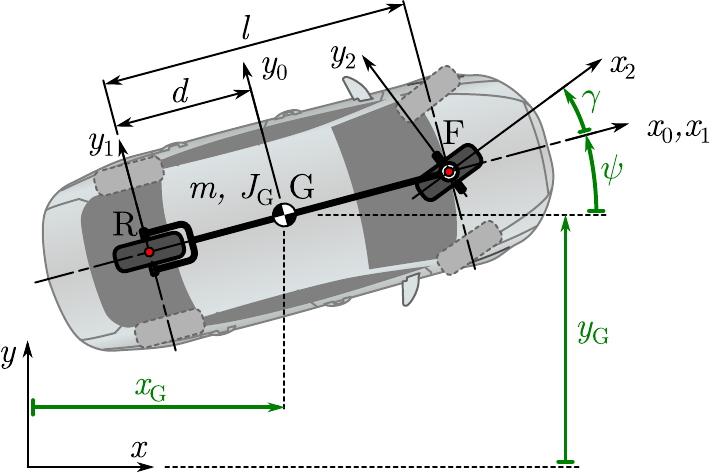}
\end{center}
\caption{Single track (also called bicycle) model of an automobile with geometry and coordinate frames indicated. \label{fig:bicycle}}
\end{figure}

\begin{figure}[!t]
\begin{center}
\includegraphics[scale=1]{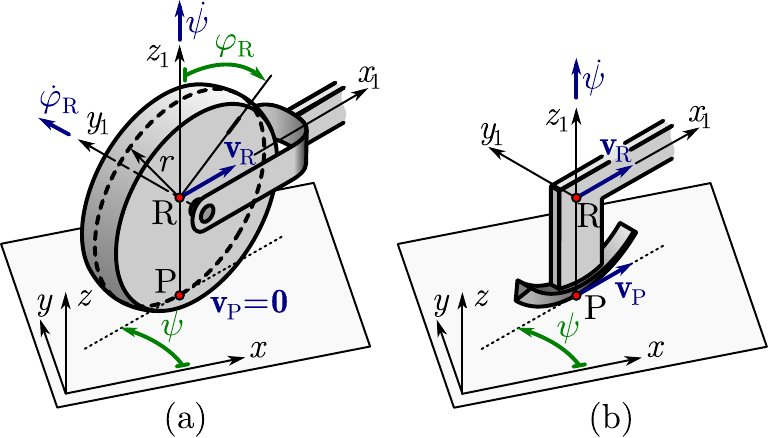}
\end{center}
\caption{Models of rigid rolling wheel (a) and skate (b) with geometry and kinematics indicated. \label{fig:wheel_skate}}
\end{figure}

In Fig.~\ref{fig:bicycle}, the dimmed part shows the top view of a four-wheeled, front-wheel-steered vehicle. By approximating the front wheel pair with a single front wheel and the rear wheel pair with a single rear wheel, we obtain an abstraction of the vehicle, called the single track or bicycle model \cite{Gil92,PopSch10,SchHilBar14,Raj12,UlsPenCak12,LimMas2018} which is emphasized by dark gray color.
The mass of the vehicle body is $m$, the wheel base is $l$, the distance between the rear axle and the center of mass G is $d$, and the moment of inertia of the vehicle body about the center of mass G is $J_{\rm G} $. Points R and F mark the center of the rear and front wheels, respectively, and the steering angle is $\gamma$.

To derive the constraints and the models, we define multiple (right-hand) coordinate systems as follows.
\begin{enumerate}
  \item ${(x,y,z)}$ is the Earth-fixed frame (denoted as $\mathcal{F}$) with the origin located at point $\rm O$;
  \item ${(x_0,y_0,z_0)}$ is the body-fixed frame of the vehicle (denoted as $\mathcal{F}_0$) with the origin located at the center of mass G of the vehicle, the $x_0$ and $y_0$ axes pointing towards the longitudinal and lateral directions;
  \item ${(x_1,y_1,z_1)}$ is the body-fixed frame of the rear wheel (denoted as $\mathcal{F}_1$) with the origin located at the center of the rear wheel $\rm R$, the $x_1$ and $y_1$ axes pointing towards the longitudinal and lateral directions;
  \item ${(x_2,y_2,z_2)}$ is the body-fixed frame of the front wheel (denoted as $\mathcal{F}_2$) with the origin located at the center of the front wheel $\rm F$, the $x_2$ and $y_2$ axes pointing towards the longitudinal and lateral directions of the front wheel.
\end{enumerate}
The basis of the frames are denoted by ${\mathbf{i}_k,\mathbf{j}_k,\mathbf{k}_k}$, where the subscript $k$ refers to the frame $\mathcal{F}_k$.
In the Earth-fixed frame $\mathcal{F}$, the yaw angle of the vehicle is $\psi$, while the position of points G, R and F are $({x}_{\rm G}, {y}_{\rm G})$, ${({x}_{\rm R} , {y}_{\rm R}  )}$ and ${({x}_{\rm F} , {y}_{\rm F} )}$, respectively.

In the bicycle model, different wheel-ground contact models can be used from simple rigid wheel assumptions \cite{DeLuca_Planning_1998,Varszegi_2019} to complex tire models \cite{Pacejka_2002,Mi_2020}. Here, we consider the first case, namely, we consider a single contact point at each wheel with no slip condition. This approach can be formulated by the consideration of a \emph{rigid wheel} or a \emph{skate}. In the following subsections, we summarize the main assumptions and the related kinematic constraints of these two different cases. In particular, we derive the kinematic constraints for the rear wheel of the vehicle. Indeed, similar formulas can be obtained for the front wheel that are also given later in the paper.

\begin{table*}[!t]
\caption{Single track mechanical models of the automobile.\label{tab:models}}
\begin{center}
\includegraphics[width=\textwidth]{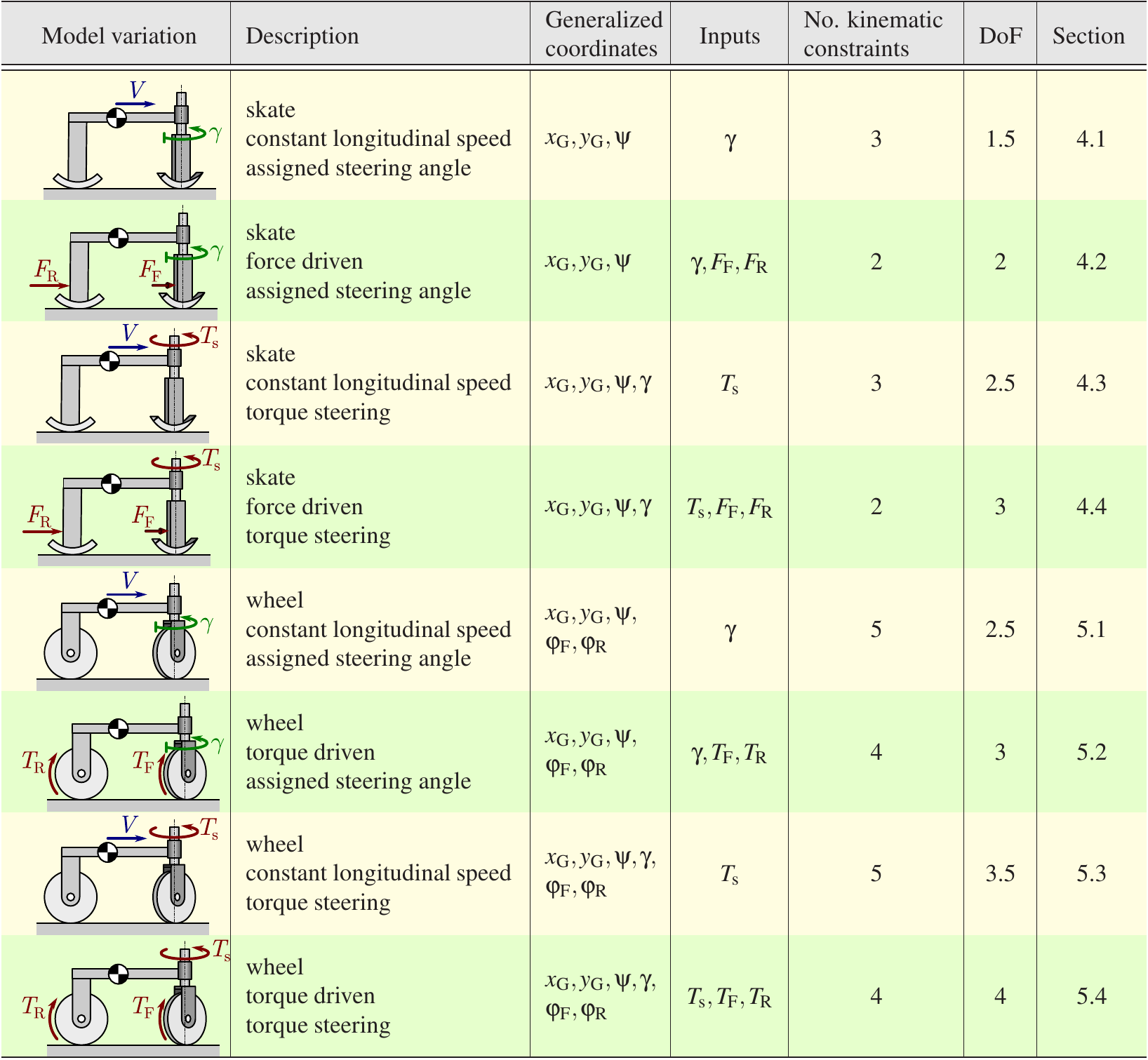}
\end{center}
\end{table*}

\subsection{Modeling Rigid Wheels}\label{sec:rigidwheel}

Let us consider a rigid wheel of radius $r$ as shown in  Fig.~\ref{fig:wheel_skate}(a). In order to describe the rotation about the wheel's symmetry axis $y_1$ we introduce the rotational angle $\varphi_{\rm R}$. Then the angular velocity vector of the rear wheel is given as
\begin{equation}\label{eq:rearwheelangvel}
  \boldsymbol{\omega}_{\rm R} =
\begin{bmatrix}
0 \\ \dot{\varphi}_{\rm R} \\ \dot{\psi}
\end{bmatrix}_{\mathcal{F}_1}.
\end{equation}
Let us denote the velocity components of the wheel center point R in $\mathcal{F}_1$ by $v^{x_1}_{{\rm R}}$, $v^{y_1}_{{\rm R}}$, $v^{z_1}_{{\rm R}}$. Since the bicycle model assumes planar motion of the vehicle body, the vertical velocity of the wheel center point is zero, i.e., $v^{z_1}_{{\rm R}}\equiv 0$. Thus, the velocity of the contact point P can be calculated as
\begin{equation}
\begin{split}
\mathbf{v}_{\rm P} &= \mathbf{v}_{\rm R}+\boldsymbol{\omega}_{\rm R}\times \mathbf{r}_{\rm RP}\\
&=\begin{bmatrix}
v^{x_1}_{{\rm R}}\\ v^{y_1}_{{\rm R}}\\ 0
\end{bmatrix}_{\mathcal{F}_1}+
\begin{bmatrix}
0 \\ \dot{\varphi}_{\rm R} \\ \dot{\psi}
\end{bmatrix}_{\mathcal{F}_1}\times
\begin{bmatrix}
0\\ 0\\ -r
\end{bmatrix}_{\mathcal{F}_1}\\
&=
\begin{bmatrix}
v^{x_1}_{{\rm R}} - r\, \dot{\varphi}_{\rm R}\\
v^{y_1}_{{\rm R}} \\
0
\end{bmatrix}_{\mathcal{F}_1}.
    \end{split}
\end{equation}
Then, the kinematic constraint of rolling ${\mathbf{v}_{\rm P}=\mathbf{0}}$ leads to
\begin{equation}\label{eq:wheelconstraints}
    v^{x_1}_{{\rm R}} - r\, \dot{\varphi}_{\rm R} = 0\ ,
\qquad
    v^{y_1}_{{\rm R}} = 0\ .
\end{equation}
These formulas can be turned into be first order scalar ordinary differential equations, when the velocity components $v^{x_1}_{{\rm R}}$ and $v^{y_1}_{{\rm R}}$ are expressed as functions of the generalized velocities and coordinates. We will manage these calculations later for both wheels, but here we rather focus on the physical meaning of \eqref{eq:wheelconstraints}.
Namely, the kinematic constraints of rolling can be rephrased in simple statements: the longitudinal speed of the wheel center point is equal to the wheel radius times the angular velocity component along the symmetry axis; and the lateral speed of wheel center point is zero.

\subsection{Modeling with Skates}\label{sec:skate}

When the dynamics of the automobile are analyzed without considering of the drivetrain dynamics, the rotational angular speed of the wheels may not be of interest. As a consequence, the first equation in \eqref{eq:wheelconstraints} is related to the longitudinal direction of wheel can be ``neglected" and the rotational angle $\varphi_{\rm R}$ can be ``eliminated" from the mechanical model. More precisely, the skate model (see Fig.~\ref{fig:wheel_skate}(b)) can be considered, which also simplifies the derivation of the kinematic constraint.

As the skate blade glides ahead, the velocity of its contact point P is parallel with the longitudinal direction of the blade, i.e., $\mathbf{v}_{\rm P} \parallel \mathbf{i}_1$, where $\mathbf{i}_1$ indicates the directions of the $x_1$-axis. In other words, the lateral speed of the contact point is zero, which also holds for the point R, that is,
\begin{equation}\label{eq:skateconstraint}
        v^{y_1}_{{\rm R}} = 0\ .
\end{equation}
This formula is identical to the second equation in \eqref{eq:wheelconstraints} and it eliminates the rotational degree of freedom of the wheel.

\subsection{Model Categorizations}\label{sec:models}

In Sections~\ref{sec:model_skate} and \ref{sec:model_wheels}, we derive multiple single track vehicle models using skates and rigid wheels, respectively. We apply the kinematic constraints \eqref{eq:skateconstraint} and \eqref{eq:wheelconstraints} for the skates and rigid wheels, respectively, and use the Appellian formalism to derive the equations of motion for these nonholonomic vehicle models.
In particular, we derive four different models while using the skate approach as shown in the first four rows of Table~\ref{tab:models}. Here we list the assumptions, number of generalized coordinates, control inputs, kinematic constraints, and degrees of freedom for each model. The related sections of the paper are also indicated in order to guide the reader.

We start with the kinematic bicycle model with constant longitudinal speed and assigned steering angle. This is described by three configuration coordinates (the position of the center of mass G and the yaw angle) and has three kinematic constraints (zero velocity components normal to the plane of the rear skate \eqref{eq:skateconstraint}, the equivalent constraint for the front skate, and constant longitudinal speed assumption). These lead to ${3-3/2 = 1.5}$ degrees of freedom, that is, three first-order ordinary differential equations. When the longitudinal speed is not constrained, we obtain a force-driven model with two kinematic constraints (zero normal velocities for both skates) yielding ${3-2/2=2}$ degrees of freedom. Keeping the longitudinal speed constant but steering the front wheel by applying a steering torque requires one more configuration coordinate, so that with the three kinematic constraints we obtain ${4-3/2=2.5}$ degrees of freedom. Finally, driving the vehicle with forces and steering it with torque lead to ${4-2/2=3}$ degrees of freedom.

In the last four rows of Table~\ref{tab:models}, we summarize the models derived using the rigid wheel approach. When comparing to the corresponding skate models, one may observe that the rigid wheel models contain two more configuration coordinates, the rotational angles of the wheels, and they also have two more kinematic constraints for the rotational speeds of the wheels; cf.~\eqref{eq:wheelconstraints}. Consequently, the degrees of freedom grow with $2-2/2 = 1$ compared to the corresponding skate models, yielding 2.5, 3, 3.5, and 4 degrees of freedom models, respectively. Another change is that while the skate models are driven by forces the rigid wheel models are driven by torques applied to the axles. In Section~\ref{sec:model_wheels}, we will discuss the equivalence between force and torque driving.

\section{Bicycle Models with Skates \label{sec:model_skate}}

\begin{figure}[t]
  \centering
  \includegraphics[scale = 1]{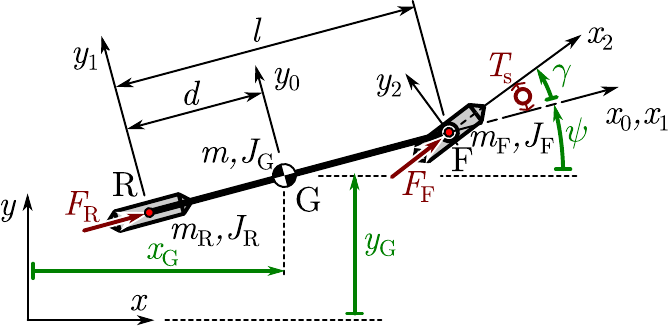}\\
  \caption{Bicycle model considering skates at the wheels. \label{fig:bicycle_skate}}
\end{figure}

In this section, we derive models with skates listed in the first four rows of Table~\ref{tab:models}. The mechanical model is shown in Fig.~\ref{fig:bicycle_skate} where skates are used to model both the rear and front wheels; see Fig.~\ref{fig:wheel_skate}(b). The masses of the skates are $m_{\rm R} $ and $m_{\rm F} $ while the mass moments of inertia about the points R and F are $J_{\rm R}$ and $J_{\rm F}$, respectively. We assume that the driving forces $F_{\rm R}$ and $F_{\rm F}$ can be applied at the rear and front wheels, respectively, while the internal steering torque is $T_{\rm s}$.

Since the vehicle is moving in two-dimensional space, three generalized coordinates are needed to describe its position and orientation. Here we choose the position ${({x}_{\rm G},{y}_{\rm G})}$ of the center of gravity G and the yaw angle $\psi$. Alternatively, one may choose the position ${({x}_{\rm R},{y}_{\rm R})}$ of the center of the rear axle R and the yaw angle $\psi$. Without any kinematic constraint these would correspond to three degrees of freedom, i.e., three second order ordinary differential equations. Below we will show how the number of degrees of freedom, i.e., the number of ordinary differential equations will be reduced due to the kinematic constraints. Moreover, the state of the steering system can be described by an additional generalized coordinate, the steering angle $\gamma$, leading to an additional degree of freedom. This degree of freedom can also be removed, however, assuming that the steering angle can be assigned.

\begin{table*}[!t]
\caption{Mechanical models with skates and their governing equations.\label{tab:modelswithskates}}
\begin{center}
\includegraphics[width=\textwidth]{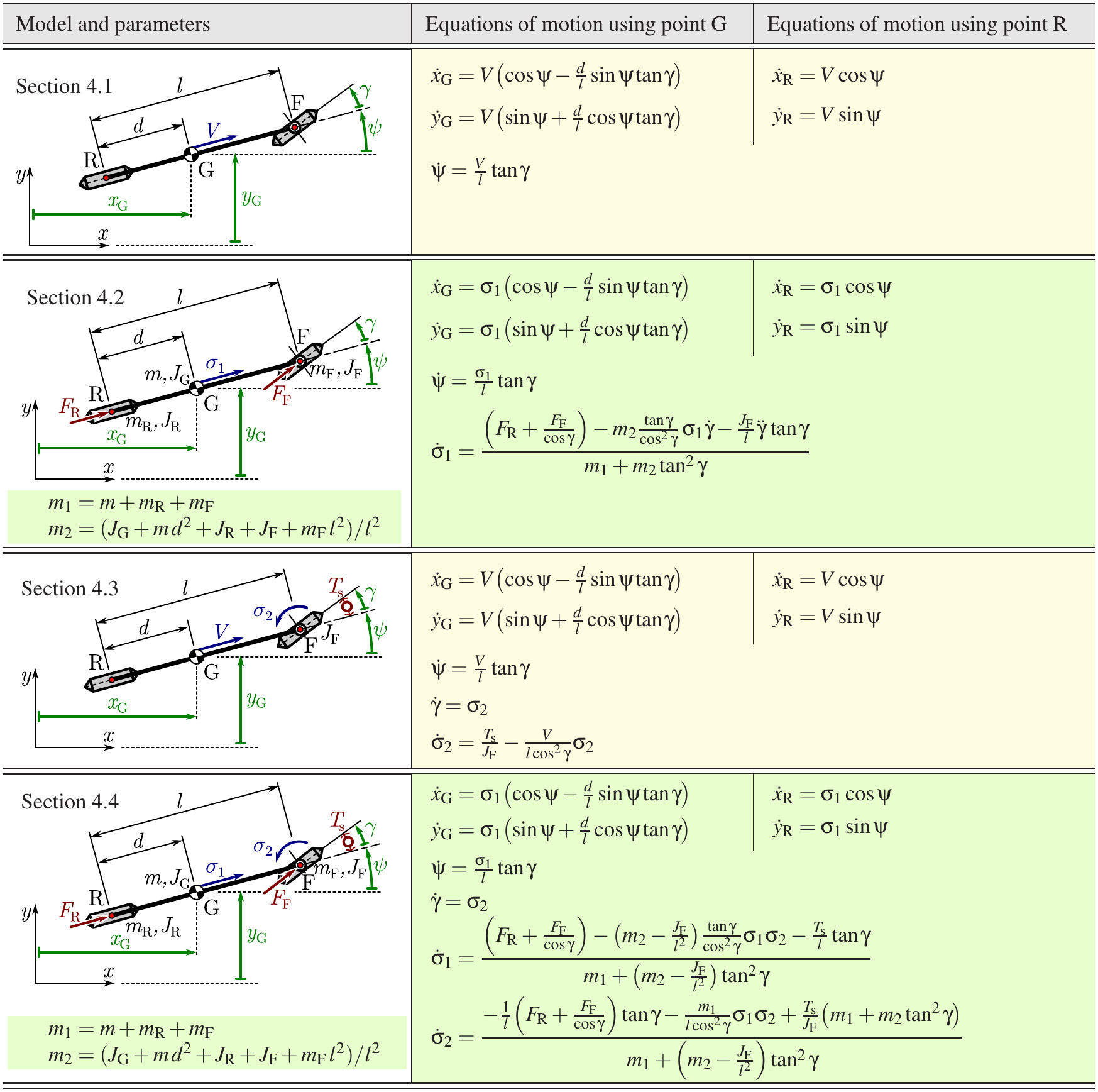}
\end{center}
\end{table*}

The kinematic constraints of the skates can be formulated based on Section~\ref{sec:skate}, namely, the lateral velocity components of the points R and F are zeros:
\begin{equation}\label{eq:kinconstraints_skates}
v^{y_1}_{\rm R}  =0\ ,
\quad
v^{y_2}_{\rm F} =0\ .
\end{equation}
The velocities of points R and F can be expressed as
\begin{equation}\label{eq:vRvF}
\begin{split}
\mathbf{v}_{\rm R}  & =
\begin{bmatrix}
\dot{x}_{\rm R}   \\
\dot{y}_{\rm R}   \\
0
\end{bmatrix}_{\mathcal{F}}
=
\begin{bmatrix}
\dot{x}_{\rm R} \cos\psi +\dot{y}_{\rm R}  \sin\psi  \\
-\dot{x}_{\rm R} \sin\psi +\dot{y}_{\rm R}  \cos\psi \\
0
\end{bmatrix}_{\mathcal{F}_1}, \\
\mathbf{v}_{\rm F}  & =
\begin{bmatrix}
\dot{x}_{\rm F}   \\
\dot{y}_{\rm F}  \\
0
\end{bmatrix}_{\mathcal{F}}
=
\begin{bmatrix}
\dot{x}_{\rm F} \cos(\psi+\gamma) +\dot{y}_{\rm F} \sin(\psi+\gamma)  \\
-\dot{x}_{\rm F} \sin(\psi+\gamma) +\dot{y}_{\rm F} \cos(\psi+\gamma) \\
0
\end{bmatrix}_{\mathcal{F}_2}
\ .
\end{split}
\end{equation}
From geometry, the positions of points R and F are
\begin{equation}\label{eqn:skate pos}
\begin{split}
{x}_{\rm R}  &= {x}_{\rm G}-d\cos\psi\ ,
\\
{y}_{\rm R}   &= {y}_{\rm G}-d\sin\psi\ ,
\\
{x}_{\rm F}  &= {x}_{\rm G}+(l-d)\cos\psi\ ,
\\
{y}_{\rm F}  &= {y}_{\rm G}+(l-d)\sin\psi\ .
\end{split}
\end{equation}
After substituting the derivatives of \eqref{eqn:skate pos} (cf.~\eqref{eqn:skate_dxdy_rf}) into \eqref{eq:vRvF}, the formulas in \eqref{eq:kinconstraints_skates} result in the kinematic constraining equations:
\begin{equation}\label{eqn:skate_consraint_no_slip}
\begin{split}
  \dot{x}_{\rm G}\sin\psi-\dot{y}_{\rm G}\cos\psi+d\,\dot{\psi} &= 0\ ,
  \\
  \dot{x}_{\rm G}\sin(\psi+\gamma)-\dot{y}_{\rm G}\cos(\psi+\gamma)-(l-d)\dot{\psi}\,\cos\gamma &= 0\ .
\end{split}
\end{equation}

As mentioned before, we consider different levels of model complexity with skates. Here we describe the four models listed in Table~\ref{tab:modelswithskates} (cf.~the first four rows of Table~\ref{tab:models})
while the detailed derivations are left for the following subsections. Readers who are not interested in these derivations may skip Sections~\ref{sec:model_skate_kinematic_asign}-\ref{sec:model_skate_dynamic_SteerTrq} and continue with Section~\ref{sec:model_wheels}.

The simplest model is given in the first row of Table~\ref{tab:modelswithskates} corresponding to the setup in the first row of Table~\ref{tab:models}. This is often referred to as the kinematic bicycle model as it contains neither mass nor moment of inertia but simply parameterized by the wheelbase $l$, the distance $d$, and the longitudinal velocity $V$. Given these parameters and assigning the steering angle $\gamma$, the position and the orientation of the vehicle can be determined by integrating the three ordinary differential equations. The first two equations can be further simplified when using the position of point R instead of the position of point G, leaving the last equation intact. Such simplification is also possible for all the other models presented in this paper. The kinematic bicycle model is often used for path planning of automated vehicles due to its simplicity and due to the fact that paths generated by this model are second order smooth. The model derivations can be found in Section~\ref{sec:model_skate_kinematic_asign}.

The model given in the second row of Table~\ref{tab:modelswithskates} uses the setup in the second row of Table~\ref{tab:models} where the longitudinal speed $\sigma_1$ is not restricted but obtained from the driving forces. The first three equations are analogous to those in the kinematic model while the longitudinal dynamics are given by the last equation that contains the masses and mass moments of inertia of the rigid bodies as well as the driving forces $F_{\rm R} $ and $F_{\rm F} $. Notice that for small steering angles one may omit the nonlinear terms and obtain ${\dot{\sigma}_1 = (F_{\rm R}  + F_{\rm F} )/m_1}$. This simplified model is often used for trajectory planning of automated vehicles as it can generate paths of second order smoothness and also allow the design of the velocity along the paths; see more details in Section~\ref{sec:control}. We remark that the nonlinear terms, that originate from the nonholonomic constraining forces, may become significant as the steering angle and the steering rate increase. Finally, we note that the masses $m_{\rm R} $, $m_{\rm F} $ and the mass moments of inertia $J_{\rm R}  $, $J_{\rm F} $ of the wheels may be neglected as these are much smaller compared to the other terms in $m_1$, $m_2$. This assumption also makes the term with $\ddot\gamma$ disappear in the last equation. The model derivations can be found in Section~\ref{sec:model_skate_dynamic_assign}.

The model in the third row of Table~\ref{tab:modelswithskates} uses the setup in the third row of Table~\ref{tab:models}. Here the longitudinal speed is restricted and the front wheel is steered by the torque $T_{\rm s} $. Thus, the steering angle $\gamma$ becomes a configuration coordinate and the two additional ordinary differential equations describe the evolution of the steering angle $\gamma$ and the steering rate $\sigma_2$. The last equation contains the steering torque $T_{\rm s} $ as well as a self alignment term that acts as a nonlinear damper in the steering dynamics (and becomes singular for $|\gamma| = \frac{\pi}{2}$). The model derivations can be found in Section~\ref{sec:model_skate_kinematic_SteerTrq}.

Finally, the model presented in the fourth row of Table~\ref{tab:modelswithskates} uses the setup in the fourth row of Table~\ref{tab:models}. Here neither the longitudinal speed nor the steering angle are assigned but they are derived from the driving forces $F_{\rm R} $, $F_{\rm F} $ and the steering torque $T_{\rm s} $, leading to a systems of six ordinary differential equations. Compared to the second model, the equation for $\sigma_1$ changes a little but the added terms are small compared to the others, that is, the longitudinal dynamics of the vehicle does not change significantly. On the other hand, compared to the third model, the equation for $\sigma_2$ changes significantly illustrating that changing the assumptions about the vehicle dynamics can have a large effects on the steering dynamics. The model derivations can be found in Section~\ref{sec:model_skate_dynamic_SteerTrq}.

\subsection{Kinematic Model
\label{sec:model_skate_kinematic_asign}}

Again to describe the vehicle we need three configuration coordinates which we choose to be the position of the center of gravity  ${x}_{\rm G}$ and ${y}_{\rm G}$ and the yaw angle $\psi$.
In this model, we assume that steering angle $\gamma$ is directly assigned and the longitudinal speed is constant $V$,  that is, $\mathbf{v}_{\rm G}\cdot \mathbf{i}_0=V$ which can be expressed as
\begin{equation}\label{eqn:skate_constraints_const_long}
\dot{x}_{\rm G}\cos\psi+\dot{y}_{\rm G}\sin\psi= V \, .
\end{equation}
The three kinematic constraints in (\ref{eqn:skate_consraint_no_slip},\ref{eqn:skate_constraints_const_long}) reduce the degrees of freedom of the system by $1.5$. Solving these equations for the generalized velocities we obtain the equation of motion:
\begin{equation}\label{eqn:EOM_000}
\begin{split}
\dot{x}_{\rm G} &=V\Big(\cos\psi-\frac{d}{l}\sin\psi\tan\gamma\Big)\ ,
\\
\dot{y}_{\rm G} &=V\Big(\sin\psi+\frac{d}{l}\cos\psi\tan\gamma\Big)\ ,
\\
\dot{\psi} &= \frac{V}{l}\tan\gamma\ .
\end{split}
\end{equation}

\subsection{Force-driven Model with Assigned Steering Angle
\label{sec:model_skate_dynamic_assign}}

Releasing the constraint of constant longitudinal speed, the driving forces $F_{\rm R} $ and $F_{\rm F} $ are applied at the rear and front skates (along their moving directions). Assuming that the steering angle is still directly assigned,
one pseudo-velocity is needed and we choose the longitudinal speed
\begin{equation}\label{eqn:010_sigma}
\sigma_{1} := \dot{x}_{\rm G}\cos\psi+\dot{y}_{\rm G}\sin\psi\ .
\end{equation}
For  different choices of pseudo-velocities we refer to Section~\ref{sec:singularities}.
Together with the kinematic constraints in \eqref{eqn:skate_consraint_no_slip}, one can express the generalized velocities as
\begin{equation}\label{eqn:010_dxdy}
\begin{split}
\dot{x}_{\rm G} &= \sigma_{1}\Big(\cos\psi-\frac{d}{l}\sin\psi\tan\gamma\Big)\ ,
\\
\dot{y}_{\rm G} &= \sigma_{1}\Big(\sin\psi+\frac{d}{l}\cos\psi\tan\gamma\Big)\ ,
\\
\dot{\psi} &= \frac{\sigma_{1}}{l}\tan\gamma\ .
\end{split}
\end{equation}

Taking the second derivative of \eqref{eqn:skate pos} (cf.~\eqref{eqn:skate_ddxddy_rf}) and utilizing the first derivative of \eqref{eqn:010_dxdy} (cf.~\eqref{eqn:010:ddxddyddpsi}), one can derive the acceleration energy
\begin{equation}
\begin{split}
S &= \frac{1}{2}m\big(\ddot{x}_{\rm G}^{2}+\ddot{y}_{\rm G}^{2}\big)+\frac{1}{2}J_{\rm G} \ddot{\psi}^{2}
\\
&+\frac{1}{2}m_{\rm R} \big(\ddot{x}_{\rm R} ^{2}+\ddot{y}_{\rm R}  ^{2}\big)+\frac{1}{2}J_{\rm R}  \,\ddot{\psi}^{2}
\\
&+\frac{1}{2}m_{\rm F} \big(\ddot{x}_{\rm F} ^{2}+\ddot{y}_{\rm F} ^{2}\big)+\frac{1}{2}J_{\rm F} (\ddot{\psi}+\ddot{\gamma})^{2}
\\
&=\frac{1}{2}\big(m_{1}+m_{2}\tan^{2}\gamma\big)\dot{\sigma}_{1}^{2}
\\
&+\Big(m_{2}\frac{\tan\gamma}{\cos^{2}\gamma}\sigma_{1}\dot{\gamma}
+\frac{J_{\rm F} }{l}\,\ddot{\gamma}\,\tan\gamma\Big)\dot{\sigma}_{1}
\\
&+(\textrm{terms\ without\ }\dot{\sigma}_{1})\ ,
\end{split}
\end{equation}
where
\begin{equation}\label{eqn:const_m1_m2}
\begin{split}
m_{1} &= m+m_{\rm R} +m_{\rm F} \ ,
\\
m_{2} &= \frac{1}{l^{2}}(J_{\rm G} +m\, d^{2}+J_{\rm R}  +J_{\rm F} +m_{\rm F} \,l^{2})\ .
\end{split}
\end{equation}

The virtual power consists of the powers of the driving forces acting on the vehicle body, that is,
\begin{equation}\label{eqn:virtPwr_skate_dyn_steerAng}
\begin{split}
\delta P &=
\begin{bmatrix}
F_{\rm R} \cos\psi & F_{\rm R} \sin\psi & 0
\end{bmatrix}_{\mathcal{F}}
\begin{bmatrix}
\delta\dot{x}_{\rm R}  \\ \delta\dot{y}_{\rm R}  \\0
\end{bmatrix}_{\mathcal{F}} \\
&+
\begin{bmatrix}
F_{\rm F} \cos(\psi+\gamma) & F_{\rm F} \sin(\psi+\gamma) & 0
\end{bmatrix}_{\mathcal{F}}
\begin{bmatrix}
\delta\dot{x}_{\rm F}  \\ \delta\dot{y}_{\rm F}  \\ 0
\end{bmatrix}_{\mathcal{F}}
\\
&= \Big(F_{\rm R} +\frac{F_{\rm F} }{\cos\gamma}\Big)\,\delta\sigma_{1}\ ,
\end{split}
\end{equation}
implying that the pseudo force is
\begin{align}\label{eqn:010_pseudo_force}
\Pi_{1} &= F_{\rm R} +\frac{F_{\rm F} }{\cos\gamma}\ .
\end{align}
The Appell equation
\begin{equation}
\frac{\partial S}{\partial \dot{\sigma}_{1}} =\Pi_{1}\ ,
\end{equation}
leads to
\begin{equation}\label{eqn:010_dsigma1}
\begin{split}
\left({m_{1}+m_{2}\tan^{2}\gamma}\right)\dot{\sigma}_{1}+ m_{2}\frac{\tan\gamma}{\cos^{2}\gamma}\,\sigma_{1}\dot{\gamma} +\frac{J_{\rm F} }{l}\ddot{\gamma}\,\tan\gamma
\\
= {F_{\rm R} +\frac{F_{\rm F} }{\cos\gamma}}\ .
\end{split}
\end{equation}
Combining (\ref{eqn:010_dxdy}) and (\ref{eqn:010_dsigma1}), the equation of motion are
\begin{equation}\label{eqn:EOM_010}
\begin{split}
\dot{x}_{\rm G} &= \sigma_{1}\Big(\cos\psi-\frac{d}{l}\sin\psi\tan\gamma\Big)\ ,
\\
\dot{y}_{\rm G} &= \sigma_{1}\Big(\sin\psi+\frac{d}{l}\cos\psi\tan\gamma\Big)\ ,
\\
\dot{\psi} &= \frac{\sigma_{1}}{l}\tan\gamma\ ,
\\
\dot{\sigma}_{1} &= \frac{\Big(F_{\rm R} +\frac{F_{\rm F} }{\cos\gamma}\Big)-m_{2}\frac{\tan\gamma}{\cos^{2}\gamma}\,\sigma_{1}\dot{\gamma}
-\frac{J_{\rm F} }{l}\ddot{\gamma}\,\tan\gamma}{m_{1}+m_{2}\tan^{2}\gamma}\ .
\end{split}
\end{equation}

\subsection{Model with Constrained Longitudinal Speed and Steering Torque
\label{sec:model_skate_kinematic_SteerTrq}}

Here we consider the steering dynamics of the front wheel and apply a steering torque $T_{\rm s} $, meanwhile we prescribe  constant longitudinal speed. Thus, we have four generalized coordinates, i.e., the position ${({x}_{\rm G},{y}_{\rm G})}$, the yaw angle $\psi$, and the steering angle $\gamma$. The two kinematic constraints in (\ref{eqn:skate_consraint_no_slip}) are used to guarantee that there is no side-slip at the skates, and the kinematic constraint \eqref{eqn:skate_constraints_const_long}  maintains the constant longitudinal speed $V$. Thus, one pseudo-velocity is needed, and we choose the steering rate
\begin{align}
\sigma_{2} &: = \dot{\gamma}\ . \label{eqn:001_sigma2}
\end{align}
Solving (\ref{eqn:skate_consraint_no_slip},\ref{eqn:skate_constraints_const_long},\ref{eqn:001_sigma2}), one can obtain
\begin{equation}\label{eqn:001_dxdy}
\begin{split}
\dot{x}_{\rm G} &=V\Big(\cos\psi-\frac{d}{l}\sin\psi\tan\gamma\Big)\ ,
\\
\dot{y}_{\rm G} &=V\Big(\sin\psi+\frac{d}{l}\cos\psi\tan\gamma\Big)\ ,
\\
\dot{\psi} &= \frac{V}{l}\tan\gamma\ ,
\\
\dot{\gamma} &= \sigma_{2}\ .
\end{split}
\end{equation}
The acceleration energy is
\begin{equation}\label{eqn:001_S}
\begin{split}
S &= \frac{1}{2}m\big(\ddot{x}_{\rm G}^{2}+\ddot{y}_{\rm G}^{2}\big)+\frac{1}{2}J_{\rm G} \ddot{\psi}^{2}
\\
&+\frac{1}{2}m_{\rm R} \big(\ddot{x}_{\rm R} ^{2}+\ddot{y}_{\rm R}  ^{2}\big)+\frac{1}{2}J_{\rm R}  \ddot{\psi}^{2}
\\
&+\frac{1}{2}m_{\rm F} \big(\ddot{x}_{\rm F} ^{2}+\ddot{y}_{\rm F} ^{2}\big)+\frac{1}{2}J_{\rm F} (\ddot{\psi}+\ddot{\gamma})^{2}
\\
&= \frac{1}{2}J_{\rm F} \,\dot{\sigma}_{2}^{2} + \frac{J_{\rm F} \,V}{l\cos^{2}\gamma}\sigma_{2}\dot{\sigma}_{2}
\\
&+(\textrm{terms\ without\ } \dot{\sigma}_{2})\ ,
\end{split}
\end{equation}
where we used the second derivative of \eqref{eqn:skate pos} (cf.~\eqref{eqn:skate_ddxddy_rf}) and the first derivative of \eqref{eqn:001_dxdy} (cf.~\eqref{eqn:001:ddxddyddpsiddgamma}).

The virtual power consists of the powers of the steering torque acting on the vehicle body and the front wheel, that is,
\begin{equation}\label{eqn:001_virtualpower}
\begin{split}
\delta P &=
\begin{bmatrix}
0 & 0& T_{\rm s}
\end{bmatrix}_{\mathcal{F}_2}
\begin{bmatrix}
0 \\ \delta\dot{\varphi}_{\rm F} \\ \delta\dot{\psi}+\delta\dot{\gamma}
\end{bmatrix}_{\mathcal{F}_2}
+\begin{bmatrix}
0 & 0& -T_{\rm s}
\end{bmatrix}_{\mathcal{F}_0}
\begin{bmatrix}
0 \\ 0 \\ \delta\dot{\psi}
\end{bmatrix}_{\mathcal{F}_0}\\
&= T_{\rm s} \ \delta\dot{\gamma}
= T_{\rm s} \ \delta \sigma_{2}\ ,
\end{split}
\end{equation}
implying that the pseudo-force is
\begin{equation}\label{eqn:001_pseudo_force}
\Pi_{2} = T_{\rm s} \ .
\end{equation}

The Appell equation
\begin{align}
\frac{\partial S}{\partial \dot{\sigma}_{2}} &=\Pi_{2}\ ,
\end{align}
leads to
\begin{equation}\label{eqn:001_dsigma2}
{J_{\rm F} } \,\dot{\sigma}_{2} + \frac{J_{\rm F} \, V }{l\cos^{2}\gamma}\sigma_{2} = {T_{\rm s} }\ .
\end{equation}
According to (\ref{eqn:001_dxdy}) and (\ref{eqn:001_dsigma2}) the equations of motion are
\begin{equation}\label{eqn:EOM_001}
\begin{split}
\dot{x}_{\rm G} &=V\Big(\cos\psi-\frac{d}{l}\sin\psi\tan\gamma\Big)\ ,
\\
\dot{y}_{\rm G} &=V\Big(\sin\psi+\frac{d}{l}\cos\psi\tan\gamma\Big)\ ,
\\
\dot{\psi} &= \frac{V}{l}\tan\gamma\ ,
\\
\dot{\gamma} &= \sigma_{2}\ ,
\\
\dot{\sigma}_{2} &= \frac{T_{\rm s} }{J_{\rm F} }-\frac{V\sigma_{2}}{l\cos^{2}\gamma}\ .
\end{split}
\end{equation}

\subsection{Force-driven Model with Steering Torque \label{sec:model_skate_dynamic_SteerTrq}}

Here we consider the steering and longitudinal dynamics together, that is, we apply a steering torque $T_{\rm s} $ and the driving forces $F_{\rm R} $ and $F_{\rm F} $ at the rear and front skates. Again, we need four generalized coordinates, the position ${({x}_{\rm G},{y}_{\rm G})}$, the heading angle $\psi$, and the steering angle $\gamma$. As the kinematic constraints  (\ref{eqn:skate_consraint_no_slip}) for the skates are still considered, we need two pseudo-velocities, i.e., the longitudinal speed and the steering rate:
\begin{align}
\begin{split}\label{eqn:011_sigma12}
\sigma_{1} &:= \dot{x}_{\rm G}\cos\psi+\dot{y}_{\rm G}\sin\psi,\\
\sigma_{2} &:= \dot{\gamma}\ .
\end{split}
\end{align}
Solving (\ref{eqn:skate_consraint_no_slip},\ref{eqn:011_sigma12}) we can express the generalized velocities as
\begin{align}
\begin{split}\label{eqn:011_dxdy}
\dot{x}_{\rm G} &= \sigma_{1}\Big(\cos\psi-\frac{d}{l}\sin\psi\tan\gamma\Big),\\
\dot{y}_{\rm G} &= \sigma_{1}\Big(\sin\psi+\frac{d}{l}\cos\psi\tan\gamma\Big),\\
\dot{\psi} &= \frac{\sigma_{1}}{l}\tan\gamma,\\
\dot{\gamma} &= \sigma_{2}.
\end{split}
\end{align}

The acceleration energy is
\begin{equation}\label{eqn:011_S}
\begin{split}
S &= \frac{1}{2}m\big(\ddot{x}_{\rm G}^{2}+\ddot{y}_{\rm G}^{2}\big)+\frac{1}{2}J_{\rm G} \ddot{\psi}^{2}
\\
&+\frac{1}{2}m_{\rm R}
\big(\ddot{x}_{\rm R} ^{2}+\ddot{y}_{\rm R}  ^{2}\big)+\frac{1}{2}J_{\rm R}  \ddot{\psi}^{2}
\\
&+\frac{1}{2}m_{\rm F} \big(\ddot{x}_{\rm F} ^{2}+\ddot{y}_{\rm F} ^{2}\big)+\frac{1}{2}J_{\rm F} (\ddot{\psi}+\ddot{\gamma})^{2}
\\
&= \frac{1}{2}\big(m_{1}+m_{2}\tan^{2}\gamma\big)\dot{\sigma}_{1}^{2}+\frac{1}{2}J_{\rm F} \,\dot{\sigma}_{2}^{2}
+\frac{J_{\rm F} }{l}\tan\gamma\,\dot{\sigma}_{1}\dot{\sigma}_{2}
\\
& +m_{2}\frac{\tan\gamma}{\cos^{2}\gamma}\sigma_{1}\sigma_{2}\dot{\sigma}_{1} +\frac{J_{\rm F} }{l\cos^{2}\gamma}\sigma_{1}\sigma_{2}\dot{\sigma}_{2}
\\
&+(\textrm{terms\ without\ }\dot{\sigma}_{1}\ {\rm and}\ \dot{\sigma}_{2})\ .
\end{split}
\end{equation}
where we used the second derivative of \eqref{eqn:skate pos} (cf.~\eqref{eqn:skate_ddxddy_rf}) and the first derivative of \eqref{eqn:011_dxdy} (cf.~Appendix~\ref{app:deriv}). Here $m_{1}$ and $m_{2}$ are still given by \eqref{eqn:const_m1_m2}.
The virtual power consists of the powers of the driving forces and steering torque acting on the vehicle body and the front wheel:
\begin{equation}\label{eqn:virtPwr_skate_dyn_steerTrq}
\begin{split}
\delta P &=
\begin{bmatrix}
F_{\rm R} \cos\psi & F_{\rm R} \sin\psi &0
\end{bmatrix}_{\mathcal{F}}
\begin{bmatrix}
\delta\dot{x}_{\rm R}  \\ \delta\dot{y}_{\rm R}  \\0
\end{bmatrix}_{\mathcal{F}} \\
&+
\begin{bmatrix}
F_{\rm F} \cos(\psi+\gamma) & F_{\rm F} \sin(\psi+\gamma) &0
\end{bmatrix}_{\mathcal{F}}
\begin{bmatrix}
\delta\dot{x}_{\rm F}  \\ \delta\dot{y}_{\rm F} \\0
\end{bmatrix}_{\mathcal{F}}\\
&+\begin{bmatrix}
0 & 0& T_{\rm s}
\end{bmatrix}_{\mathcal{F}_2}
\begin{bmatrix}
0 \\ \delta\dot{\varphi}_{\rm F} \\ \delta\dot{\psi}+\delta\dot{\gamma}
\end{bmatrix}_{\mathcal{F}_2}
+\begin{bmatrix}
0 & 0& -T_{\rm s}
\end{bmatrix}_{\mathcal{F}_0}
\begin{bmatrix}
0 \\ 0 \\ \delta\dot{\psi}
\end{bmatrix}_{\mathcal{F}_0} \\
&=  \Big(F_{\rm R} +\frac{F_{\rm F} }{\cos\gamma}\Big)\,\delta \sigma_{1}+T_{\rm s} \ \delta \sigma_{2}\ ,
\end{split}
\end{equation}
implying the pseudo forces
\begin{equation}\label{eqn:011_pseudo_force}
\Pi_{1} = F_{\rm R} +\frac{F_{\rm F} }{\cos\gamma}\ ,
\qquad
\Pi_{2} = T_{\rm s} \ .
\end{equation}

The Appell equations
\begin{equation}
\frac{\partial S}{\partial \dot{\sigma}_{1}} =\Pi_{1}\ ,
\qquad
\frac{\partial S}{\partial \dot{\sigma}_{2}} =\Pi_{2}\ ,
\end{equation}
yield
\begin{equation}\label{eqn:011_dsigma12}
\begin{split}
\begin{bmatrix}
\big(m_{1}+m_{2}\tan^{2}\gamma\big) & \dfrac{J_{\rm F} }{l}\tan\gamma
\\
\dfrac{J_{\rm F} }{l}\tan\gamma & J_{\rm F}
\end{bmatrix}
\begin{bmatrix}
\dot{\sigma}_{1}
\\
\dot{\sigma}_{2}
\end{bmatrix}
+
\begin{bmatrix}
m_{2}\dfrac{\tan\gamma}{\cos^{2}\gamma}
\\
\dfrac{J_{\rm F} }{l\cos^{2}\gamma}
\end{bmatrix}
\sigma_{1}\sigma_{2}
\\
 =
\begin{bmatrix}
F_{\rm R} +\dfrac{F_{\rm F} }{\cos\gamma}
\\
T_{\rm s}
\end{bmatrix}\ .
\end{split}
\end{equation}
Combining (\ref{eqn:011_dxdy}) and (\ref{eqn:011_dsigma12}) gives the equation of motion
\begin{equation}\label{eqn:EOM_011}
\begin{split}
\dot{x}_{\rm G} &= \sigma_{1}\Big(\cos\psi-\frac{d}{l}\sin\psi\tan\gamma\Big)\ ,
\\
\dot{y}_{\rm G} &= \sigma_{1}\Big(\sin\psi+\frac{d}{l}\cos\psi\tan\gamma\Big)\ ,
\\
\dot{\psi} &= \frac{\sigma_{1}}{l}\tan\gamma\ ,
\\
\dot{\gamma} &= \sigma_{2}\ ,
\\
\dot{\sigma}_{1} &= \frac{\Big(F_{\rm R} +\frac{F_{\rm F} }{\cos\gamma}\Big)-\big(m_{2}-\frac{J_{\rm F} }{l^{2}}\big)
\frac{\tan\gamma}{\cos^{2}\gamma}\sigma_{1}\sigma_{2}-\frac{T_{\rm s} }{l}\tan\gamma}{m_{1}
+\big(m_{2}-\frac{J_{\rm F} }{l^{2}}\big)\tan^{2}\gamma}\ ,
\\
\dot{\sigma}_{2} &= \frac{-\frac{1}{l}\Big(F_{\rm R} +\frac{F_{\rm F} }{\cos\gamma}\Big)\tan\gamma
-\frac{m_{1}}{l\cos^{2}\gamma}\sigma_{1}\sigma_{2} }{m_{1}+ \big(m_{2}-\frac{J_{\rm F} }{l^{2}}\big)\tan^{2}\gamma}
\ldots
\\
&\ldots\frac{+\frac{T_{\rm s} }{J_{\rm F} } \big(m_{1} +m_{2}\tan^{2}\gamma\big)}{}
\ .
\end{split}
\end{equation}

\section{Bicycle Models with Rigid Wheels\label{sec:model_wheels}}

\begin{figure}[t!]
  \centering
  \includegraphics[scale = 1]{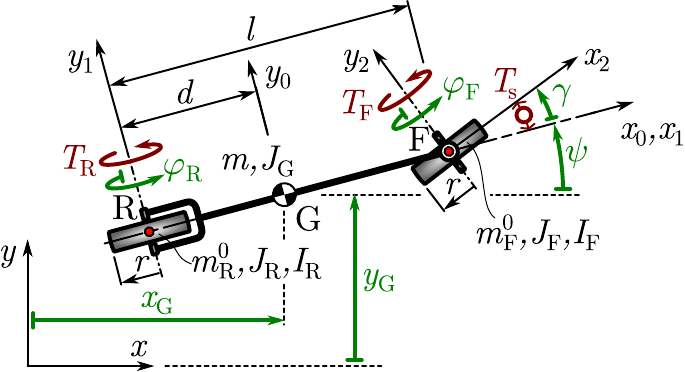}\\
  \caption{Bicycle model considering rolling wheels. \label{fig:bicycle_wheels}}
\end{figure}

In this section, we derive models with wheels listed in the last four rows of Table~\ref{tab:models}. The mechanical model is depicted in Fig.~\ref{fig:bicycle_wheels} where both wheels are considered to be rigid (see Fig.~\ref{fig:wheel_skate}(a)) and the rotation angles of the wheels around the $y_{1}$ and $y_2$ axes are denoted by $\varphi_{\rm R}$ and $\varphi_{\rm F}$, respectively. These will be added to generalized coordinates ${x}_{\rm G},  {y}_{\rm G}, \psi$ to describe the configuration of the vehicle. In case of torque steering these are augmented with the steering angle $\gamma$. The radii of the wheels are equal and are denoted by $r$. The masses of the rear and front wheels are $m_{\rm R} ^{0}$ and $m_{\rm F} ^{0}$, respectively, while their mass moment of inertia tensors about points R and F are given by
\begin{equation}\label{eq:momentofinertia}
            \mathbf{J}_{\rm R} =
            \begin{bmatrix}
                J_{\rm R}   & 0 & 0\\
                0 & I_{\rm R} & 0\\
                0 & 0 & J_{\rm R}
            \end{bmatrix}_{\mathcal{F}_1},
            \qquad
            \mathbf{J}_{\rm F} =
            \begin{bmatrix}
                J_{\rm F}  & 0 & 0\\
                0 & I_{\rm F} & 0\\
                0 & 0 & J_{\rm F}
            \end{bmatrix}_{\mathcal{F}_2},
\end{equation}
respectively. That is, $J_{\rm R}  $ and $J_{\rm F} $ denote the mass moments of inertia with respect to the $x$ and $z$ axes, while $I_{\rm R}$ and $I_{\rm F}$ are the mass moments of inertia with respect to the symmetry axes $y_1$ and $y_2$ of the wheels.
The driving torques $T_{\rm R} $ and $T_{\rm F} $ act on the rear wheel and the front wheel about their symmetry axes, respectively. In case of torque steering, the front wheel is steered by the internal steering torque $T_{\rm s} $, which acts between the vehicle body and the wheel about the vertical axis.

Denote the wheel-ground contact points by P and Q  for the rear and front wheels, respectively. We assume rolling without slipping, i.e., the velocities of the contact points are zeros: ${\mathbf{v}_{\rm P}=\mathbf{0}}$ and ${\mathbf{v}_{\rm Q}=\mathbf{0}}$. Based on Section~\ref{sec:rigidwheel}, the rolling constraints can be expressed as
\begin{equation}\label{eq:bothwheelsconstraint}
    \begin{split}
        v^{x_1}_{{\rm R}} - r\, \dot{\varphi}_{\rm R} = 0\ ,&\qquad
        v^{y_1}_{{\rm R}} = 0\ ,
        \\
        v^{x_2}_{{\rm F}} - r\, \dot{\varphi}_{\rm F} = 0\ ,&\qquad
        v^{y_2}_{{\rm F}} = 0\ .
    \end{split}
\end{equation}
Substituting the velocity components from \eqref{eq:vRvF} into these formulas, and using the derivatives of \eqref{eqn:skate pos} (cf.~\eqref{eqn:skate_dxdy_rf}), the kinematic constraining equations  become
\begin{equation}\label{eqn:wheel_constraints}\footnotesize
\begin{split}
\dot{x}_{\rm G}\sin\psi-\dot{y}_{\rm G}\cos\psi+d\,\dot{\psi} &= 0\ ,
\\
\dot{x}_{\rm G}\sin(\psi+\gamma)-\dot{y}_{\rm G} \cos(\psi+\gamma)-(l-d)\dot{\psi}\cos\gamma &= 0\ ,
\\
\dot{x}_{\rm G}\cos\psi+\dot{y}_{\rm G}\sin\psi-r\,\dot{\varphi}_{\rm R} &= 0\ ,
\\
\dot{x}_{\rm G}\cos(\gamma+\psi)+\dot{y}_{\rm G}\sin(\gamma+\psi)
+(l-d)\dot{\psi}\sin\gamma-r\,\dot{\varphi}_{\rm F} &= 0\ .
\end{split}
\end{equation}
Note that the first two equations in \eqref{eqn:wheel_constraints} are the same as those in \eqref{eqn:skate_consraint_no_slip}.

Again, we consider different levels of model complexity with rigid wheels. Here we describe the four models listed in Table~\ref{tab:modelswithwheels}; cf.~the last four rows of Table~\ref{tab:models}. The detailed derivations are left for Sections~\ref{sec:model_wheels_kinematic_assign}-\ref{sec:model_wheels_dynamic_SteerTrq}, which may be skipped if they are outside of the reader's interest.

The kinematic bicycle model given in the first row of Table~\ref{tab:modelswithwheels} corresponds to the setup given in the fifth row of Table~\ref{tab:models}. In order to understand the role of the wheels, it is worth comparing this model to the one in the first row of Table~\ref{tab:modelswithskates}. A color scheme is added to emphasize that the first three equations of the model with wheels are identical to those of the model with skates. One may observe that while the wheels increase the degrees of freedom with one, the two additional ordinary differential equations written for the rotation angles $\varphi_{\rm R}$ and $\varphi_{\rm F}$ are decoupled from the rest of the system. That is, someone not interested in these quantities may omit the last two equations. The model derivations can be found in Section~\ref{sec:model_wheels_kinematic_assign}.

\begin{table*}[!t]
\caption{Mechanical models with wheels and their governing equations.\label{tab:modelswithwheels}}
\begin{center}
\includegraphics[width=\textwidth]{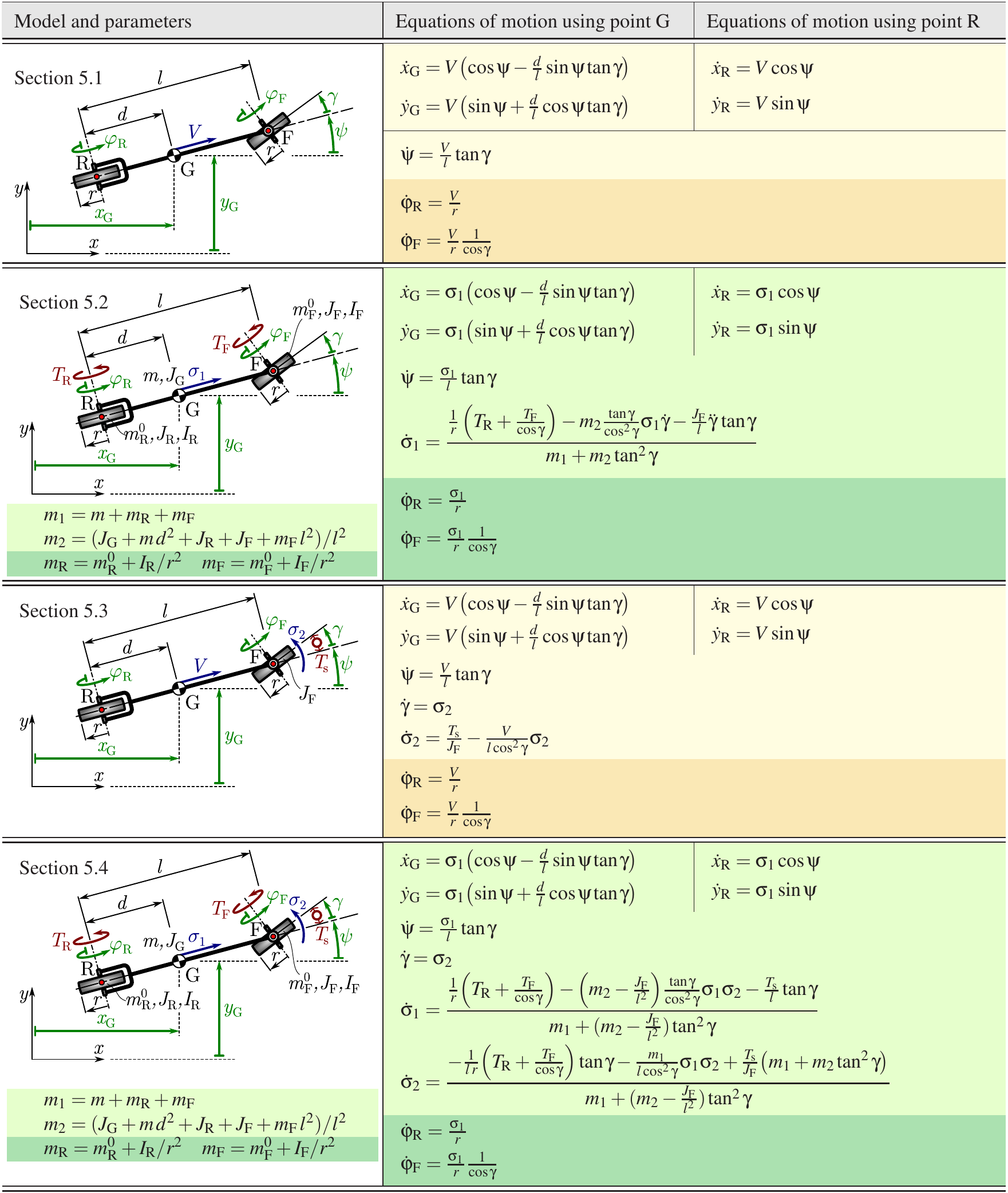}
\end{center}
\end{table*}

Similar structure is observed for the torque-driven model in the second row of Table~\ref{tab:modelswithwheels}, which corresponds to the setup given in the sixth row of Table~\ref{tab:models}. Comparing this model to the one in the second row of Table~\ref{tab:modelswithskates}, two additional ordinary differential equations appear that are decoupled from the rest of the system. In addition, the effective masses used for the skates can be calculated from the masses and mass moments of inertia of the wheels. Finally, the first four equations of the model with wheels are equivalent to those of the model with skates considering the relationships
\begin{equation}\label{eqn:model_comp_F_T_match}
  F_{\rm R}   = \dfrac{T_{\rm R} }{r}\ ,
 \qquad
  F_{\rm F}   = \dfrac{T_{\rm R} }{r}\ ,
\end{equation}
between the driving forces and driving torques. The model derivations can be found in Section~\ref{sec:model_wheels_dynamic_assign}.

The model in the third row of Table~\ref{tab:modelswithwheels} corresponds to the setup given in the seventh row of Table~\ref{tab:models}. Comparing this to the model in the third row of Table~\ref{tab:modelswithskates} one may identify the identical parts. The model derivations can be found in Section~\ref{sec:model_wheels_kinematic_SteerTrq}.

Finally, the fourth row of Table~\ref{tab:modelswithwheels} corresponds to the eighth row of Table~\ref{tab:models}. One may  compare this to the fourth row of Table~\ref{tab:modelswithskates} and identify the equivalent parts with the help of \eqref{eqn:model_comp_F_T_match}. The model derivations can be found in Section~\ref{sec:model_wheels_dynamic_SteerTrq}.

\subsection{Kinematic Model with Wheels
\label{sec:model_wheels_kinematic_assign}}

Similar to Section~\ref{sec:model_skate_kinematic_asign}, we assume that steering angle $\gamma$ is directly assigned and the longitudinal velocity $V$ is constant, i.e., the constraint \eqref{eqn:skate_constraints_const_long} is considered in addition to the rolling constraints (\ref{eqn:wheel_constraints}).

In order to describe the system we need five configuration coordinates. Here we choose ${{x}_{\rm G},  {y}_{\rm G}, \psi, \varphi_{\rm R}, \varphi_{\rm F}}$.
The five kinematic constraints reduce the degrees of freedom by 2.5, and the resulting 2.5 degree of freedom dynamic system is described by five ordinary differential equations.
Solving the constraints (\ref{eqn:skate_constraints_const_long},\ref{eqn:wheel_constraints}) for the generalized velocities lead to the equations of motion
\begin{equation}\label{eqn:EOM_100}
\begin{split}
\dot{x}_{\rm G} &=V\Big(\cos\psi-\frac{d}{l}\sin\psi\tan\gamma\Big)\ ,
\\
\dot{y}_{\rm G} &=V\Big(\sin\psi+\frac{d}{l}\cos\psi\tan\gamma\Big)\ ,
\\
\dot{\psi} &= \frac{V}{l}\tan\gamma\ ,
\\
\dot{\varphi}_{\rm R} &= \frac{V}{r}\ ,
\\
\dot{\varphi}_{\rm F} &= \frac{V}{r}\frac{1}{\cos\gamma}\ .
\end{split}
\end{equation}
We remark that the first three equations are decoupled from that last two equations and that the first three equations are identical to those in the skate model (\ref{eqn:EOM_000}) developed in Section~\ref{sec:model_skate_kinematic_asign}.

\subsection{Torque-driven Model with Assigned Steering Angle \label{sec:model_wheels_dynamic_assign}}

Now, we release the constraint of constant longitudinal speed.
Instead, the driving torques $T_{\rm R} $ and $T_{\rm F} $ are applied on the rear and front axles, respectively.
However, the steering angle is still directly assigned.
Thus, we have four kinematic constraints and five generalized coordinates, which requires one pseudo-velocity to be chosen. As formerly, we use the longitudinal speed:
\begin{align}\label{eqn:110:sigma1}
\sigma_{1} &:= \dot{x}_{\rm G}\cos\psi+\dot{y}_{\rm G}\sin\psi\ .
\end{align}
Solving (\ref{eqn:wheel_constraints},\ref{eqn:110:sigma1}),
one can express the generalized velocities as functions of the pseudo velocity:
\begin{equation}\label{eqn:110:appell:dxdydpsidphi1dphi2dgamma}
\begin{split}
\dot{x}_{\rm G} &= \sigma_{1}\,\Big(\cos\psi-\frac{d}{l}\sin\psi\tan\gamma\Big)\ ,
\\
\dot{y}_{\rm G} &= \sigma_{1}\,\Big(\sin\psi+\frac{d}{l}\cos\psi\tan\gamma\Big)\ ,
\\
\dot{\psi} &= \frac{\sigma_{1}}{l}\,\tan\gamma\ ,
\\
\dot{\varphi}_{\rm R} &= \frac{\sigma_{1}}{r}\ ,
\\
\dot{\varphi}_{\rm F} &= \frac{\sigma_{1}}{r\,\cos\gamma }\ .
\end{split}
\end{equation}

The acceleration energy of the system reads
\begin{equation}\label{eqn:110:S}
\begin{split}
S &= \frac{1}{2}m\big(\ddot{x}_{\rm G}^{2}+\ddot{y}_{\rm G}^{2}\big)+\frac{1}{2}J_{\rm G} \ddot{\psi}^{2}
\\
&+\frac{1}{2}m_{\rm R} ^{0}\big(\ddot{x}_{\rm R} ^{2}+\ddot{y}_{\rm R}  ^{2}\big) +\frac{1}{2}\boldsymbol{\alpha}_{\rm R} \cdot \mathbf{J}_{\rm R}\boldsymbol{\alpha}_{\rm R}
+\boldsymbol{\alpha}_{\rm R}\cdot(\boldsymbol{\omega}_{\rm R}\times \mathbf{H}_{\rm R})
\\
&+\frac{1}{2}m_{\rm F} ^{0}\big(\ddot{x}_{\rm F} ^{2}+\ddot{y}_{\rm F} ^{2}\big) +\frac{1}{2}\boldsymbol{\alpha}_{\rm F} \cdot \mathbf{J}_{\rm F}\boldsymbol{\alpha}_{\rm F}
+\boldsymbol{\alpha}_{\rm F}\cdot(\boldsymbol{\omega}_{\rm F}\times \mathbf{H}_{\rm F})\ ,
\end{split}
\end{equation}
where $\boldsymbol{\omega}_{\rm R}$, $\boldsymbol{\alpha}_{\rm R}$ and $\mathbf{H}_{\rm R}$ represent the angular velocity, angular acceleration and angular momentum vectors of the rear wheel, while $\boldsymbol{\omega}_{\rm F}$, $\boldsymbol{\alpha}_{\rm F}$ and $\mathbf{H}_{\rm F}$ are the angular velocity, angular acceleration and angular momentum vectors of the front wheel. Since the moment of inertia tensors $\mathbf{J}_{\rm R}$ and $\mathbf{J}_{\rm F}$ are expressed in frames $\mathcal{F}_1$ and $\mathcal{F}_2$, respectively (cf.~\ref{eq:momentofinertia}), the angular velocities should also be expressed in the corresponding frames:
\begin{equation}\label{eqn:rigid_disk_angularvelocity}
\boldsymbol{\omega}_{\rm R} =
\begin{bmatrix}
0\\
\dot{\varphi}_{\rm R}\\
\dot{\psi}
\end{bmatrix}_{\mathcal{F}_1},
\qquad
\boldsymbol{\omega}_{\rm F} =
\begin{bmatrix}
0\\
\dot{\varphi}_{\rm F}\\
\dot{\psi}+\dot{\gamma}
\end{bmatrix}_{\mathcal{F}_2}.
\end{equation}
By taking the derivative of (\ref{eqn:rigid_disk_angularvelocity}), one can obtain
\begin{equation}\label{eqn:rigid_disk_angularacceleration}
\begin{split}
\boldsymbol{\alpha}_{\rm R} &=
\accentset{\circ}{\boldsymbol{\omega}}_{\rm R}
+\boldsymbol{\omega}_{\mathcal{F}_1}\times\boldsymbol{\omega}_{\rm R}
\\
&=\begin{bmatrix}
0\\
\ddot{\varphi}_{\rm R}\\
\ddot{\psi}
\end{bmatrix}_{\mathcal{F}_1}
+\begin{bmatrix}
0\\
0\\
\dot{\psi}
\end{bmatrix}
_{\mathcal{F}_1}\times
\begin{bmatrix}
0\\
\dot{\varphi}_{\rm R}\\
\dot{\psi}
\end{bmatrix}_{\mathcal{F}_1}
\\
&=\begin{bmatrix}
-\dot{\psi}\dot{\varphi}_{\rm R}\\
\ddot{\varphi}_{\rm R}\\
\ddot{\psi}
\end{bmatrix}_{\mathcal{F}_1} ,
\\
\boldsymbol{\alpha}_{\rm F} &=
\accentset{\diamond}{\boldsymbol{\omega}}_{\rm F}
+\boldsymbol{\omega}_{\mathcal{F}_2}\times\boldsymbol{\omega}_{\rm F}
\\
&=
\begin{bmatrix}
0\\
\ddot{\varphi}_{\rm F}\\
\ddot{\psi}+\ddot{\gamma}
\end{bmatrix}_{\mathcal{F}_2}+
\begin{bmatrix}
0\\
0\\
\dot{\psi}+\dot{\gamma}
\end{bmatrix}_{\mathcal{F}_2}\times
\begin{bmatrix}
0\\
\dot{\varphi}_{\rm F}\\
\dot{\psi}+\dot{\gamma}
\end{bmatrix}_{\mathcal{F}_2}
\\
&=
\begin{bmatrix}
-(\dot{\psi}+\dot{\gamma})\dot{\varphi}_{\rm F}\\
\ddot{\varphi}_{\rm F}\\
\ddot{\psi}+\ddot{\gamma}
\end{bmatrix}_{\mathcal{F}_2} ,
\end{split}
\end{equation}
where $\accentset{\circ}{\square}$ and $\accentset{\diamond}{\square}$ represent the frame derivatives with respect to the body-fixed frames $\mathcal{F}_1$ and $\mathcal{F}_2$, respectively, and $\omega_{\mathcal{F}_1}$ and $\omega_{\mathcal{F}_2}$ represent the angular velocities of the frames $\mathcal{F}_1$ and $\mathcal{F}_2$, respectively.
Moreover, the angular momentum of the rear wheel and the front wheel are
\begin{equation}\label{eqn:rigid_disk_angularmomentum}
\begin{split}
\mathbf{H}_{\rm R} &= \mathbf{J}_{\rm R}\,\boldsymbol{\omega}_{\rm R}=
\begin{bmatrix}
0\\
I_{\rm R}\dot{\varphi}_{\rm R}\\
J_{\rm R}  \dot{\psi}
\end{bmatrix}_{\mathcal{F}_1} ,
\\
\mathbf{H}_{\rm F} &= \mathbf{J}_{\rm F}\,\boldsymbol{\omega}_{\rm F}=
\begin{bmatrix}
0\\
I_{\rm F}\dot{\varphi}_{\rm F}\\
J_{\rm F} (\dot{\psi}+\dot{\gamma})
\end{bmatrix}_{\mathcal{F}_2} .
\end{split}
\end{equation}
By taking the second derivative of \eqref{eqn:skate pos} (cf.~\eqref{eqn:skate_ddxddy_rf}) and then utilizing the first derivative of (\ref{eqn:110:appell:dxdydpsidphi1dphi2dgamma}) (cf.~(\ref{eqn:010:ddxddyddpsi},\ref{eqn:110:appell:ddxddyddthetaddphi1ddphi2})) along with (\ref{eqn:rigid_disk_angularvelocity},\ref{eqn:rigid_disk_angularacceleration},\ref{eqn:rigid_disk_angularmomentum}), the acceleration energy (\ref{eqn:110:S}) can be rewritten as
\begin{equation}
\begin{split}
S
&=\frac{1}{2}\big(m_{1}+m_{2}\tan^{2}\gamma\big)\dot{\sigma}_{1}^{2}
\\
&+\Big(m_{2}\frac{\tan\gamma}{\cos^{2}\gamma}\sigma_{1}\dot{\gamma}
+\frac{J_{\rm F} }{l}\,\ddot{\gamma}\,\tan\gamma\Big)\dot{\sigma}_{1}
\\
&+(\textrm{terms\ without\ }\dot{\sigma}_{1})\ .
\end{split}
\end{equation}
Here $m_{1}$ and $m_{2}$ are defined in \eqref{eqn:const_m1_m2}, and the effective mass of the rear and front wheels are
\begin{equation}\label{eqn:wheel_effect_mass}
m_{\rm R}  = m_{\rm R} ^{0}+\frac{{I}_{\rm R}}{r^{2}}\ ,
\qquad
m_{\rm F}  = m_{\rm F} ^{0}+\frac{{I}_{\rm F}}{r^{2}}\ .
\end{equation}

The virtual power consist of the powers of the driving torques acting between the wheels and the vehicle body:
\begin{equation}
\begin{split}
\delta P &
=\begin{bmatrix}
0 & T_{\rm R} & 0
\end{bmatrix}_{\mathcal{F}_1}
\begin{bmatrix}
0 \\ \delta\dot{\varphi}_{\rm R} \\ \delta\dot{\psi}
\end{bmatrix}_{\mathcal{F}_1}
+\begin{bmatrix}
0 & -T_{\rm R} & 0
\end{bmatrix}_{\mathcal{F}_0}
\begin{bmatrix}
0 \\ 0 \\ \delta\dot{\psi}
\end{bmatrix}_{\mathcal{F}_0} \\
&+\begin{bmatrix}
0 & T_{\rm F} & 0
\end{bmatrix}_{\mathcal{F}_2}
\begin{bmatrix}
0 \\ \delta\dot{\varphi}_{\rm F} \\ \delta\dot{\psi}+\delta\dot{\gamma}
\end{bmatrix}_{\mathcal{F}_2}
+\begin{bmatrix}
0 & -T_{\rm F} & 0
\end{bmatrix}_{\mathcal{F}_0}
\begin{bmatrix}
0 \\ 0 \\ \delta\dot{\psi}
\end{bmatrix}_{\mathcal{F}_0} \\
&= T_{\rm R} \, \delta\dot{\varphi}_{\rm R}+T_{\rm F} \, \delta\dot{\varphi}_{\rm F}\\
&=\bigg(\dfrac{T_{\rm R} }{r}+\dfrac{T_{\rm F} }{r\,\cos\gamma}\bigg)\delta\sigma_{1}\ ,
\end{split}
\end{equation}
yielding the pseudo-force
\begin{align}
\Pi_{1} &= \dfrac{1}{r}\bigg(T_{\rm R} +\dfrac{T_{\rm F} }{\cos\gamma}\bigg)\ .
\end{align}
Using the Appell equation
\begin{align}
\frac{\partial S}{\partial \dot{\sigma}_{1}} &=\Pi_{1}\ ,
\end{align}
one can obtain
\begin{equation}\label{eqn:110:dsigma1}
\begin{split}
{\left(m_{1}+m_{2}\tan^{2}\gamma\right)}\dot{\sigma}_{1} +m_{2}\frac{\tan\gamma}{\cos^{2}\gamma}\sigma_{1}\dot{\gamma}
+\frac{J_{\rm F} }{l}\ddot{\gamma}\,\tan\gamma
\\
= {\frac{1}{r}\left(T_{\rm R} +\frac{T_{\rm F} }{\cos\gamma}\right)}\, .
\end{split}
\end{equation}
Therefore, the equations of motion consist of \eqref{eqn:110:appell:dxdydpsidphi1dphi2dgamma} and \eqref{eqn:110:dsigma1}:
\begin{equation}\label{eqn:EOM_110}
\begin{split}
\dot{x}_{\rm G} &= \sigma_{1}\Big(\cos\psi-\frac{d}{l}\sin\psi\tan\gamma\Big)\ ,
\\
\dot{y}_{\rm G} &= \sigma_{1}\Big(\sin\psi+\frac{d}{l}\cos\psi\tan\gamma\Big)\ ,
\\
\dot{\psi} &= \frac{\sigma_{1}}{l}\tan\gamma\ ,
\\
\dot{\sigma}_{1} &= \frac{\frac{1}{r}\Big(T_{\rm R} +\frac{T_{\rm F} }{\cos\gamma}\Big)-m_{2}\frac{\tan\gamma}{\cos^{2}\gamma}\sigma_{1}\dot{\gamma}
-\frac{J_{\rm F} }{l}\ddot{\gamma}\,\tan\gamma}{m_{1}+m_{2}\tan^{2}\gamma}\, ,
\\
\dot{\varphi}_{\rm R} &= \frac{\sigma_{1}}{r}\ ,
\\
\dot{\varphi}_{\rm F} &= \frac{\sigma_{1}}{r}\frac{1}{\cos\gamma}\ .
\end{split}
\end{equation}
We remark that the first four equations are decoupled from the last two equations and that the first four equations are equivalent to those in the skate model \eqref{eqn:EOM_010} developed in Section~\ref{sec:model_skate_dynamic_assign}; cf.~\eqref{eqn:model_comp_F_T_match}.

\subsection{Wheeled Model with Constrained Longitudinal Speed and Steering Torque \label{sec:model_wheels_kinematic_SteerTrq}}

Considering the steering dynamics and applying the steering torque $T_{\rm s}$, we have six generalized coordinates, ${x_{\rm G}, y_{\rm G}, \psi, \varphi_{\rm R}, \varphi_{\rm F}, \gamma}$. The four kinematic constraints \eqref{eqn:wheel_constraints} are used to guarantee that there is no slip at the wheel-ground contact points. Moreover, we consider the kinematic constraint \eqref{eqn:skate_constraints_const_long} to maintain constant longitudinal speed. Thus, one pseudo-velocity is needed, and as before we choose the steering rate
\begin{equation}\label{eqn:101:sigma2}
\sigma_{2} = \dot{\gamma}\ .
\end{equation}
Solving (\ref{eqn:skate_constraints_const_long},\ref{eqn:wheel_constraints},\ref{eqn:101:sigma2}), for the generalized velocities lead to
\begin{equation}\label{eqn:101:dxdydpsidphi1dphi2dgamma}
\begin{split}
\dot{x}_{\rm G} &=V\Big(\cos\psi-\frac{d}{l}\sin\psi\tan\gamma\Big)\ ,
\\
\dot{y}_{\rm G} &=V\Big(\sin\psi+\frac{d}{l}\cos\psi\tan\gamma\Big)\ ,
\\
\dot{\psi} &= \frac{V}{l}\tan\gamma\ ,
\\
\dot{\gamma} &= \sigma_{2}\ .
\\
\dot{\varphi}_{\rm R} &= \frac{V}{r}\ ,
\\
\dot{\varphi}_{\rm F} &= \frac{V}{r\cos\gamma}\ .
\end{split}
\end{equation}

The general expression of acceleration energy, and calculation of angular velocity, angular acceleration and angular momentum are the same as (\ref{eqn:110:S},\ref{eqn:rigid_disk_angularvelocity},\ref{eqn:rigid_disk_angularacceleration},\ref{eqn:rigid_disk_angularmomentum}), respectively.
By taking the second derivative of \eqref{eqn:skate pos} (cf.~\eqref{eqn:skate_ddxddy_rf}) and then utilizing the first derivative of (\ref{eqn:101:dxdydpsidphi1dphi2dgamma}) (cf.~(\ref{eqn:001:ddxddyddpsiddgamma},\ref{eqn:101:appell:ddxddyddthetaddphi1ddphi2})) along with (\ref{eqn:rigid_disk_angularvelocity},\ref{eqn:rigid_disk_angularacceleration},\ref{eqn:rigid_disk_angularmomentum}), the acceleration energy (\ref{eqn:110:S}) can be rewritten as
\begin{equation}
S = \frac{1}{2}J_{\rm F} \,\dot{\sigma}_{2}^{2}+\frac{J_{\rm F} \,V}{l\cos^{2}\gamma}\sigma_{2}\dot{\sigma}_{2}+(\textrm{terms\ with\ out\ } \dot{\sigma}_{2})\ .
\end{equation}
The virtual power consists of the powers of the steering torque acting between the front wheel and the vehicle body, that is,
\begin{equation}
\begin{split}
\delta P &=
\begin{bmatrix}
0 & 0& T_{\rm s}
\end{bmatrix}_{\mathcal{F}_2}
\begin{bmatrix}
0 \\ \delta\dot{\varphi}_{\rm F} \\ \delta\dot{\psi}+\delta\dot{\gamma}
\end{bmatrix}_{\mathcal{F}_2}
+\begin{bmatrix}
0 & 0& -T_{\rm s}
\end{bmatrix}_{\mathcal{F}_0}
\begin{bmatrix}
0 \\ 0 \\ \delta\dot{\psi}
\end{bmatrix}_{\mathcal{F}_0}\\
&=T_{\rm s} \,\delta\dot{\gamma}
=T_{\rm s} \,\delta\sigma_{2}\ ,
\end{split}
\end{equation}
implying that the pseudo-force is
\begin{equation}
\Pi_{2} = T_{\rm s} \ .
\end{equation}
The Appell equation
\begin{equation}
\frac{\partial S}{\partial \dot{\sigma}_{2}} =\Pi_{2}\ ,
\end{equation}
results in
\begin{equation}\label{eqn:101:dsigma2}
{J_{\rm F} }\,\dot{\sigma}_{2} +\frac{J_{\rm F} \, V}{l\cos^{2}\gamma}\sigma_{2} = {T_{\rm s} }\ .
\end{equation}
Therefore, the equations of motion are given by \eqref{eqn:101:dxdydpsidphi1dphi2dgamma} and \eqref{eqn:101:dsigma2} are
\begin{equation}\label{eqn:EOM_101}
\begin{split}
\dot{x}_{\rm G} &=V\Big(\cos\psi-\frac{d}{l}\sin\psi\tan\gamma\Big)\ ,
\\
\dot{y}_{\rm G} &=V\Big(\sin\psi+\frac{d}{l}\cos\psi\tan\gamma\Big)\ ,
\\
\dot{\psi} &= \frac{V}{l}\tan\gamma\ ,
\\
\dot{\gamma} &= \sigma_{2}\ ,
\\
\dot{\sigma}_{2} &= \frac{T_{\rm s} }{J_{\rm F} }-\frac{V}{l\cos^{2}\gamma}\sigma_{2}\ ,
\\
\dot{\varphi}_{\rm R} &= \frac{V}{r}\ ,
\\
\dot{\varphi}_{\rm F} &= \frac{V}{r}\frac{1}{\cos\gamma}\ .
\end{split}
\end{equation}
We remark that the first five equations are decoupled from the last two equations and that the first five equations are identical to those in the skate model  \eqref{eqn:EOM_001} developed in Section~\ref{sec:model_skate_kinematic_SteerTrq}.

\subsection{Torque-driven Model with Steering Torque
\label{sec:model_wheels_dynamic_SteerTrq}}

In the most complex model of this paper, we consider the steering and longitudinal dynamics together. That is, the steering torque $T_{\rm s} $ as well as the driving torques $T_{\rm R} $ and $T_{\rm F} $ are applied. In this case, we have six general coordinates and four kinematic constraints, therefore two pseudo-velocities are required. As above, we use the longitudinal speed and the steering rate:
\begin{equation}\label{eqn:111:sigma1sigma2}
\begin{split}
\sigma_{1} &:= \dot{x}_{\rm G}\cos\psi+\dot{y}_{\rm G}\sin\psi\ ,
\\
\sigma_{2} &:= \dot{\gamma}\ .
\end{split}
\end{equation}
Using these together with the kinematic constraints (\ref{eqn:wheel_constraints}), one can obtain
\begin{equation}\label{eqn:111:dxdydpsidphi1dphi2dgamma}
\begin{split}
\dot{x}_{\rm G} &= \sigma_{1}\Big(\cos\psi-\frac{d}{l}\sin\psi\tan\gamma\Big)\ ,
\\
\dot{y}_{\rm G} &= \sigma_{1}\Big(\sin\psi+\frac{d}{l}\cos\psi\tan\gamma\Big)\ ,
\\
\dot{\psi} &= \dfrac{\sigma_{1}}{l}\tan\gamma\ ,
\\
\dot{\gamma} &= \sigma_{2}\ ,
\\
\dot{\varphi}_{\rm R} &= \dfrac{\sigma_{1}}{r}\ ,
\\
\dot{\varphi}_{\rm F} &= \dfrac{\sigma_{1}}{r}\dfrac{1}{\cos\gamma}\ .
\end{split}
\end{equation}

Similar to the explanation in Section~\ref{sec:model_wheels_kinematic_SteerTrq}, the general expression of acceleration energy, and calculation of angular velocity, angular acceleration and angular momentum are already given in (\ref{eqn:110:S},\ref{eqn:rigid_disk_angularvelocity},\ref{eqn:rigid_disk_angularacceleration},\ref{eqn:rigid_disk_angularmomentum}).
By taking the second derivative of \eqref{eqn:skate pos} (cf.~\eqref{eqn:skate_ddxddy_rf}) and then utilizing the first derivative of (\ref{eqn:111:dxdydpsidphi1dphi2dgamma}) (cf.~Appendix~\ref{app:deriv}) along with (\ref{eqn:rigid_disk_angularvelocity},\ref{eqn:rigid_disk_angularacceleration},\ref{eqn:rigid_disk_angularmomentum}), the acceleration energy (\ref{eqn:110:S}) can be rewritten as
\begin{equation}
\begin{split}
S
&= {\frac{1}{2}}\big(m_{1}+m_{2}\tan^{2}\gamma\big)\dot{\sigma}_{1}^{2}+\frac{1}{2}J_{\rm F} \,\dot{\sigma}_{2}^{2} +\frac{J_{\rm F} }{l}\tan\gamma\, \dot{\sigma}_{1}\dot{\sigma}_{2}
\\
&+m_{2}\frac{\tan\gamma}{\cos^{2}\gamma}\sigma_{1}\sigma_{2}\dot{\sigma}_{1} +\frac{J_{\rm F} }{l\cos^{2}\gamma}\sigma_{1}\sigma_{2}\dot{\sigma}_{2}
\\
&+(\textrm{terms\ without\ }\dot{\sigma}_{1}\ \textrm{or}\ \dot{\sigma}_{2})\ .
\end{split}
\end{equation}
The virtual power consist of the virtual powers of the driving and steering torques acting on the wheels and vehicle body, that is,
\begin{equation}
\begin{split}
\delta P &
=\begin{bmatrix}
0 & T_{\rm R} & 0
\end{bmatrix}_{\mathcal{F}_1}
\begin{bmatrix}
0 \\ \delta\dot{\varphi}_{\rm R} \\ \delta\dot{\psi}
\end{bmatrix}_{\mathcal{F}_1}
+\begin{bmatrix}
0 & -T_{\rm R} & 0
\end{bmatrix}_{\mathcal{F}_0}
\begin{bmatrix}
0 \\ 0 \\ \delta\dot{\psi}
\end{bmatrix}_{\mathcal{F}_0} \\
&+\begin{bmatrix}
0 & T_{\rm F} & 0
\end{bmatrix}_{\mathcal{F}_2}
\begin{bmatrix}
0 \\ \delta\dot{\varphi}_{\rm F} \\ \delta\dot{\psi}+\delta\dot{\gamma}
\end{bmatrix}_{\mathcal{F}_2}
+\begin{bmatrix}
0 & -T_{\rm F} & 0
\end{bmatrix}_{\mathcal{F}_0}
\begin{bmatrix}
0 \\ 0 \\ \delta\dot{\psi}
\end{bmatrix}_{\mathcal{F}_0} \\
&+\begin{bmatrix}
0 & 0& T_{\rm s}
\end{bmatrix}_{\mathcal{F}_2}
\begin{bmatrix}
0 \\ \delta\dot{\varphi}_{\rm F} \\ \delta\dot{\psi}+\delta\dot{\gamma}
\end{bmatrix}_{\mathcal{F}_2}
+\begin{bmatrix}
0 & 0& -T_{\rm s}
\end{bmatrix}_{\mathcal{F}_0}
\begin{bmatrix}
0 \\ 0 \\ \delta\dot{\psi}
\end{bmatrix}_{\mathcal{F}_0}\\
&= T_{\rm R} \, \delta\dot{\varphi}_{\rm R}+T_{\rm F} \, \delta\dot{\varphi}_{\rm F}+T_{\rm s} \delta\dot{\gamma} \\
&=\bigg(\dfrac{T_{\rm R} }{r}+\dfrac{T_{\rm F} }{r\,\cos\gamma}\bigg)\delta\sigma_{1}+T_{\rm s} \delta\sigma_{2}\ ,
\end{split}
\end{equation}
implying that the pseudo-forces are
\begin{equation}
\Pi_{1} = \dfrac{1}{r}\bigg(T_{\rm R} +\dfrac{T_{\rm F} }{\cos\gamma}\bigg)\ ,
\qquad
\Pi_{2} = T_{\rm s} \ .
\end{equation}
The Appell equations
\begin{equation}
\frac{\partial S}{\partial \dot{\sigma}_{1}} =\Pi_{1}\ ,
\qquad
\frac{\partial S}{\partial \dot{\sigma}_{2}} =\Pi_{2}\ ,
\end{equation}
lead to
\begin{equation}\label{eqn:111:dsigma1dsigma2}
\begin{split}
\begin{bmatrix}
\big(m_{1}+m_{2}\tan^{2}\gamma\big) & \dfrac{J_{\rm F} }{l}\tan\gamma
\\
\dfrac{J_{\rm F} }{l}\tan\gamma & J_{\rm F}
\end{bmatrix}
\begin{bmatrix}
\dot{\sigma}_{1}
\\
\dot{\sigma}_{2}
\end{bmatrix}
+
\begin{bmatrix}
m_{2}\dfrac{\tan\gamma}{\cos^{2}\gamma}
\\
\dfrac{J_{\rm F} }{l\cos^{2}\gamma}
\end{bmatrix}
\sigma_{1}\sigma_{2}
\\ =
\begin{bmatrix}
\dfrac{1}{r}\bigg(T_{\rm R} +\dfrac{T_{\rm F} }{\cos\gamma}\bigg)
\\
T_{\rm s}
\end{bmatrix}\ ,
\end{split}
\end{equation}
where $m_{1}$ and $m_{2}$ are given in (\ref{eqn:const_m1_m2},\ref{eqn:wheel_effect_mass}). Combining (\ref{eqn:111:dxdydpsidphi1dphi2dgamma}) with (\ref{eqn:111:dsigma1dsigma2}) yields the equations of motion
\begin{equation}\label{eqn:EOM_111}
\begin{split}
\dot{x}_{\rm G} &= \sigma_{1}\Big(\cos\psi-\frac{d}{l}\sin\psi\tan\gamma\Big)\ ,
\\
\dot{y}_{\rm G} &= \sigma_{1}\Big(\sin\psi+\frac{d}{l}\cos\psi\tan\gamma\Big)\ ,
\\
\dot{\psi} &= \frac{\sigma_{1}}{l}\tan\gamma\ ,
\\
\dot{\gamma} &= \sigma_{2}\ ,
\\
\dot{\sigma}_{1} &= \frac{\frac{1}{r}\Big(T_{\rm R} +\frac{T_{\rm F} }{\cos\gamma}\Big)
-\Big(m_{2}-\frac{J_{\rm F} }{l^{2}}\Big)\frac{\tan\gamma}{\cos^{2}\gamma}\sigma_{1}\sigma_{2}-\frac{T_{\rm s} }{l}\tan\gamma}{m_{1}+\big(m_{2}-\frac{J_{\rm F} }{l^{2}}\big)\tan^{2}\gamma}\, ,
\\
\dot{\sigma}_{2} &= \frac{-\frac{1}{l\,r}\left(T_{\rm R} +\frac{T_{\rm F} }{\cos\gamma}\right)\tan\gamma
-\frac{m_{1}}{l\cos^{2}\gamma}\sigma_{1}\sigma_{2}}{m_{1} +\big(m_{2}-\frac{J_{\rm F} }{l^{2}}\big)\tan^{2}\gamma}\ldots
\\
&\ldots\frac{+\frac{T_{\rm s} }{J_{\rm F} }\big(m_{1} +m_{2}\tan^{2}\gamma\big)}{}\ ,
\\
\dot{\varphi}_{\rm R} &= \frac{\sigma_{1}}{r}\ ,
\\
\dot{\varphi}_{\rm F} &= \frac{\sigma_{1}}{r}\frac{1}{\cos\gamma}\ .
\end{split}
\end{equation}
We remark that the first six equations are decoupled from the last two equations and that the first six equations are equivalent to those in the skate model \eqref{eqn:EOM_011} developed in Section~\ref{sec:model_skate_dynamic_SteerTrq}; cf.~\eqref{eqn:model_comp_F_T_match}.

\section{Discussion \label{sec:discussion}}

In this section, we highlight some of the important properties of the models and the corresponding equations of motion developed in Sections~\ref{sec:model_skate} and \ref{sec:model_wheels}. We also show some possible extensions of our mechanical models that may be helpful in a practical point of view.

\subsection{Pseudo Velocities and Singularities}\label{sec:singularities}

When introducing the Appellian approach in Section~\ref{sec:Appell}, we emphasized that the pseudo velocities were chosen such that the coefficient matrix $\mathbf{C}$ in \eqref{eqn:pseudo_and_kinconstraints} is not singular; cf.~\eqref{eqn:nonsingular}. Then, one may uniquely express the generalized velocities in terms of the pseudo velocities. Singularity may occur for certain values of the configuration coordinates due to the inappropriate selection of pseudo velocities or due to the physical structure of the mechanical model \cite{Varszegi_2019,YonaOr2019}. Here, we focus on the former case. We remark that in this paper, pseudo velocities are chosen not simply to avoid singularity, but also to have velocity components with clear physical meaning.

Let us investigate the models of Sections~\ref{sec:model_skate} and \ref{sec:model_wheels} in terms of the singularities. In Section~\ref{sec:model_skate_dynamic_assign}, we chose the longitudinal velocity of the vehicle as pseudo velocity $\sigma_1$ (cf.~\eqref{eqn:010_sigma}), which together with the kinematic constraints (\ref{eqn:skate_consraint_no_slip}) form the linear system
\begin{equation}\label{eqn:singularity_lin_eqn}
  \underbrace{\begin{bmatrix}
    \sin\psi & -\cos\psi & d
    \\
    \sin(\psi+\gamma) & -\cos(\psi+\gamma) & -(l-d)\cos\gamma
    \\
    \cos\psi & \sin\psi & 0
  \end{bmatrix}}_{=\mathbf{C}}
  \begin{bmatrix}
    \dot{x}_{\rm G} \\ \dot{y}_{\rm G} \\ \dot{\psi}
  \end{bmatrix}
  =\begin{bmatrix}
    0 \\ 0 \\ \sigma_{1}
  \end{bmatrix}\ .
\end{equation}
The determinant of the coefficient matrix reads $\det \mathbf{C} = l\cos\gamma$, which is only singular at ${|\gamma| =\pi/2}$. Since the steering angle $\gamma$ does not reach $\pi/2$ for conventional automobiles, $\sigma_{1}$ is an appropriate choice.

For the same model, one may also use one of the generalized velocities $\dot{x}_{\rm G}$, $\dot{y}_{\rm G}$ or $\dot{\psi}$ as pseudo velocity. This generic idea corresponds to the Lagrangian approach introduced in Section~\ref{sec:Lagrange}, which naturally chooses pseudo velocities as generalized velocities. For example, let us consider the choice
\begin{equation}\label{eqn:010_sigma_singular}
  \overline{\sigma}_1 := \dot{\psi},
\end{equation}
as pseudo velocity. This results in the linear system \begin{equation}\label{eqn:singularity_lin_eqn_singular}
  \underbrace{\begin{bmatrix}
    \sin\psi & -\cos\psi & d
    \\
    \sin(\psi+\gamma) & -\cos(\psi+\gamma) & -(l-d)\cos\gamma
    \\
    0 & 0 & 1
  \end{bmatrix}}_{=\overline{\mathbf{C}}}
  \begin{bmatrix}
    \dot{x}_{\rm G} \\ \dot{y}_{\rm G} \\ \dot{\psi}
  \end{bmatrix}
  =\begin{bmatrix}
    0 \\ 0 \\ \overline{\sigma}_1
  \end{bmatrix},
\end{equation}
where the determinant of the coefficient matrix is $\det \overline{\mathbf{C}} = \sin\gamma$, which is singular at ${\gamma = 0}$. This means that the corresponding equation of motion are singular for the rectilinear motion of the vehicle, i.e., they cannot describe the most common maneuver.
The choices $\overline{\overline{\sigma}}_1 := \dot{x}_{\rm G}$ and ${\widehat{\widehat{\sigma}}_1 := \dot{y}_{\rm G}}$ lead to the determinants
$\det \overline{\overline{\mathbf{C}}} = l\cos\psi\cos\gamma-d\sin\psi\sin\gamma$
and
$\det \widehat{\widehat{\mathbf{C}}} = l\sin\psi\cos\gamma+d\cos\psi\sin\gamma$,
respectively. That is, moving straight forward along the $y$-axis (${\psi=\pi/2}$ and ${\gamma=0}$) and along the $x$-axis (${\psi=0}$ and ${\gamma=0}$) lead to singular models when choosing $\overline{\overline{\sigma}}_1$ and $\widehat{\widehat{\sigma}}_1$, respectively. These singularities highlight the fundamental limitation of the Lagrangian approach compared to the Appellian approach: in Lagrangian case the choice of generalized coordinates predetermines the generalized velocities, while in the Appellian case the pseudo velocities can be chosen by the modeller.

One may notice that singularities can be completely eliminated by choosing the velocity of center of the front wheel along the wheel direction as pseudo velocity, i.e.,
\begin{equation}\label{eqn:010_sigma_non_singular}
  \widehat{\sigma}_1 := \dot{x}_{\rm G}\cos(\psi+\gamma) + \dot{y}_{\rm G}\sin(\psi+\gamma)+(l-d)\;\dot{\psi}\sin\gamma\ .
\end{equation}
This yields the linear equation
\begin{equation}\label{eqn:singularity_lin_eqn_non_singular}
  \underbrace{\begin{bmatrix}
    \sin\psi & -\cos\psi & d\\
    \sin(\psi+\gamma) & -\cos(\psi+\gamma) & -(l-d)\cos\gamma\\
    \cos(\psi+\gamma) & \sin(\psi+\gamma) & (l-d)\sin\gamma
  \end{bmatrix}}_{=\widehat{\mathbf{C}}}
  \begin{bmatrix}
    \dot{x}_{\rm G} \\ \dot{y}_{\rm G} \\ \dot{\psi}
  \end{bmatrix}
  =\begin{bmatrix}
    0 \\ 0 \\ \widehat{\sigma}_1
  \end{bmatrix}.
\end{equation}
The determinant of the coefficient matrix becomes $\det \widehat{\mathbf{C}}= l$, which is not singular for any value of the generalized coordinates and the steering angle. Consequently, the equations of motion will change to
\begin{equation}\label{eqn:EOM_010_frontwheel}
\begin{split}
\dot{x}_{\rm G} &= \widehat{\sigma}_1\Big(\cos\psi\cos\gamma-\frac{d}{l}\sin\psi\sin\gamma\Big)\ ,
 \\
\dot{y}_{\rm G} &= \widehat{\sigma}_1\Big(\sin\psi\cos\gamma+\frac{d}{l}\cos\psi\sin\gamma\Big)\ ,
\\
\dot{\psi} &= \frac{\widehat{\sigma}_1}{l}\sin\gamma\ ,
\\
\dot{\widehat{\sigma}}_{1} &= \frac{F_{\rm R} \cos\gamma+F_{\rm F}+(m_{1}-m_{2})\,\widehat{\sigma}_1\dot{\gamma}\sin\gamma\cos\gamma}{m_{1}\cos^{2}\gamma+m_{2}\sin^{2}\gamma}
\\
&  -\frac{\frac{J_{\rm F} }{l}\ddot{\gamma}\,\sin\gamma}{m_{1}\cos^{2}\gamma+m_{2}\sin^{2}\gamma}\,
\end{split}
\end{equation}
cf.~\eqref{eqn:EOM_010}.

We emphasize that when selecting pseudo velocities there is no ``recipe" how to avoid singularity. Only after the choice is made one may check the determinant of the coefficient matrix. In addition, eliminating singularity completely might not be the best choice. Instead, one may choose pseudo velocities which have clear physical interpretation, eliminate singularities under normal working conditions, and provide insights for control design. Considering these aspects, the longitudinal speed \eqref{eqn:010_sigma} may be a better choice for pseudo velocity than the speed \eqref{eqn:010_sigma_non_singular}.

\subsection{Constraining Forces}\label{sec:constrainforces}

All mechanical models used in this paper consider no-slip conditions at the wheel-ground contact points. To investigate the validity of these assumptions, one shall analyze the nonholonomic constraining forces. Since the Appellian approach does not provide information about these forces, one shall revert to the Lagrangian or the Newtonian approach. As an example, we consider the model developed in Section~\ref{sec:model_skate_dynamic_assign} and derive the lateral constraining forces acting perpendicular to the skate blades, which we denote by $\tilde{F}_{\rm R}$ and $\tilde{F}_{\rm F}$). If these exceed some critical limits (determined by the friction coefficient and the normal force), the wheels start to slip.

In Appendices~\ref{app:Lagrange} and \ref{app:model_newton}, we present the derivation of the equations of motion for the model presented in Section~\ref{sec:model_skate_dynamic_assign} using the Lagrangian and Newtonian methods, respectively. The derivations result in the lateral constraining forces \eqref{eqn:skate_F_lat_gen} expressed as function of the accelerations $\ddot{x}_{\rm G}$, $\ddot{y}_{\rm G}$, and $\ddot{\psi}$. These forces exhibit singularity at ${\gamma=0}$, corresponding to the discussion in Section~\ref{sec:singularities}. However, choosing the pseudo velocity appropriately this singularity can be avoided. In particular, one can eliminate $\ddot{x}_{\rm G}$, $\ddot{y}_{\rm G}$, and $\ddot{\psi}$ in \eqref{eqn:skate_F_lat_gen} by plugging in the first derivative of \eqref{eqn:010_dxdy} (cf.~\eqref{eqn:010:ddxddyddpsi}), and obtain formulas that depend on the velocities $\dot{x}_{\rm G}$, $\dot{y}_{\rm G}$, $\dot{\psi}$ and $\dot{\sigma}_{1}$. Then, substituting \eqref{eqn:EOM_010}, we can obtain the lateral constraining forces
\begin{equation}\label{eqn:010:F1F2}
\begin{split}
\tilde{F}_{\rm R} &=-\dfrac{(m_{2}-m_{4}) \tan\gamma}{m_{1}+m_{2} \tan^{2} \gamma}\,\bigg(F_{\rm R} +\dfrac{F_{\rm F} }{\cos\gamma}\bigg)
\\
&+(m_{1} - m_{4})\dfrac{\sigma_{1}^{2}}{l}\tan\gamma+\dfrac{m_{4} \sigma_{1}\dot{\gamma}}{\cos^{2} \gamma}
\\
&-\dfrac{m_{1}+m_{4}\tan^{2} \gamma}{m_{1}+m_{2} \tan^{2} \gamma}\bigg(\dfrac{m_{2} \sigma_{1}\dot{\gamma}}{\cos^{2} \gamma}+\dfrac{J_{\rm F} }{l} \ddot{\gamma}\bigg)\ ,
\\
\tilde{F}_{\rm F} &= \dfrac{1}{m_{1}+m_{2} \tan^{2} \gamma}\bigg(m_{2} F_{\rm R}  \dfrac{\tan\gamma}{\cos\gamma}+(m_{2}-m_{1}) F_{\rm F}  \tan\gamma
\\
&+m_{1} \dfrac{m_{2} \sigma_{1}\dot{\gamma} }{\cos^{3} \gamma} +m_{1} \dfrac{J_{\rm F} }{l}\dfrac{\ddot{\gamma}}{\cos\gamma} \bigg)
+m_{4} \dfrac{\sigma_{1}^{2}}{l}\dfrac{\tan\gamma}{\cos\gamma}\ ,
\end{split}
\end{equation}
which are singular at ${|\gamma|=\pi/2}$, corresponding to the choice of the pseudo velocity $\sigma_1$; cf.~the discussion after \eqref{eqn:singularity_lin_eqn}. Notations $m_{1}$, $m_{2}$ are the same as in \eqref{eqn:const_m1_m2}, and
\begin{align}
m_{4} &=m_{\rm F} + \frac{d}{l}m\ .
\end{align}

\subsection{Different Types of Drivetrain}

The skate models (\ref{eqn:EOM_010}) and (\ref{eqn:EOM_011}) developed in Sections~\ref{sec:model_skate_dynamic_assign} and \ref{sec:model_skate_dynamic_SteerTrq} include the driving forces $F_{\rm R} $ and $F_{\rm F} $ at the rear and front. This enables us to consider different drivetrains. In case of front wheel drive (FWD) vehicles, the active driving force at the rear is zero (except for braking), i.e., ${F_{\rm R}  =0}$ shall be used in the equations of motion. Analogously, for rear wheel drive (RWD) vehicles one shall substitute ${F_{\rm F}  =0}$ into the equations.

For all wheel drive (AWD) vehicles, the driving force distribution at the rear and front wheels are controlled by torque vectoring differentials. If the resultant active driving force on the wheels is $F_{\rm res}$ and torque split ratio is ${0\le \beta \le 1}$, then the active forces
\begin{equation}\label{eqn:trq_split}
  F_{\rm R}   = \beta\, F_{\rm res}\ ,
  \qquad
  F_{\rm F}   = (1-\beta) F_{\rm res}\ ,
\end{equation}
shall be substituted into the equations of motion \eqref{eqn:EOM_010} and \eqref{eqn:EOM_011}.

As mentioned before, the vehicle models developed in Sections~\ref{sec:model_skate_dynamic_assign} and \ref{sec:model_skate_dynamic_SteerTrq} with skates and driving forces are equivalent to the models with rigid wheels and driving torques developed in Sections~\ref{sec:model_wheels_dynamic_assign} and \ref{sec:model_wheels_dynamic_SteerTrq}. In particular, the formulas (\ref{eqn:EOM_010}) and (\ref{eqn:EOM_011}) can be matched with (\ref{eqn:EOM_110}) and (\ref{eqn:EOM_111}) using \eqref{eqn:model_comp_F_T_match}. Consequently, the above explained drivetrain scenarios can be adapted to these cases.

\subsection{Resistance Forces}

In the models developed above the resistance forces were neglected. Nevertheless in many driving scenarios these forces play an important role. Here we include these in the force driven models with skates (\ref{eqn:EOM_010}) and (\ref{eqn:EOM_011}), but analogously one may do the same for the torque driven models with rigid wheels (\ref{eqn:EOM_110}) and (\ref{eqn:EOM_111}).

When incorporating the road inclination, rolling resistance, and air drag in the model, the pseudo-force $\Pi_{1}$ given in (\ref{eqn:010_pseudo_force}) and (\ref{eqn:011_pseudo_force}) changes to
\begin{equation}
\begin{split}\label{eqn:01x_pseudo_force}
\Pi_{1} & = F_{\rm R} +\dfrac{F_{\rm F} }{\cos \gamma}-\zeta\, m_{1} g \cos\theta -m_{1} g \sin\theta
\\
&- \rho (v_{\rm w}+\sigma_{1})^{2},
\end{split}
\end{equation}
where $\zeta$ is rolling resistance coefficient, $\rho$ is air drag coefficient, $g$ is gravitational constant, $\theta$ is the inclination angle, and $v_{\rm w}$ is the velocity of the headwind. Thus, the equation for $\sigma_{1}$ in \eqref{eqn:EOM_010} and the equations for $\sigma_{1}$ and $\sigma_{2}$ in \eqref{eqn:EOM_011} change accordingly.

\subsection{Path-following Problem}

\begin{figure}[t]
\begin{center}
\includegraphics[scale=1]{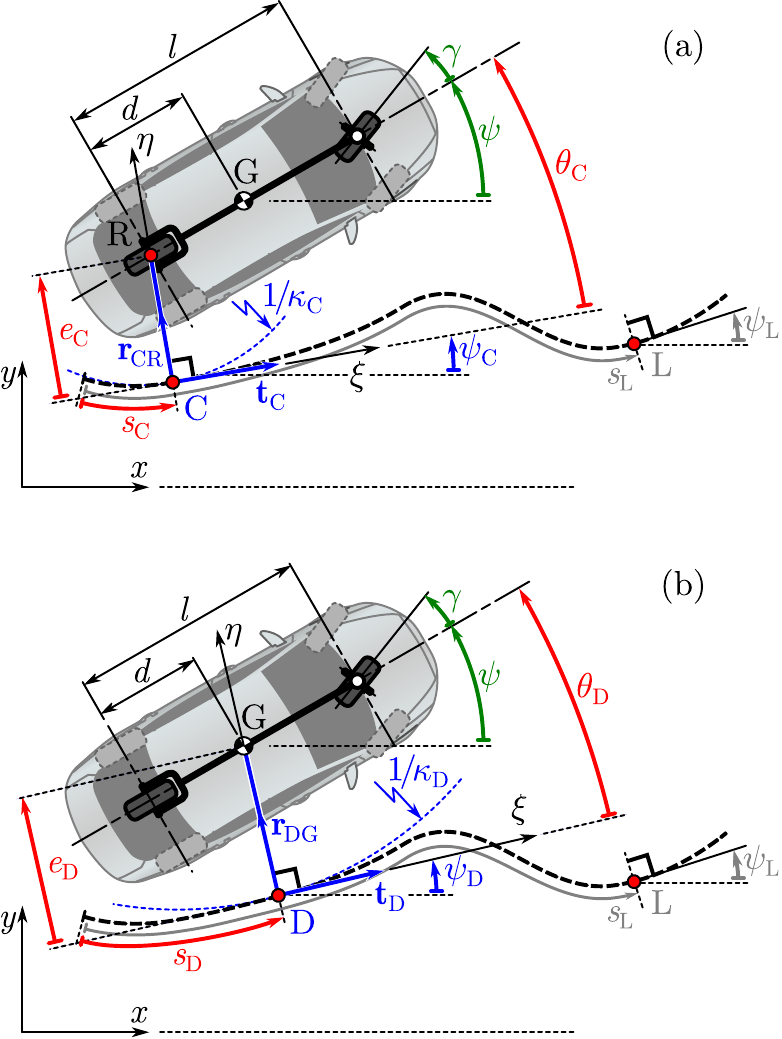}
\end{center}
\caption{Path-following problem when the information is about the point at the rear axle center point R (a) and the center of mass G (b). \label{fig:steer_ctrl}}
\end{figure}

An important utilization of the presented mechanical models relates to the path planning and path-following control of automated vehicles \cite{paden2016survey,gonzalez2016review,kellynagy2003}. It will be beneficial to transform the models, which are based on absolute position and orientation in an Earth-fixed frame, to relative position and orientation with respect to the given path \cite{RucNotHau2015,LeeTse2018,GohGoeGer2020,vanHoek2021}. Here we present an analytical method that can be used to execute this transformation for any planar vehicle model and we apply this to the nonholonomic models developed above.

Let us consider the scenario shown in Fig.~\ref{fig:steer_ctrl}(a), where the vehicle aims to follow a given path depicted by the black dashed curve. More precisely, our goal is to ensure the rear axle center point R can follow the black dashed path. We use the positions (${{x}_{\rm R} , {y}_{\rm R}  }$) of the rear axle center point R and the yaw angle $\psi$ as states to localize the vehicle in the ${(x,y)}$ plane. Point C marks the closest point to R along the desired path.
We assume the given path is second-order smooth and that closest point C is unique.
If $\mathbf{r}_{\rm CR}$ denotes the vector pointing from C to R and $\mathbf{t}_{\rm C}$ is the unit tangential vector of the path at point C, then ${\mathbf{t}_{\rm C} \perp \mathbf{r}_{\rm CR}}$. The angle $\psi_{\rm C}$ indicates the direction of tangential vector $\mathbf{t}_{\rm C}$ while the curvature of the path at point C is denoted by $\kappa_{\rm C}$. The path-reference frame ${(\xi,\eta)}$ travels along the path as the vehicle moves forward, and thus, the angle $\psi_{\rm C}(s_{\rm C})$ and the curvature $\kappa_{\rm C}(s_{\rm C})$ depends on the arclength coordinate $s_{\rm C}$ of the path. We assume that the information about the desired path is known, namely, the tuple ${(x_{\rm C},\, y_{\rm C},\, \psi_{\rm C},\, \kappa_{\rm C})}$ is given as a function of the arc length $s_{\rm C}$.

\begin{table*}[!t]
\caption{Path-reference frame models with skates and their governing equations. \label{tab:modelswithskates_path_ref_frame}}
\begin{center}
\includegraphics[scale=1]{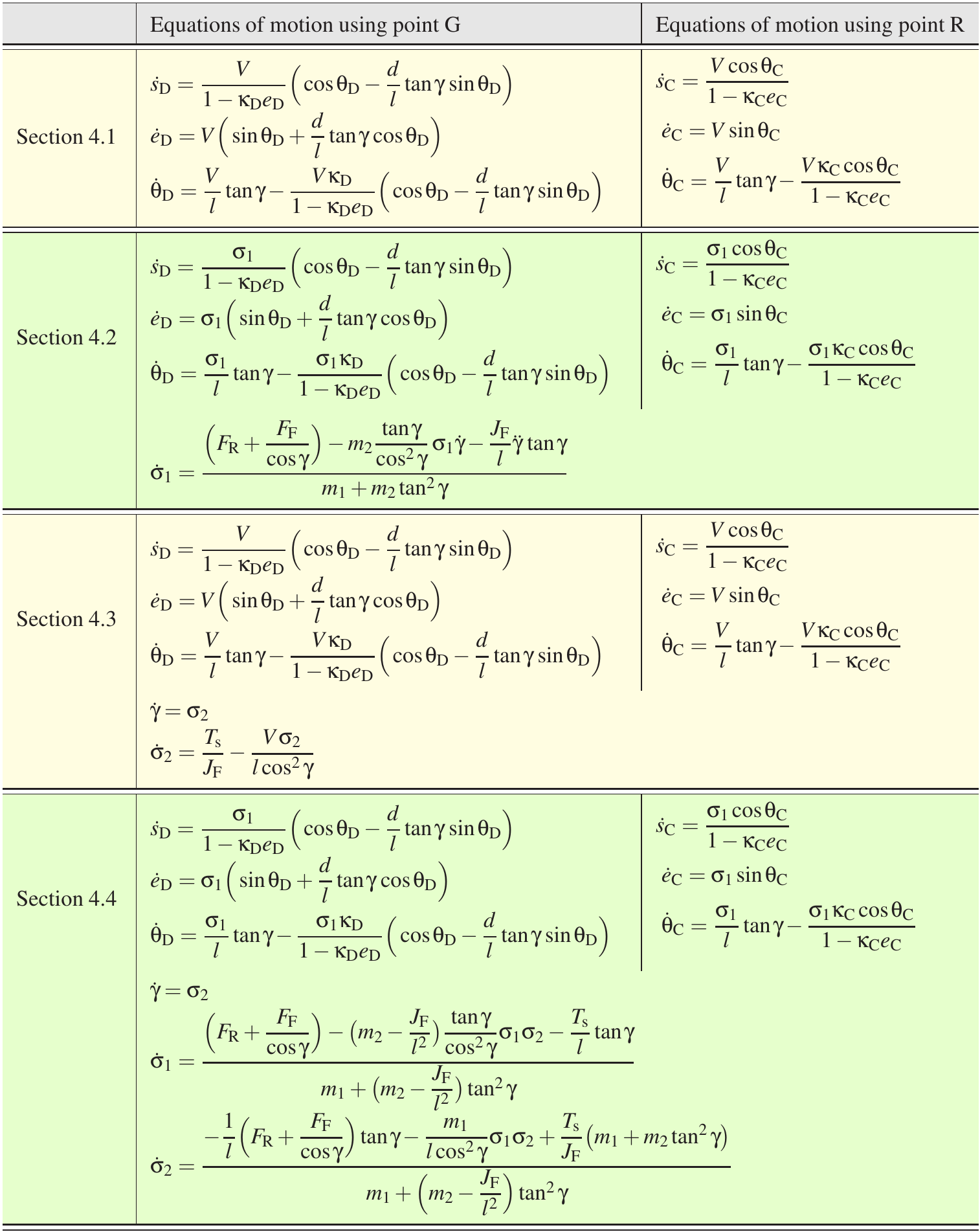}
\end{center}
\end{table*}

To follow the desired path, a controller has to correct the lateral deviation and the relative yaw angle with respect to the path. The lateral deviation can be defined as
\begin{align}
e_{\rm C}&=(\mathbf{t}_{\rm C} \times \mathbf{r}_{\rm CR})\cdot \mathbf{k}\ ,
\end{align}
which is positive/negative when point R is on the left/right hand side of the path.
Then, by expressing the tangential vector $\mathbf{t}_{\rm C}$ with the angle $\psi_{\rm C}$, we can obtain the lateral deviation as
\begin{align}\label{eqn:lat_err}
  e_{\rm C}
    &=-({x}_{\rm R} -x_{\rm C})\sin\psi_{\rm  C}+({y}_{\rm R}  -y_{\rm C})\,\cos\psi_{\rm  C}\ .
\end{align}
Similarly, one can define the relative yaw angle as
\begin{align}\label{eqn:heading_err}
  \theta_{\rm C}&=\psi-\psi_{\rm C}\ .
\end{align}

We remark that to ensure ${\theta_{\rm C}\in [-\pi,\,\pi)}$, one can generalize this definition as ${\theta_{\rm C}=\psi-\psi_{\rm C}-2\pi\left[\frac{\psi-\psi_{\rm C}}{2\pi}\right]}$, where $[\cdot]$ refers to the round function that rounds to the nearest integer. We will use this generalized definition in the simulations presented in Section~\ref{sec:control}.

To facilitate path-following controller design, we transform the absolute position and orientation ${(x,y,\psi)}$ expressed in the Earth-fixed frame to the relative position and orientation $(s_{\rm C}, e_{\rm C}, \psi_{\rm C})$ with respect to the path; see \cite{Samson_ICARCV_1992,Samson_TAC_1995,DeLuca_Planning_1998,Snider_2009}. First, we derive the coordinate transformation \eqref{eqn:transf_s_eta_deriv_2} for an arbitrary point in differential form in Appendix~\ref{append:coord_trans}. Then, we apply this to the rear axle center point R (see Fig.~\ref{fig:steer_ctrl}(a)), i.e., substitute ${x={x}_{\rm R}}$, ${y={y}_{\rm R}}$, ${\xi \equiv 0}$, ${\eta = e_{\rm C}}$ and ${{\mathrm{\Omega}} \equiv {\rm C}}$, which leads to
\begin{equation}\label{eqn:transf_s_eta_deriv_3}
\begin{split}
 \dot{s}_{\rm C}&=\dfrac{\cos\psi_{\rm C} }{1-\kappa_{\rm C}e_{\rm C}}\dot{x}_{\rm R} +\dfrac{  \sin \psi_{\rm C}}{1-\kappa_{\rm C}e_{\rm C}}\dot{y}_{\rm R}\ ,
 \\
 \dot{e}_{\rm C} & =-\dot{x}_{\rm R} \sin\psi_{\rm C}+\dot{y}_{\rm R}  \cos \psi_{\rm C}\ .
\end{split}
\end{equation}
The evolution of relative yaw angle can be determined by differentiating \eqref{eqn:heading_err} with respect to time and using \eqref{eqn:EOM_ref_path_dt} for ${\mathrm{\Omega}} \equiv {\rm C}$  with \eqref{eqn:transf_s_eta_deriv_3}:
\begin{equation}\label{eqn:transf_theta_deriv}
\dot{\theta}_{\rm C} = -\dfrac{\kappa_{\rm C}\cos\psi_{\rm C} }{1-\kappa_{\rm C}e_{\rm C}}\dot{x}_{\rm R} -\dfrac{\kappa_{\rm C}  \sin \psi_{\rm C}}{1-\kappa_{\rm C}e_{\rm C}}\dot{y}_{\rm R}+\dot{\psi}\ .
\end{equation}

We remark that \eqref{eqn:transf_s_eta_deriv_3} and \eqref{eqn:transf_theta_deriv} define a nonlinear transformation from absolute state ${({x}_{\rm R},\, {y}_{\rm R},\,\psi )}$ to relative state ${(s_{\rm C},\, e_{\rm C},\, \theta_{\rm C})}$. The inverse transformation in differential form is
\begin{equation}
\begin{split}
 \dot{x}_{\rm R} &=(1-\kappa_{\rm C}e_{\rm C})\dot{s}_{\rm C}\cos\psi_{\rm C}-\dot{e}_{\rm C}\sin\psi_{\rm C}\ ,
 \\
 \dot{y}_{\rm R}  &=(1-\kappa_{\rm C}e_{\rm C})\dot{s}_{\rm C}\sin\psi_{\rm C}+\dot{e}_{\rm C}\cos\psi_{\rm C}\ ,
 \\
 \dot{\psi} & = \kappa_{\rm C}\dot{s}_{\rm C}+\dot{\theta}_{\rm C}\ .
\end{split}
\end{equation}
One can also obtain the inverse transform in coordinate form as
\begin{equation}\label{eqn:xy_transf}
  \begin{split}
    {x}_{\rm R}  & =x_{\rm C} -e_{\rm C}\sin\psi_{\rm C}\ ,
    \\
    {y}_{\rm R}   & =y_{\rm C} +e_{\rm C}\cos\psi_{\rm C}\ ,
    \\
    \psi & = \psi_{\rm C} +\theta_{\rm C}\ ,
  \end{split}
\end{equation}
by solving (\ref{eqn:xy_s_eta_transf}, \ref{eqn:heading_err}) where ${x={x}_{\rm R}}$, ${y={y}_{\rm R}}$, ${\xi \equiv 0}$, ${\eta = e_{\rm C}}$ and ${\mathrm{\Omega} \equiv {\rm C}}$ are substituted into \eqref{eqn:xy_s_eta_transf}.

Now, let us use the kinematic model of Section~\ref{sec:model_skate_kinematic_asign} as an example. Substituting the governing equations of the model (see the first row of Table~\ref{tab:modelswithskates}) into \eqref{eqn:transf_s_eta_deriv_3} and \eqref{eqn:transf_theta_deriv}, we obtain
\begin{equation}\label{eqn:EOM_s_err}
\begin{split}
 \dot{s}_{\rm C}&=\dfrac{V\cos\theta_{\rm C}}{1-\kappa_{\rm C}e_{\rm C}}\ ,
 \\
 \dot{e}_{\rm C} & =V\sin\theta_{\rm C}\ ,
 \\
 \dot{\theta}_{\rm C} &=\dfrac{V}{l}\tan\gamma-\dfrac{V\kappa_{\rm C}\cos\theta_{\rm C}}{1-\kappa_{\rm C}e_{\rm C}}\ ,
\end{split}
\end{equation}
where the first equation characterizes the longitudinal motion of the point C along the path, while the last two equations provide the evolution of lateral deviation and relative yaw angle with respect to the path.
Note again that the curvature $\kappa_{\rm C}(s_{\rm C})$ depends on the path coordinate $s_{\rm C}$, that is, the differential equations in \eqref{eqn:EOM_s_err} are all coupled.

The transformation described above can be applied to all mechanical models presented in this paper. We summarize the related formulas in Table~\ref{tab:modelswithskates_path_ref_frame} for the models of Section~\ref{sec:model_skate}. In this table, we also present the case when the relative position and orientation are calculated at the center of mass G with respect to the path; see Fig.~\ref{fig:steer_ctrl}(b). As shown, the transformed formulas are more complicated for this latter case. Note that the differential equations in Table~\ref{tab:modelswithskates_path_ref_frame} can also be used to describe the dynamics of the mechanical models of Section~\ref{sec:model_wheels} in terms of relative position and orientation. The remaining governing equations can be collected from Table~\ref{tab:modelswithwheels}; see the parts highlighted by dark shading.

In the next section, we study path-following control design using the transformed nonlinear dynamics \eqref{eqn:EOM_s_err}. The states are the longitudinal position $s_{\rm C}$, the lateral deviation $e_{\rm C}$, the relative yaw angle $\theta_{\rm C}$, and the input is the steering angle $\gamma$. We assume that  $e_{\rm C}$ and $\theta_{\rm C}$ can be measured in real-time with the help of optical sensors. Thus, the control objective is to design the input $\gamma$ based on the outputs $e_{\rm C}$ and $\theta_{\rm C}$  to ensure that the vehicle  approaches a path (of second order smoothness), and then traces it perfectly while keeping both errors $e_{\rm C}$ and $\theta_{\rm C}$ zero at the same time. In other words, the closed-loop system must possess a stable steady-state motion ${s_{\rm C}=Vt}$, ${e_{\rm C}\equiv0}$, ${\theta_{\rm C}\equiv0}$.

Note that such perfect tracking is possible when the rear axle center point R is used, since the nonlinear control system is differentially flat
\cite{MarMurRou2003,AgrParCosRosAmePan2021,MurRatSlu1995,FliLevMarRou1995,Lev2009}. In other words, one may define flat outputs and a nonlinear transformation such that, in an extended state space, the system can be described by linear differential equations.
Once using the center of mass G, it is not possible to keep the errors $e_{\rm D}$ and $\theta_{\rm D}$ zero simultaneously, except when the desired path is straight; see \cite{Wubing_TIV_2022}. Indeed, the corresponding control system (see the first row of Table~\ref{tab:modelswithskates_path_ref_frame}) is not differentially flat. While proving that a system is not differentially flat is far from trivial, this means physically that when the vehicle follows a curve, the center of mass G has nonzero lateral velocity.

\section{Controller Design and Simulations}\label{sec:control}

In this section, we focus on the path-following control problem of automated vehicles. We acknowledge that path-following control has a long history starting in the 1950s \cite{Segel_1956} and that many different sophisticated control techniques have been utilized over the years including model predictive control \cite{Borrelli_IJVAS_2005,Falcone_VSD_2008,BerQuiUnoCai2020,Lietal2020}, Lyapunov-based control \cite{Rossetter_2006,Talvala_Gerdes_2011}, sliding mode control \cite{Choi_Hedrick_2015_ECC},
look-ahead/preview control \cite{Andersen_AIM_2016,XuPenTan_2021}, and machine learning-based control \cite{Bae2020,Ave2021}, just to mention a few. Here we propose a low-complexity nonlinear controller and demonstrate its high performance on the model developed above. The controller consists of a feedforward term and a nonlinear feedback term. In case of small lateral deviation and relative yaw angle, the latter one
is equivalent to widely used linear controllers; see, for example, \cite{Chatzikomis_2009}.

We start with the kinematic bicycle model developed in Section~\ref{sec:model_skate_kinematic_asign}; see the first rows of Table~\ref{tab:modelswithskates} and Table~\ref{tab:modelswithskates_path_ref_frame} with the original and path-reference states, respectively. In this model, the steering angle $\gamma$ is assigned, that is, it can track any desired value perfectly:
\begin{align}\label{eqn:steer_angle_track}
  \gamma &= \gamma_{\rm des}\ .
\end{align}
The desired steering angle $\gamma_{\rm des}$ is determined by the path-following controller, the goal of which is to drive the rear axle point R along a given path while making the relative yaw angle zero; see Fig.~\ref{fig:steer_ctrl}(a).

Below we design a controller that rely on feedforward and feedback actions \cite{Astrom_Murray_2008}. In particular, we propose
\begin{align}\label{eqn:steer_controller}
  \gamma_{\rm des} & = \gamma_{\rm ff} + \gamma_{\rm fb}\ ,
\end{align}
which consists of the feedforward control
\begin{align}
  \gamma_{\rm ff} &= \arctan (\kappa_{\rm C}\,l)\ , \label{eqn:steer_controller_ff}
\end{align}
and the feedback control
\begin{align}
  \gamma_{\rm fb} &= g \Big( k_{1}\big(\theta_{\rm C}+\arctan (k_{2}\,e_{\rm C})\big)\Big)\ . \label{eqn:steer_controller_fb}
\end{align}
Here $k_{1}$ and $k_{2}$ are the tunable control gains, and $g (x)$ denotes a wrapper function with the following properties:
\begin{itemize}
  \item[$\bullet$] It is continuously differentiable and monotonically increasing over $\mathbb{R}$.
  \item[$\bullet$] It is an odd function, i.e., ${g(x)=-g(-x)}$ for ${x\in \mathbb{R}_{\geq0}}$.
  \item[$\bullet$] It is bounded by $g_{\rm sat}$, i.e.,
  ${g(x) \le g_{\rm sat}}$ for ${x\in \mathbb{R}_{\geq0}}$.
  \item[$\bullet$] Its derivative decreases monotonically for ${x\in \mathbb{R}_{\geq0}}$ such that ${g'(0)=1}$ and ${\lim_{x \to \infty}  g'(x) = 0}$.
\end{itemize}
In this section, we use the wrapper function
\begin{equation}
    g(x) =\dfrac{2\,g_{\rm sat}}{\pi}\arctan \Big(\dfrac{\pi}{2\,g_{\rm sat}} x\Big)\ , \label{eqn:satfunction}
\end{equation}
which is selected from a larger family of wrapper functions as described in Appendix~\ref{app:wrapper}.

In Sections~\ref{sec:feedforward} and \ref{sec:feedback}, we present the details of the feedforward control law \eqref{eqn:steer_controller_ff} and feedback control law \eqref{eqn:steer_controller_fb}. Readers, who are not interested in these details, may choose to jump to Section~\ref{sec:stab_analysis} for the stability analysis or to Section~\ref{sec:sim_res} for simulation results.

\subsection{Feedforward Control Design}\label{sec:feedforward}

The feedforward control given in \eqref{eqn:steer_controller_ff} is similar to pure pursuit algorithm \cite{Snider_2009,Park_CAS_2014,Andersen_AIM_2016}. To determine the optimal feedforward term, one can use the governing equations \eqref{eqn:EOM_s_err} of the model given in the path-reference frame. Consider that the center of the rear axle R follows the path precisely such that the lateral deviation and relative yaw angle are zeros, i.e., ${e_{\rm C}\equiv 0}$, ${\theta_{\rm C}\equiv 0}$; cf.~Fig.~\ref{fig:steer_ctrl}(a). Substituting these and $\gamma=\gamma_{\rm ff}$ into \eqref{eqn:EOM_s_err} leads to
\begin{equation}\label{eqn:ff_calc0}
\begin{split}
    \dot{s}_{\rm C} &= V\ ,
    \\
    0 & = 0\ ,
    \\
 0 &=\dfrac{V}{l}\tan\gamma_{\rm ff}-V\kappa_{\rm C}\ ,
\end{split}
\end{equation}
where the third equation can be satisfied with the feedforward control law
\eqref{eqn:steer_controller_ff}. We remark that, in case of  path-following with respect to the center of gravity G (cf.~Fig.~\ref{fig:steer_ctrl}(b)), one should determine the feedforward control action by finding the steady-state solution of the governing equations given for point G in the first row of Table~\ref{tab:modelswithskates_path_ref_frame}; see \cite{Wubing_TIV_2022}.

The feedforward controller may predict the desired steering angle perfectly to ensure fast response but cannot correct the errors caused by the initial state and/or disturbances. This requires the usage of  feedback control as discussed below.

\subsection{Feedback Control Design}\label{sec:feedback}

The nonlinear feedback control law \eqref{eqn:steer_controller_fb} allows the vehicle to correct the steering angle for both small and large values of the errors $e_{\rm C}$ and $\theta_{\rm C}$. For small errors one may neglect the nonlinearities and obtain the linear controller
\begin{align}
  \gamma^{0}_{\rm fb} & = k_1  \theta_{\rm C} + k_1 k_2 e_{\rm C}\ , \label{eqn:steer_linear_fb}
\end{align}
which is widely used in the literature \cite{Chatzikomis_2009}. However,
as demonstrated below, for larger errors this linear feedback controller may produce unwanted behaviors.

\begin{figure}[!t]
  \centering
  \includegraphics[scale=1]{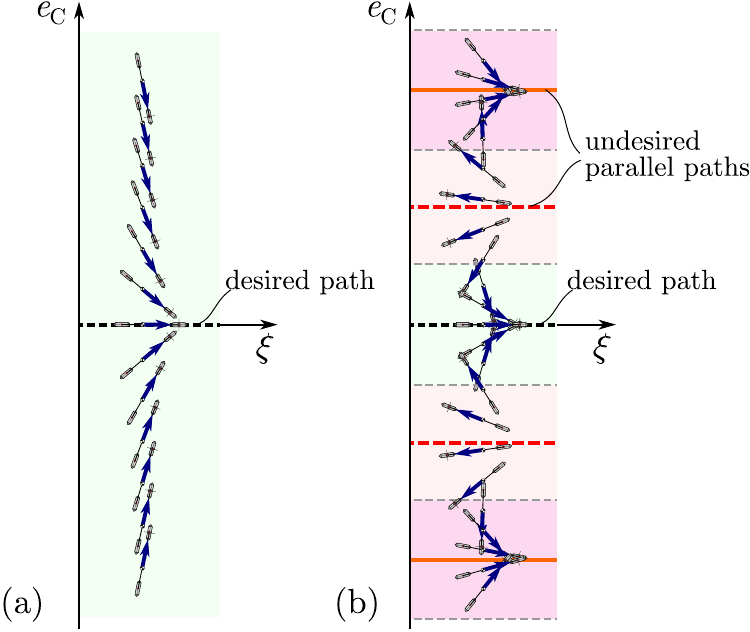}\\
  \caption{(a) Desired vehicle heading with nonlinear controller \eqref{eqn:steer_nonlinear_fb}. (b) Desired vehicle heading with linear controller \eqref{eqn:steer_linear_fb}.
  \label{fig:desiredtheta}}
\end{figure}

Substituting the wrapper function with the identity map in \eqref{eqn:steer_controller_fb}
yields
\begin{align}
  \gamma^{1}_{\rm fb}  =k_{1}\big(\theta_{\rm C}+\arctan (k_{2}\,e_{\rm C})\big)\ , \label{eqn:steer_nonlinear_fb}
\end{align}
see also \cite{Lee_2013}. One may interpret this controller as trying to achieve the desired relative yaw angle
\begin{align}\label{eqn:desired_heading}
  \theta^{\rm des}_{\rm C}  = - \arctan (k_{2}\,e_{\rm C})\ ,
\end{align}
depending on the lateral deviation $e_{\rm C}$. Fig.~\ref{fig:desiredtheta}(a) depicts the desired heading of the vehicle as a function of the lateral deviation $e_{\rm C}$ where the desired path is given by the black dashed line. Notice that when the vehicle is far from the path, the desired heading points toward the path since
$ e_{\rm C}\rightarrow \pm \infty$ yields $\theta^{\rm des}_{\rm C} \rightarrow \mp \pi/2$.

Similarly, one may interpret the linear controller \eqref{eqn:steer_linear_fb} as trying to achieve the desired relative yaw angle
\begin{align}
  \theta^{\rm des,0}_{\rm C}  = - k_{2}\,e_{\rm C}\ ,
\end{align}
depending on the lateral deviation $e_{\rm C}$. The pictographs of the vehicle depicted in Fig.~\ref{fig:desiredtheta}(b) show that this may result in wrong desired headings when the vehicle is far from the path due to the $2\pi$ periodicity of the angle. For example, the linear controller requires the vehicle to drive parallel to the path when the error is $e_{\rm C} = j \pi/k_2$, $j=\pm 1, \pm 2, \ldots$.

Fig.~\ref{fig:desiredtheta} illustrates that the nonlinear controller \eqref{eqn:steer_nonlinear_fb} is able to provide the appropriate steering effort for both small and large errors. This is a large improvement compared to most path-following controllers, which require the lateral deviation and the relative yaw angle to be small once the controller is engaged. As it will be demonstrated below, due to the proper handling of large errors, our controller can be used to follow a large variety of paths with varying curvature.

\begin{figure}[!t]
  \centering
  \includegraphics[scale=0.9]{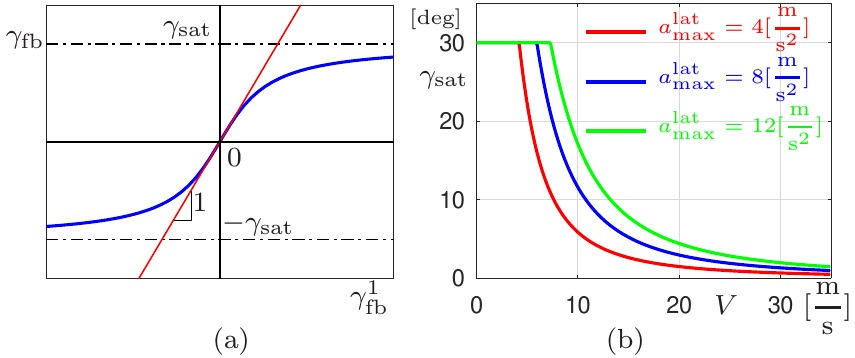}\\
  \caption{(a) The wrapper function  \eqref{eqn:satfunction}. (b) Maximum allowable steering angle \eqref{eqn:max_allow_steer_ang} at different speed.
  \label{fig:fun_arctan_steer_max}}
\end{figure}

The nonlinear controller \eqref{eqn:steer_nonlinear_fb} uses the same control gains for small and large errors. However, we prefer larger gains for small errors to ensure good tracking performance and prefer smaller gains for large errors to avoid ``overreaction" and potential oscillations. Thus, we apply the wrapper function $g(x)$ defined in \eqref{eqn:satfunction} with $g_{\rm sat}=\gamma_{\rm sat}$ yielding the nonlinear controller \eqref{eqn:steer_controller_fb}; see Fig.~\ref{fig:fun_arctan_steer_max}(a). Since the derivative of the wrapper function decreases with $|x|$, the effective gains are reduced as the errors increase. More details about wrapper functions can be found in Appendix~\ref{app:wrapper}.

When setting the allowable steering angle $\gamma_{\rm sat}$ of the feedback controller, one can simply choose a value that is smaller than the physical steering angle limit of the vehicle. However, this may lead to passenger discomfort and even to roll over hazards for high speed. In order to avoid this, the allowable steering angle can be determined from the lateral acceleration. Using the equations of motion given in the first row of Table~\ref{tab:modelswithskates}, one can obtain the lateral acceleration of the rear axle center point R as
\begin{align}
  a_{\rm R}^{\rm lat} &= -\ddot{x}_{\rm R} \sin\psi + \ddot{y}_{\rm R}  \cos\psi = \dfrac{V^{2}}{l}\tan\gamma\, ,
\end{align}
which only depends on the longitudinal speed and the steering angle. Note, that this property still holds when longitudinal speed is not a constant. Thus, we set the maximum allowable steering angle as
\begin{align}
  \gamma_{\rm sat} = \min\left\{\gamma_{\max},\, \arctan\Big(\dfrac{a_{\max}^{\rm lat}\,l}{V^2}\Big)\right\}\ ,
  \label{eqn:max_allow_steer_ang}
\end{align}
where $\gamma_{\max}$ is the physical steering angle limit, $a_{\max}^{\rm lat}$ is the maximum allowed lateral acceleration. Fig.~\ref{fig:fun_arctan_steer_max}(b) shows the maximum allowable steering angle as a function of the longitudinal speed for different lateral acceleration limits $a_{\max}^{\rm lat}$ when $l=2.57$~m and $\gamma_{\max}=30$~deg.

\subsection{Stability Analysis\label{sec:stab_analysis}}

In this part, we analyze the linear stability of the proposed controller. We use the governing equations \eqref{eqn:EOM_s_err} of the model given in the path-reference frame. By substituting the controller (\ref{eqn:steer_controller},\ref{eqn:steer_controller_ff},\ref{eqn:steer_controller_fb}) with the wrapper function \eqref{eqn:satfunction} into \eqref{eqn:EOM_s_err}, we obtain the closed-loop dynamics
\begin{equation}\label{eqn:closed_reduced_nonlinear_dyn}
\begin{split}
 \dot{s}_{\rm C}&=\dfrac{V\cos\theta_{\rm C}}{1-\kappa_{\rm C}e_{\rm C}}\ ,
 \\
 \dot{e}_{\rm C} &= V\sin\theta_{\rm C}\ ,
 \\
 \dot{\theta}_{\rm C} &=-\dfrac{V\kappa_{\rm C}\cos\theta_{\rm C}}{1-\kappa_{\rm C}e_{\rm C}}+\dfrac{V}{l}\tan\bigg( \arctan(\kappa_{\rm C}l)
 \\
 &+\dfrac{2\,\gamma_{\rm sat}}{\pi}\arctan\Big(\dfrac{k_{1}\pi}{2\,\gamma_{\rm sat}}\big(\theta_{\rm C}+\arctan (k_{2}\,e_{\rm C})\big) \Big) \bigg)\ .
\end{split}
\end{equation}

One can verify that \eqref{eqn:closed_reduced_nonlinear_dyn} possesses the desired steady-state solution
\begin{equation}\label{eqn:equil}
  s_{\rm C}^{\ast}  =V t\ ,
  \qquad
  e_{\rm C}^{\ast}  =0\ ,
  \qquad
  \theta_{\rm C}^{\ast}  =0\ ,
\end{equation}
which corresponds to the vehicle following the path perfectly. We assume that the road curvature $\kappa$ varies around a constant value $\kappa^{\ast}$. By defining the input perturbation as
\begin{equation}
 \tilde{\kappa}_{\rm C} = \kappa_{\rm C}-\kappa^{*}\ ,
\end{equation}
and the state perturbations as
\begin{equation}
 \tilde{s}_{\rm C} = s_{\rm C}-s_{\rm C}^{*}\ ,
 \quad
 \tilde{e}_{\rm C} = e_{\rm C}-e_{\rm C}^{*}\ ,
 \quad
 \tilde{\theta}_{\rm C} = \theta_{\rm C}-\theta_{\rm C}^{*}\ ,
\end{equation}
we can derive the linearized dynamics of \eqref{eqn:closed_reduced_nonlinear_dyn} as
\begin{equation}\label{eqn:closed_lin_dyn}
\begin{split}
 \dot{\tilde{s}}_{\rm C}&= V \kappa^{*}\tilde{e}_{\rm C}\ ,
 \\
 \dot{\tilde{e}}_{\rm C} &= V\tilde{\theta}_{\rm C}\ ,
 \\
 \dot{\tilde{\theta}}_{\rm C} &= \frac{V}{l}\big(k_{1}k_{2}+k_{1}k_{2}\kappa^{*2}l^{2}-\kappa^{*2}l\big)\tilde{e}_{\rm C}\\
 &+\frac{V}{l}k_{1}\big(1+\kappa^{*2}l^{2}\big)\tilde{\theta}_{\rm C}\ .
\end{split}
\end{equation}
Notice that \eqref{eqn:closed_lin_dyn} is free from the input perturbation $\tilde{\kappa}_{\rm C}$, implying that the linearized system possesses zero input-to-output response. In other words, as long as the system \eqref{eqn:closed_lin_dyn} is stable, the controller ensures that the vehicle can perfectly track given paths with varying curvatures as well. This is due to the fact that \eqref{eqn:equil} is always a solution to \eqref{eqn:closed_reduced_nonlinear_dyn} regardless of the given path.
Observe that the last two equations in \eqref{eqn:closed_lin_dyn} are decoupled from the first one
and they can be written as
\begin{equation}\label{eqn:lin_model}\footnotesize
\begin{split}
  \begin{bmatrix}
  \dot{\tilde{e}}_{\rm C} \\ \dot{\tilde{\theta}}_{\rm C}
  \end{bmatrix}
  \!=\!
\underbrace{
\begin{bmatrix}
  0 & V
  \\
  \frac{V}{l}\big(k_{1}k_{2}+k_{1}k_{2}\kappa^{*2}l^{2}-\kappa^{*2}l\big) & \frac{V}{l}k_{1}\big(1+\kappa^{*2}l^{2}\big)
  \end{bmatrix}
  }_{\mathbf{A}}
  \!
  \begin{bmatrix}
  \tilde{e}_{\rm C} \\ \tilde{\theta}_{\rm C}
  \end{bmatrix}.
\end{split}
\end{equation}
The corresponding characteristic equation becomes
\begin{equation}
\begin{split}
  \det&(s\,\mathbf{I}-\mathbf{A}) = s^{2} -\frac{V k_{1}}{l}\big(1+\kappa^{*2}l^{2}\big)s
  \\
  &-\frac{V^{2}}{l}\big(k_{1}k_{2}+k_{1}k_{2}\kappa^{*2}l^{2}-\kappa^{*2}l\big)=0\ ,
  \label{eqn:char_eqn_straight}
\end{split}
\end{equation}
where $s\in \mathbb{C}$ denotes the characteristic root. To ensure that system \eqref{eqn:lin_model} is stable, the root of \eqref{eqn:char_eqn_straight} must be in the left half complex plane. Applying the Routh-Hurwitz criteria, we obtain the stability condition
\begin{equation}\label{eqn:stab_cond_0}
  k_{1} <0\ ,
  \qquad
  k_{1}k_{2} <\dfrac{\kappa^{*2}l}{1+\kappa^{*2}l^{2}}\ .
\end{equation}
Notice that
\begin{equation}
 0\le \dfrac{\kappa^{*2}l}{1+\kappa^{*2}l^{2}} < \dfrac{1}{l}\ ,
\end{equation}
leads to the sufficient condition
\begin{equation}\label{eqn:stab_cond_1}
  k_{1} <0\ ,
  \qquad
  k_{2} >0\ ,
\end{equation}
which is independent of the constant path curvature $\kappa^{*}$.

\subsection{Simulation Results\label{sec:sim_res}}

Above we showed that by choosing appropriate control gains the closed-loop system can be stabilized when the vehicle  follows a path of constant curvature ${\kappa_{\rm C}\equiv\kappa^{*}}$. Therefore, in this section, we first show the performance of the controller when the path is either a straight line (${\kappa^{*}=0}$) or a circle of radius $\rho$ (${\kappa^{*} =1/\rho}$). Then we design a closed path with varying curvature and demonstrate that the proposed controller is capable of following such path as well. The parameters used for the simulations in this section are provided in Table~\ref{tab:veh_param} and we use the longitudinal speed $V=20$~m/s.

\begin{table}[!ht]
\begin{center}
\begin{tabular}{c||l|c}
\hline\hline
\rowcolor{Gray}  & Parameter & Value 
\\
\hline\hline
 \cellcolor{white}& $l$ [m]& $2.57$ 
 \\
 \cellcolor{white}& $d$ [m] & $1.54$ 
 \\
 \cellcolor{white}& $m$ [kg]& $1770$ 
 \\
 \cellcolor{white}Original& $m_{\rm R} $ [kg]& $10$ 
 \\
 \cellcolor{white}Physical& $m_{\rm F} $ [kg] & $10$ 
 \\
 \cellcolor{white}Parameters& $J_{\rm G} $ [$\rm kg\cdot m^{2}$] & $1343$ 
 \\
 \cellcolor{white}& $J_{\rm R}  $ [$\rm kg\cdot m^{2}$]& $0.25$ 
 \\
 \cellcolor{white}& $J_{\rm F} $ [$\rm kg\cdot m^{2}$] & $0.25$ 
 \\
 \cellcolor{white}&$\gamma_{\max}$ [deg]& $30$ 
 \\
\hline\hline
 \cellcolor{white}Derived & $m_{1}$ [kg]& $1790$ 
 \\
 \cellcolor{white}Physical& $m_{2}$ [kg]& $848.9$ 
 \\
 \cellcolor{white}Parameters& $m_{4}$ [kg]& $1070.6$ 
 \\
\hline\hline
 \cellcolor{white}& $k_{1}$ [m/s]& $-0.5$ 
 \\
 \cellcolor{white}& $k_{2}$ [m$^{-1}$]& $0.02$ 
 \\
 \cellcolor{white}& $a_{\max}^{\rm lat}$ [m/s$^{2}$] & $4$ 
 \\
 \cellcolor{white}& $k_{\rm s}$ [$\rm N\cdot m$]& $-6$ 
 \\
 \cellcolor{white}Control & $T_{\rm sat}$ [$\rm N\cdot m$]& $1$ 
 \\
 \cellcolor{white}Parameters & $V$ [m/s]& $20$ 
 \\
 \cellcolor{white}& $k_{\rm a}$ [s$^{-1}$]& $-5$ 
 \\
 \cellcolor{white}& $a_{\max}^{\rm long}$ [m/s$^{2}$]& $6$ 
 \\
 \cellcolor{white}& $v_{\max}$ [m/s]& $30$ 
 \\
\hline\hline
 \cellcolor{white}Path & $s_{\rm T}$ [m]& $250$ 
 \\
 \cellcolor{white}Parameters& $N$ & $4$ 
 \\
 \cellcolor{white}& $\kappa_{\max}$ [m$^{-1}$]& $0.004\pi$ 
 \\
\hline\hline
\end{tabular}
\end{center}
\caption{Parameters used in the simulation. The physical parameters are from a Kia Soul 2016 vehicle \cite{Orosz_MEM_2017}. \label{tab:veh_param}}
\end{table}

\begin{figure}[!t]
  \centering
  \includegraphics[scale=1]{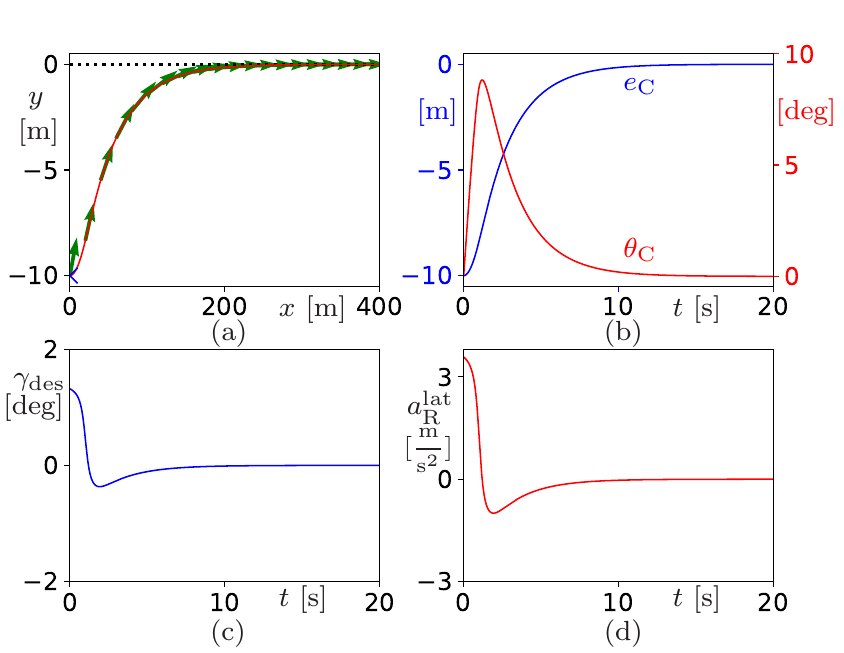}\\
  \caption{(a) Vehicle following a straight path along the $x$-axis. The blue cross marks the starting point with errors ${e_{\rm C}(0)=-10}$~m and ${\theta_{\rm C}(0) =0}$~deg. (b) Lateral deviation $e_{\rm C}$ and heading angle error $\theta_{\rm C}$. (c) Desired steering angle $\gamma_{\rm des}$. (d) Lateral acceleration at the center of rear axle $a_{\rm R}^{\rm lat}$. \label{fig:line_path_err}}
\end{figure}

Fig.~\ref{fig:line_path_err} shows that the controller allows the vehicle to follow a straight path. In panel (a) the dotted black line indicates the desired path, while the solid red curve represents the position of the rear axle center point R. The green arrows indicate the desired heading given by \eqref{eqn:desired_heading}. Panel (b) shows the time profiles of lateral deviation and relative yaw angle, indicating that the vehicle settles down to the path after a few seconds. Panels (c) and (d) depict the time profiles of the desired steering angle and the lateral acceleration, respectively. Notice that despite the relatively large initial lateral deviation, neither overshoot nor oscillations appear as the vehicle approaches the desired path. Last but not least, the observed lateral acceleration would not cause passenger discomfort.

\begin{figure}[!t]
  \centering
  \includegraphics[scale=1]{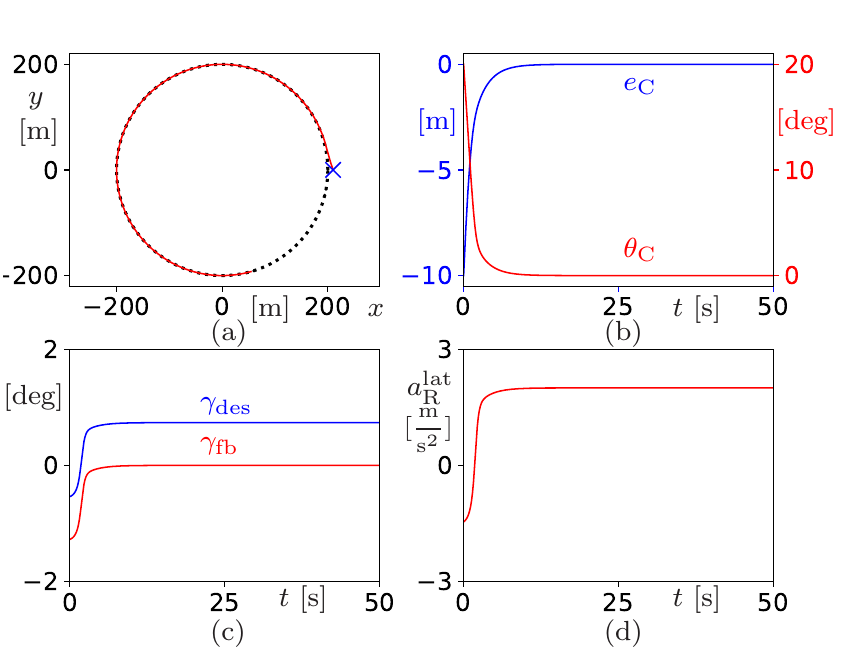}\\
  \caption{(a) Vehicle following circular path with radius ${\rho=200}$~m and initial errors ${e_{\rm C}(0)=-10}$~m and ${\theta_{\rm C}(0) =20}$~deg. (b) Lateral deviation $e_{\rm C}$ and heading angle error $\theta_{\rm C}$. (c) Steering angles $\gamma_{\rm des}$ and $\gamma_{\rm fb}$. (d) Lateral acceleration at the center of rear axle $a_{\rm R}^{\rm lat}$. \label{fig:circ_path_err}}
\end{figure}

Fig.~\ref{fig:circ_path_err} demonstrates that the controller also allows the vehicle to follow a circular path of radius ${\rho=200}$~m plotted in panel (a) by the dotted black curve. The rest of the notation is the same as in Fig.~\ref{fig:line_path_err}. In panels (c) and (d) one may notice that the feedback term goes to zero while the steering angle and the lateral acceleration approach constant values.

In order to demonstrate the performance of the controller  we consider a path where the curvature varies as function of the arclength according to
\begin{align}\label{eqn:curv_s_func}
 \kappa(s) &= \dfrac{\kappa_{\max}}{2}\left(1-\cos\bigg(\dfrac{2\pi}{s_{\rm T}}s\bigg)\right)\ ,
\end{align}
where $\kappa_{\max}$ is the maximum curvature along the path, and $s_{\rm T}$ is the period in arclength.
From differential geometry, by solving the differential equations
\begin{equation}\label{eqn:EOM_ref_path_ds}
\begin{split}
     \dfrac{\textrm{d} x}{\textrm{d} s} &=\cos\psi\ ,
     \\
    \dfrac{\textrm{d} y}{\textrm{d} s} &=\sin\psi\ ,
    \\
    \dfrac{\textrm{d} \psi}{\textrm{d} s} &=\kappa\ ,
\end{split}
\end{equation}
one can obtain the path $x(s)$, $y(s)$ and $\psi(s)$. This requires the initial configuration and in the rest of the paper we use ${x(0)=0}$, ${y(0)=0}$ and ${\psi(0)=0}$.
One can show that setting
\begin{equation}\label{eqn:curv_close_cond}
 \kappa_{\max}\,s_{\rm T}=\dfrac{4\pi}{N}\ , \quad N=2, 3, \ldots \ ,
\end{equation}
a closed path with $N$ corners and perimeter $Ns_{\rm T}$ is obtained.
For point C, these lead to  ${\kappa_{\rm C}=\kappa(s_{\rm C})}$, ${x_{\rm C}=x(s_{\rm C})}$, ${y_{\rm C}=y(s_{\rm C})}$ and ${\psi_{\rm C}=\psi(s_{\rm C})}$.

\begin{figure}[!t]
  \centering
  \includegraphics[scale=1]{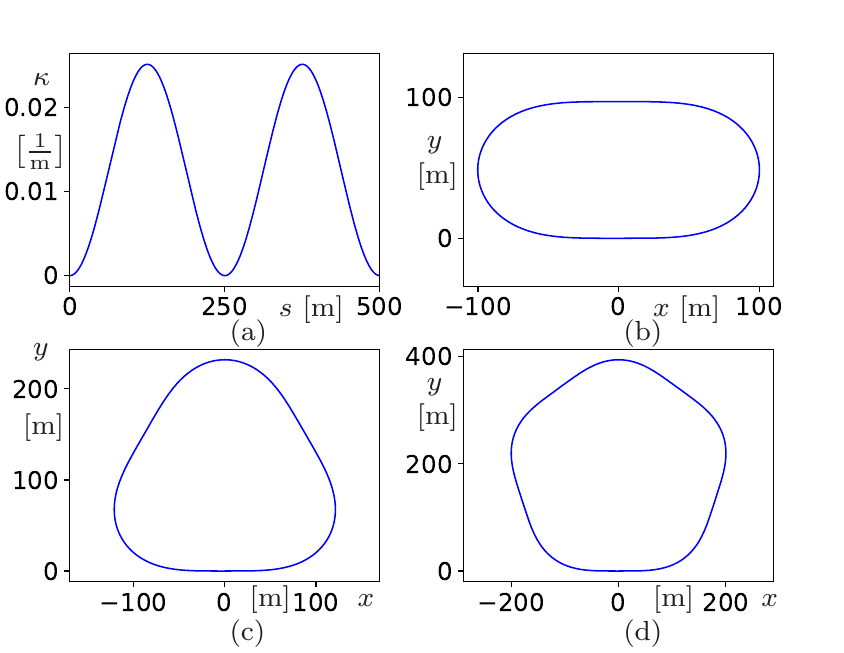}\\
  \caption{(a) Curvature \eqref{eqn:curv_s_func} with ${N=2}$ and ${s_{\rm T} = 250}$~m. (b) Path corresponding to the curvature in panel (a). (c) Path corresponding to curvature with ${N=3}$ and ${s_{\rm T} = 250}$~m. (d) Path corresponding to curvature with ${N=5}$ and ${s_{\rm T} = 250}$~m.  \label{fig:curv_path_N}}
\end{figure}

Fig.~\ref{fig:curv_path_N}(a,b) show the path described by (\ref{eqn:curv_s_func},\ref{eqn:EOM_ref_path_ds},\ref{eqn:curv_close_cond}) when ${N=2}$ and ${s_{\rm T} = 250}$~m. Panel (a) plots curvature as a function of the arclength, while panel (b) depicts the path in the ${(x,y)}$ plane with the origin corresponding to ${s_{\rm C}=0}$. Fig.~\ref{fig:curv_path_N}(c,d) shows the paths when ${N=3}$ and ${N=5}$.

\begin{figure}[!t]
  \centering
  \includegraphics[scale=1]{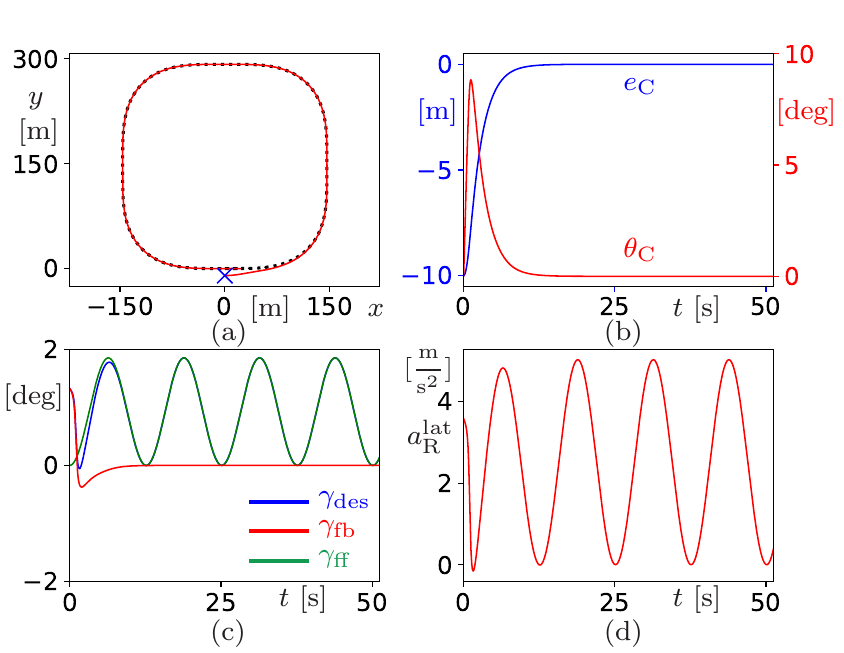}\\
  \caption{(a) Vehicle following a closed path of varying curvature, starting with errors ${e_{\rm C}=-10}$~m and ${\theta_{\rm C} =0}$~deg. (b) Lateral deviation $e_{\rm C}$ and heading angle error $\theta_{\rm C}$. (c) Steering angles $\gamma_{\rm des}$, $\gamma_{\rm fb}$ and $\gamma_{\rm ff}$. (d) Lateral acceleration at the center of rear axle $a_{\rm R}^{\rm lat}$. \label{fig:curv_path_err}}
\end{figure}

In the remainder of this paper, we consider the path with $N=4$ and $s_{\rm T} = 250$~m which yields the minimum turning radius ${1/\kappa_{\rm max} \approx 80}$~m. This path is used in Fig.~\ref{fig:curv_path_err} to showcase the tracking performance of the controller. Here the same notations are used as in Figs.~\ref{fig:line_path_err} and \ref{fig:circ_path_err}. In panel (c) at initial stage the feedback term $\gamma_{\rm fb}$ is noticeable but eventually this term converges to zero and the feedforward term $\gamma_{\rm ff}$ becomes dominant. In panel (d) there are instances when the lateral acceleration exceeds the limit $a_{\max}^{\rm lat} = 4 [\frac{\rm m}{\rm s}]$, since $a_{\max}^{\rm lat}$ is used to bound the feedback term $\gamma_{\rm fb}$, but here the feedforward term $\gamma_{\rm ff}$ dominates the lateral acceleration.

\subsection{Including Steering Dynamics}\label{sec:sim_ext_steering}

The path-following concepts above were explained using the kinematic bicycle model for simplicity, but these are indeed applicable to other models too. Here we extend the controller to the model developed in Section~\ref{sec:model_skate_kinematic_SteerTrq}, where the steering dynamics was considered. We add a lower-level controller on steering torque to make the steering angle track the desired steering angle, that is,
\begin{align}
  T_{\rm s}  &= g \big( k_{\rm s}(\gamma-\gamma_{\rm des}) \big)\ . \label{eqn:Ts_nonlinear}
\end{align}
Here $\gamma_{\rm des}$ is given by  (\ref{eqn:steer_controller},\ref{eqn:steer_controller_ff},\ref{eqn:steer_controller_fb}), $k_{\rm s}$ represent the steering gain, and in the wrapper function $g(x)$, given by \eqref{eqn:satfunction},  we set $g_{\rm sat} = T_{\rm sat}$ to represent the maximum allowable steering torque.

Following the same procedure as in Section~\ref{sec:stab_analysis}, one can find that the closed-loop system possesses the desired steady-state solution, that is,
\begin{equation}
\begin{split}\label{eqn:equil_steer_dyn}
  s_{\rm C}^{\ast}  &=V t\ ,\qquad
  e_{\rm C}^{\ast}  =0\ , \qquad
  \theta_{\rm C}^{\ast}  =0\ ,\\
  \gamma^{\ast} &= \arctan (l\kappa^{\ast})\ ,\qquad\qquad
  \sigma_{2}^{\ast} = 0,
\end{split}
\end{equation}
when the nominal value of road curvature is $\kappa^\ast$.
We remark that \eqref{eqn:equil_steer_dyn} is not a solution to the closed-loop system when the road curvature varies, unlike \eqref{eqn:equil} that is always a solution to \eqref{eqn:closed_reduced_nonlinear_dyn}. This implies that variations on road curvature lead to variations on tracking errors.
By defining the perturbations
\begin{equation}
\begin{split}\label{eqn:perturb_steer_dyn}
 \tilde{s}_{\rm C} &= s_{\rm C}-s_{\rm C}^{*}\ ,
 \enskip
 \tilde{e}_{\rm C} = e_{\rm C}-e_{\rm C}^{*}\ ,
 \enskip
 \tilde{\theta}_{\rm C} = \theta_{\rm C}-\theta_{\rm C}^{*}\ ,
  \\
 \tilde{\gamma} &= \gamma-\gamma^{*}\ ,
 \quad
 \tilde{\sigma}_2 = \sigma_{2}-\sigma_{2}^{*}\ ,
 \enskip
 \tilde{\kappa}_{\rm C} = \kappa_{\rm C}-\kappa^{*}\ ,
 \end{split}
\end{equation}
we obtain the linearized dynamics
\begin{equation}\label{eqn:closed_lin_dyn_steer_ctrl}
\begin{split}
 \dot{\tilde{s}}_{\rm C}&= V \kappa^{*}\tilde{e}_{\rm C}\ ,
 \\
 \dot{\tilde{e}}_{\rm C} &= V\tilde{\theta}_{\rm C}\ ,
 \\
 \dot{\tilde{\theta}}_{\rm C} &= -V \kappa^{*2}\tilde{e}_{\rm C}  +\frac{V}{l}\big(1+\kappa^{*2}l^{2}\big)\tilde{\gamma}
 -V \tilde{\kappa}_{\rm C}\ ,
 \\
 \dot{\tilde{\gamma}} &=\tilde{\sigma}_{2}\ ,
 \\
 \dot{\tilde{\sigma}}_{2} &=-\frac{k_{\rm s} k_{1} k_{2}}{J_{\rm F}} \tilde{e}_{\rm C}-\frac{k_{\rm s} k_{1} }{J_{\rm F}} \tilde{\theta}_{\rm C}
 +\frac{k_{\rm s}}{J_{\rm F}} \tilde{\gamma}
 -\frac{V}{l}\big(1+\kappa^{*2}l^{2}\big)\tilde{\sigma}_{2}
  \\
 &-\frac{k_{\rm s} l}{J_{\rm F} \big(1+\kappa^{*2}l^{2}\big)} \tilde{\kappa}_{\rm C}\ ,
\end{split}
\end{equation}
where $\tilde{\kappa}_{\rm C}$ serves as the disturbance input. Note that the first equation characterizes the longitudinal motion, while the latter four equations govern the lateral motion, which is decoupled from the first one. By calculating the characteristic equation of the linearized system \eqref{eqn:closed_lin_dyn_steer_ctrl}, one can derive stability conditions. Also, calculating the transfer function from the disturbance input $\tilde{\kappa}_{\rm C}$ to lateral deviation $\tilde{e}_{\rm C}$ (or relative yaw angle error $\tilde{\theta}_{\rm C}$), one can analyze the performance of this controller while following paths with varying curvatures; see \cite{Wubing_TIV_2022}. We skip these details here, but instead, we run simulations using a set of gains that can stabilize the system and achieve good tracking performance when road curvature varies.

\begin{figure}[!t]
  \centering
  \includegraphics[scale=1]{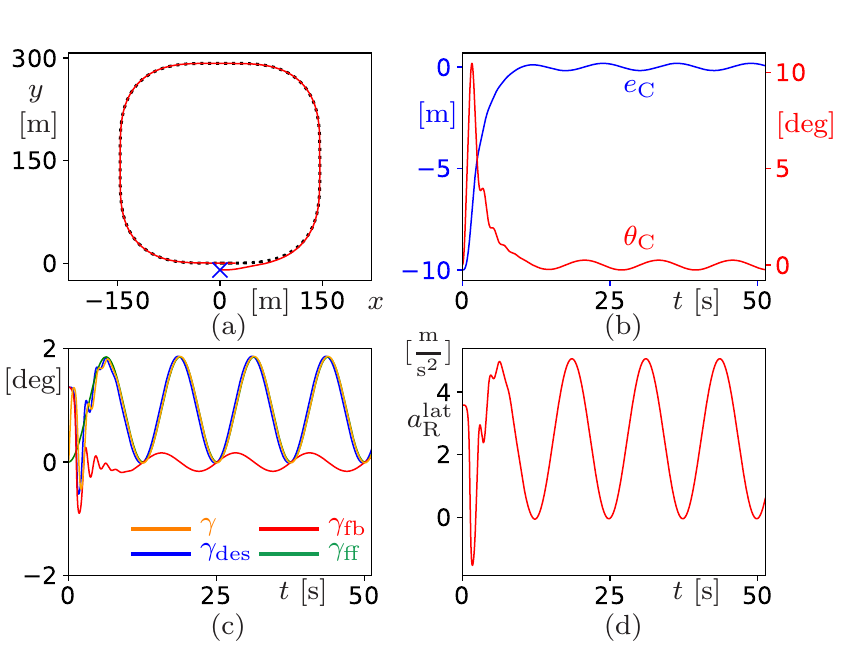}\\
  \caption{(a) Vehicle following a closed path of varying curvature, starting with errors ${e_{\rm C}=-10}$~m and ${\theta_{\rm C} =0}$~deg when including the steering dynamics. (b) Lateral deviation $e_{\rm C}$ and heading angle error $\theta_{\rm C}$. (c) Steering angles $\gamma$, $\gamma_{\rm des}$, $\gamma_{\rm fb}$ and $\gamma_{\rm ff}$. (d) Lateral acceleration at the center of rear axle $a_{\rm R}^{\rm lat}$. \label{fig:curv_path_err_trq_steer}}
\end{figure}

Fig.~\ref{fig:curv_path_err_trq_steer} shows the simulation results when the vehicle follows the closed path (\ref{eqn:curv_s_func},\ref{eqn:EOM_ref_path_ds},\ref{eqn:curv_close_cond}) with varying curvature. The same parameter values are used as in Fig.~\ref{fig:curv_path_err}, and the additional parameters can be found in Table~\ref{tab:veh_param}.
After transients decay, fluctuations in the tracking error can be observed on panel (b). This can be explained by the feedforward and feedback terms on panel (c): the feedforward term $\gamma_{\rm ff}$ varies along with the path while the
feedback term $\gamma_{\rm fb}$ makes efforts to correct the tracking errors and it does not converge to zero. This is due to the steering dynamics: the actual steering angle $\gamma$ is tracking the desired steering angle $\gamma_{\rm des}$ with some phase lag.

One can compensate the phase lags caused by the steering dynamics using the concept of look-ahead control. Namely, instead of using the curvature $\kappa_{\rm C}$ of the closest point C along the path in the feedforward controller we can use the curvature $\kappa_{\rm L}$ of the look-ahead point L; see Fig.~\ref{fig:steer_ctrl}(a). That is, instead of \eqref{eqn:steer_controller_ff}
we define the feedforward term
\begin{equation}\label{eqn:ff_preview}
  \gamma_{\rm ff}  = \arctan\left(\kappa_{\rm L}\,l\right)\ ,
\end{equation}
where ${\kappa_{\rm L} = \kappa(s_{\rm L})}$, the look-ahead distance is given by
\begin{equation}\label{eqn:arc_prev_steer_ff}
  s_{\rm L} = s_{\rm C} + V\,t_{\rm L}\ ,
\end{equation}
and $t_{\rm L}$ is called the look-ahead time.
One may verify that the equilibrium \eqref{eqn:equil_steer_dyn} remains unchanged. Using the same definitions of perturbations as \eqref{eqn:perturb_steer_dyn}, one can obtain almost the same linearized dynamics as \eqref{eqn:closed_lin_dyn_steer_ctrl} except that the last equation changes to
\begin{equation}\label{eqn:closed_lin_dyn_steer_ctrl_look_ahead}
\begin{split}
 \dot{\tilde{\sigma}}_{2} &=-\frac{k_{\rm s} k_{1} k_{2}}{J_{\rm F}} \tilde{e}_{\rm C}-\frac{k_{\rm s} k_{1} }{J_{\rm F}} \tilde{\theta}_{\rm C}
 +\frac{k_{\rm s}}{J_{\rm F}} \tilde{\gamma}  -\frac{V}{l}\big(1+\kappa^{*2}l^{2}\big)\tilde{\sigma}_{2}
 \\
 & -\frac{k_{\rm s} l}{J_{\rm F} \big(1+\kappa^{*2}l^{2}\big)} (\tilde{\kappa}_{\rm C}+V t_{\rm L}\,\tilde{\kappa}_{\rm C}')\ ,
\end{split}
\end{equation}
where $\tilde{\kappa}_{\rm C}':=\dfrac{{\rm d} \kappa}{{\rm d} s}(s_{\rm C}^{\ast})$. One can analyze the system with the aforementioned approaches.

\begin{figure}[!t]
  \centering
  \includegraphics[scale=1]{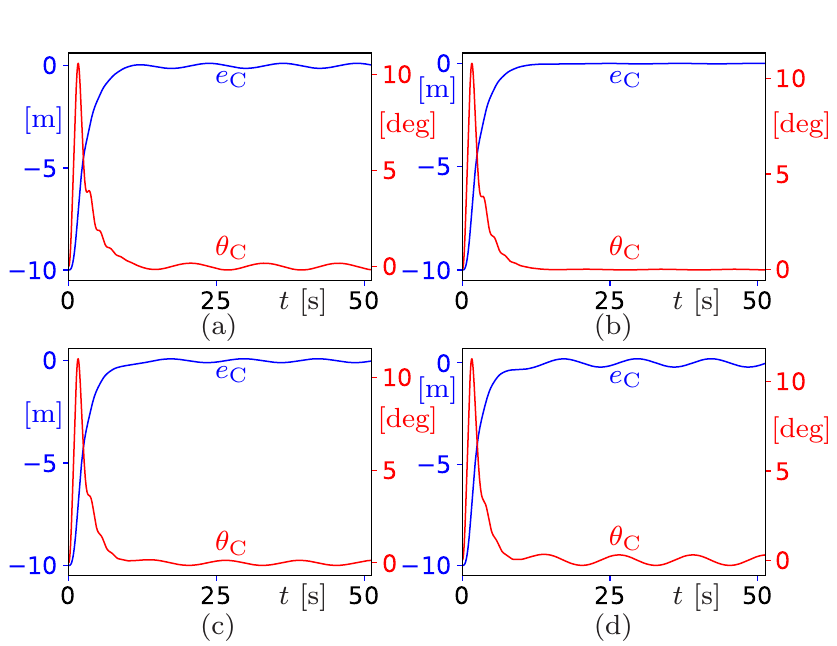}\\
  \caption{Lateral deviation $e_{\rm C}$ and relative yaw angle $\theta_{\rm C}$ when including the steering dynamics for different values of the look-ahead time: (a) ${t_{\rm L}=0.1}$~s, (b) ${t_{\rm L}=0.3}$~s, (c) ${t_{\rm L}=0.5}$~s, (d) ${t_{\rm L}=0.7}$~s. \label{fig:err_prev_time_diff}}
\end{figure}

Fig.~\ref{fig:err_prev_time_diff} shows the responses for different values of the look-ahead times $t_{\rm L}$; cf.~Fig.~\ref{fig:curv_path_err_trq_steer}(b)  where $t_{\rm L}=0$~s. Notice that as $t_{\rm L}$ is increased the tracking error first decreases and then increases. The tracking error is minimal around $0.3$~seconds which is close to the phase lag in the steering dynamics.

\subsection{Including Longitudinal Dynamics}\label{sec:sim_ext_longdyn}

In this section we extend the use of the path-following controller to the model developed in Section~\ref{sec:model_skate_dynamic_assign} that includes the longitudinal dynamics. We demonstrate that this model allows one to integrate the path-following control with  longitudinal control. We apply the path-following controller (\ref{eqn:steer_controller},\ref{eqn:steer_controller_ff},\ref{eqn:steer_controller_fb}), while changing the speed from $V$ to $\sigma_1$ in \eqref{eqn:max_allow_steer_ang}, and construct a longitudinal controller that adjusts $\sigma_1$ to a desired speed that depends on the road curvature ahead.

We consider a rear wheel drive vehicle, that is,  ${F_{\rm F}  = 0}$.  We apply feedback linearization to the longitudinal dynamics given in the second row of Table~\ref{tab:modelswithskates} with the original configuration coordinates and in the second row of Table~\ref{tab:modelswithskates_path_ref_frame} in the path-reference frame. This results in the differential equations
\begin{equation}\label{eqn:EOM_010_rear_closed_loop}
\begin{split}
\dot{x}_{\rm R}  &= \sigma_{1}\cos\psi\ ,
\\
\dot{y}_{\rm R}   &= \sigma_{1}\sin\psi\ ,
\\
\dot{\psi} &= \frac{\sigma_{1}}{l}\tan\gamma\ ,
\\
\dot{\sigma}_{1} &= a_{\rm des}\ ,
\end{split}
\end{equation}
or alternatively
\begin{equation}\label{eqn:EOM_010_rear_closed_loop2}
\begin{split}
\dot{s}_{\rm C}&=\dfrac{\sigma_{1} \cos\theta_{\rm C}}{1-\kappa_{\rm C}e_{\rm C}}\ ,
\\
\dot{e}_{\rm C}  &=\sigma_{1} \sin\theta_{\rm C}\ ,
\\
\dot{\theta}_{\rm C} &=\dfrac{\sigma_{1} }{l}\tan\gamma-\dfrac{\sigma_{1} \kappa_{\rm C}\cos\theta_{\rm C}}{1-\kappa_{\rm C}e_{\rm C}}\ ,
\\
\dot{\sigma}_{1} &= a_{\rm des}\ ,
\end{split}
\end{equation}
where ${\gamma = \gamma_{\rm des}}$ is given by the path-following controller (\ref{eqn:steer_controller},\ref{eqn:steer_controller_ff},\ref{eqn:steer_controller_fb}), and $a_{\rm des}$ is given by the longitudinal controller described below.

\begin{figure}[!t]
  \centering
  \includegraphics[scale=1]{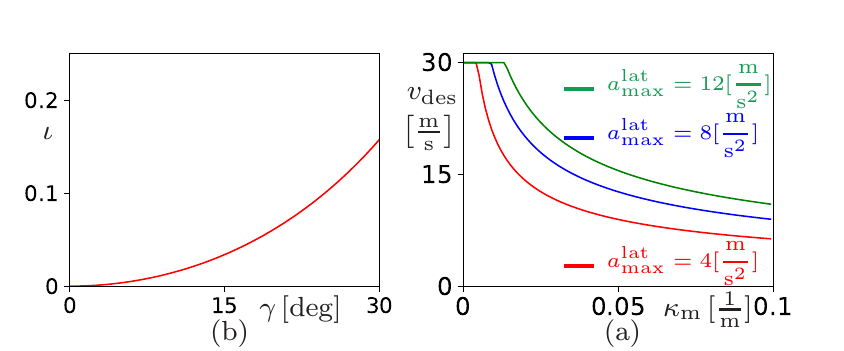}\\
  \caption{(a) Ratio $\iota$ as a function of the steering angle $\gamma$. (b)  Target speed $v_{\rm des}$ as a function of the road curvature $\kappa_{\rm m}$. \label{fig:vdes_mu}}
\end{figure}

The longitudinal driving force is given by
\begin{equation}\label{eqn:fbck_linearization}
  F_{\rm R} =\big(m_{1}+m_{2}\tan^{2}\gamma\big)a_{\rm des}+m_{2}\frac{\tan\gamma}{\cos^{2}\gamma}\,\dot{\gamma}\,\sigma_{1}
+\frac{J_{\rm F} }{l}\ddot{\gamma}\,\tan\gamma\ ,
\end{equation}
which we can rewrite as
\begin{align} \label{eqn:fbck_linearization_rewrite}
    F_{\rm R}  & = m_{1} \big((1 +\iota) a_{\rm des} + a_{1} + a_{2}\big)\ ,
\end{align}
where
\begin{equation}\label{eqn:ratios}
\begin{split}
    \iota &= \dfrac{m_{2}}{m_{1}}\tan^{2}\gamma\ ,
    \\
    a_{1} &=\dfrac{m_{2}}{m_{1}}\frac{\sin\gamma}{\cos^{3}\gamma}\,\dot{\gamma}\,\sigma_{1} \ ,
    \\
    a_{2} &=\frac{J_{\rm F} }{m_{1} l}\ddot{\gamma}\,\tan\gamma\ .
\end{split}
\end{equation}
The constant $\iota$ is plotted in Fig.~\ref{fig:vdes_mu} as a function of the steering angle $\gamma$. Notice that this only becomes significant for larger values of the steering angle. Below we also show the constants $a_1$ and $a_2$ for the numerical simulations and the derivatives $\dot\gamma$ and $\ddot\gamma$ are calculated in Appendix~\ref{append:deriv_steer_angle}.

The lateral constraining forces $\tilde{F}_{\rm R}$ and $\tilde{F}_{\rm F}$ given in \eqref{eqn:010:F1F2} can be used to define the force-to-weight ratios
\begin{align}\label{eqn:lat_force_norm}
    \mu_{\rm R} &= \dfrac{\tilde{F}_{\rm R}l}{m_{1}g(l-d)}\ , &
    \mu_{\rm F} &= \dfrac{\tilde{F}_{\rm F}l}{m_{1}gd}\ .
\end{align}
These correspond to the friction coefficients needed to ensure that the kinematic constraints hold, assuming static weight distribution, i.e., no load transfer. Note that these expressions also contain  the derivatives $\dot\gamma$ and $\ddot\gamma$ given in Appendix~\ref{append:deriv_steer_angle}.

In order to assign the longitudinal acceleration we propose the controller
\begin{align}
  a_{\rm des} = g\big(k_{\rm a}(\sigma_{1}-v_{\rm des})\big)\ , \label{eqn:ades}
\end{align}
where $k_{\rm a}$ is the feedback gain and the wrapper function $g(x)$ is given by \eqref{eqn:satfunction} with $g_{\rm sat}=a_{\max}^{\rm long}$.
Moreover, we assign the target speed $v_{\rm des}$ according to
\begin{align}\label{eqn:tgt_speed}
  v_{\rm des} &= \min\left\{v_{\rm max},\,\sqrt{\dfrac{a_{\max}^{\rm lat}}{\kappa_{\rm m}}}\right\}\ ,
\end{align}
where $v_{\max}$ is the maximum speed set, $a_{\max}^{\rm lat}$ is the maximum lateral acceleration allowed, and $\kappa_{\rm m}$ is the maximum curvature of the path between the closest point C and the look-ahead point L, i.e.,
\begin{align}
  \kappa_{\rm m}& = \max_{ s \in [s_{\rm C}, s_{\rm L}]} |\kappa(s)|\ .
\end{align}
For simplicity, here we use the constant preview distance ${s_{\rm L}-s_{\rm C}=50}$~m as opposed to using the look-ahead time as in \eqref{eqn:arc_prev_steer_ff}. In Fig.~\ref{fig:vdes_mu} the desired velocity \eqref{eqn:tgt_speed} is plotted as a function of the curvature  for different lateral acceleration limits $a_{\max}^{\rm lat}$. This is in correspondence with the maximum allowable steering angle shown in Fig.~\ref{fig:fun_arctan_steer_max}(b).

Considering a path of  larger curvature such that $v_{\rm des}$ does not saturate,  the closed-loop system (\ref{eqn:steer_controller},\ref{eqn:steer_controller_ff},\ref{eqn:steer_controller_fb}, \ref{eqn:EOM_010_rear_closed_loop2},\ref{eqn:ades},\ref{eqn:tgt_speed}) possesses the equilibrium
\begin{equation}
\begin{split}\label{eqn:equil_long_dyn}
  s_{\rm C}^{\ast}  &=\sqrt{\dfrac{a_{\max}^{\rm lat}}{|\kappa^{\ast}|}}\, t\ ,\quad
  e_{\rm C}^{\ast}  =0\ , \quad
  \theta_{\rm C}^{\ast}  =0\ ,\quad
  \sigma_{1}^{\ast} = \sqrt{\dfrac{a_{\max}^{\rm lat}}{|\kappa^{\ast}|}},
\end{split}
\end{equation}
when the nominal value of road curvature is $\kappa^\ast$.
By defining the state and input perturbations as
\begin{equation}
\begin{split}\label{eqn:perturb_long_dyn}
 \tilde{s}_{\rm C} &= s_{\rm C}-s_{\rm C}^{*}\ ,
 \enskip
 \tilde{e}_{\rm C} = e_{\rm C}-e_{\rm C}^{*}\ ,
 \enskip
 \tilde{\theta}_{\rm C} = \theta_{\rm C}-\theta_{\rm C}^{*}\ ,
 \\
 \tilde{\sigma}_{1} &= \sigma_{1}-\sigma_{1}^{*}\ ,
 \enskip
  \tilde{\kappa}_{\rm m} = \kappa_{\rm m}-\kappa^{*}\ ,
 \end{split}
\end{equation}
we obtain the linearized dynamics
\begin{equation}\label{eqn:closed_lin_dyn_long_dyn}
\begin{split}
 \dot{\tilde{s}}_{\rm C}&= \sigma_{1}^{\ast} \kappa^{*}\tilde{e}_{\rm C}+\tilde{\sigma}_{1}\ ,
 \\
 \dot{\tilde{e}}_{\rm C} &= \sigma_{1}^{\ast}\tilde{\theta}_{\rm C}\ ,
 \\
 \dot{\tilde{\theta}}_{\rm C} &= \frac{\sigma_{1}^{\ast}}{l}\big(k_{1}k_{2}+k_{1}k_{2}\kappa^{*2}l^{2}-\kappa^{*2}l\big)\tilde{e}_{\rm C}
 \\
 &+\frac{\sigma_{1}^{\ast}}{l}k_{1}\big(1+\kappa^{*2}l^{2}\big)\tilde{\theta}_{\rm C}\ ,
 \\
 \dot{\tilde{\sigma}}_{1} &=k_{\rm a}\, \tilde{\sigma}_{1}+\frac{k_{\rm a}}{|\kappa^\ast|}\sqrt{\dfrac{a_{\max}^{\rm lat}}{|\kappa^{\ast}|}}\, \tilde{\kappa}_{\rm m}\,.
\end{split}
\end{equation}
Again, one may follow the aforementioned approach to derive stability conditions and analyze performance in the presence of curvature disturbances $\tilde{\kappa}_{\rm m}$.

\begin{figure}[!t]
  \centering
  \includegraphics[scale=1]{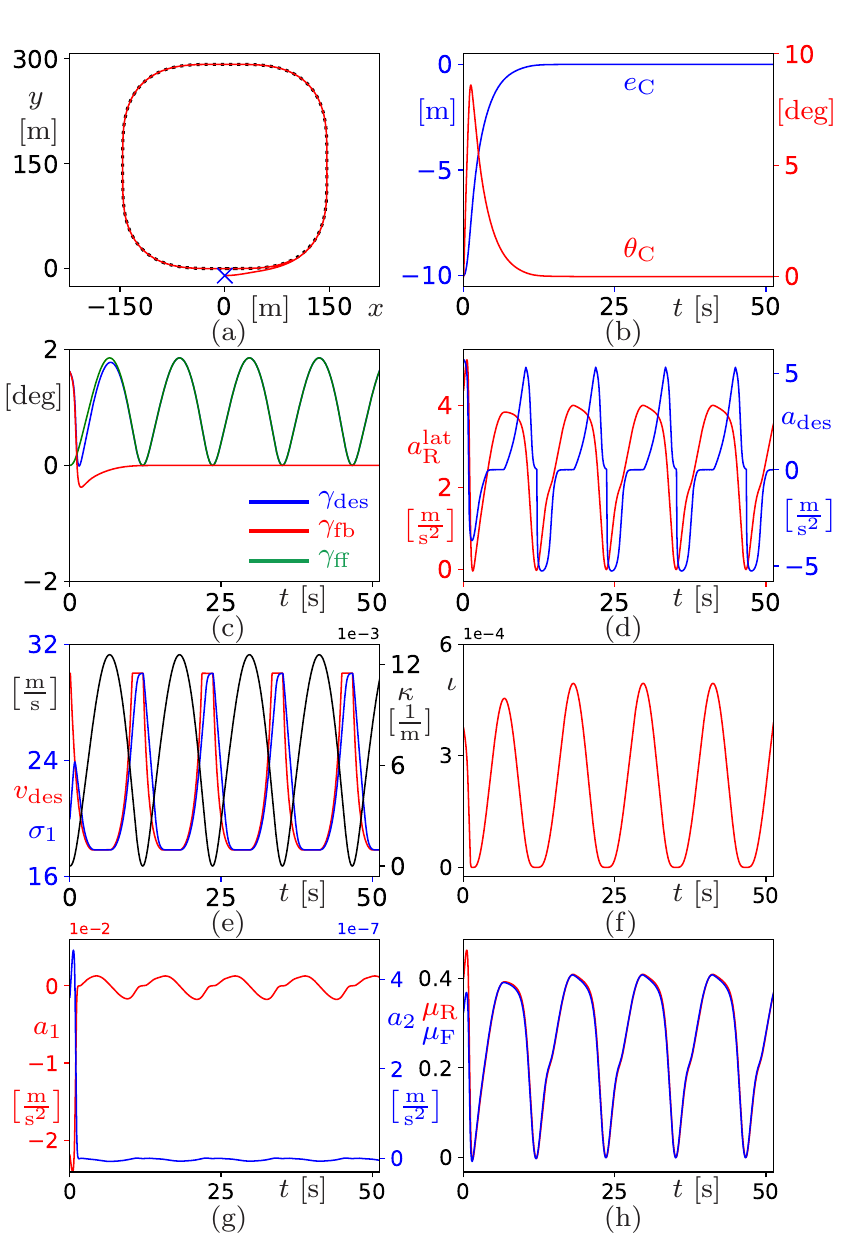}\\
  \caption{(a) Vehicle following a closed path of varying curvature, with initial errors ${e_{\rm C}=-10}$~m and ${\theta_{\rm C} =0}$~deg and speed ${\sigma_{1}(0)=20}$~m/s. (b) Lateral deviation $e_{\rm C}$ and relative yaw angle $\theta_{\rm C}$. (c) Steering angles $\gamma_{\rm des}$, $\gamma_{\rm fb}$ and $\gamma_{\rm ff}$. (d) Lateral acceleration at the center of rear axle $a_{\rm R}^{\rm lat}$ and desired longitudinal acceleration $a_{\rm des}$. (e) Desired speed $v_{\rm des}$, longitudinal velocity $\sigma_{1}$, and road curvature $\kappa_{\rm C}$. (f) The ratio $\iota$. (g) Acceleration terms $a_{1}$ and $a_{2}$. (h) Lateral force-to-weight ratios $\mu_{\rm R}$ and $\mu_{\rm F}$. \label{fig:curv_path_err_trq_steer_long_ctrl}}
\end{figure}

\begin{figure}[!t]
  \centering
  \includegraphics[scale=1]{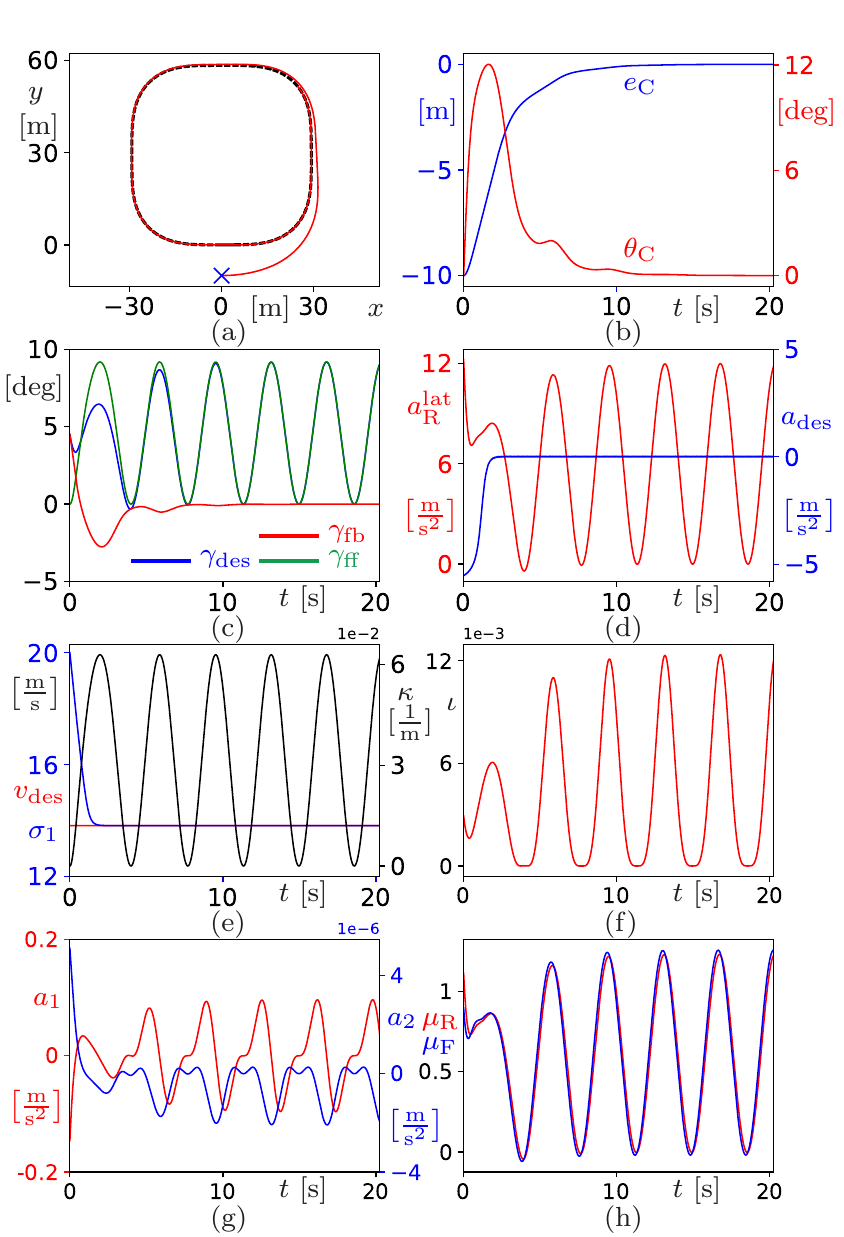}\\
  \caption{Vehicle following a closed path of varying curvature, with initial errors ${e_{\rm C}=-10}$~m and ${\theta_{\rm C} =0}$~deg and speed ${\sigma_{1}(0)=20}$~m/s. To create sharp turns we use the reduced value $s_{\rm T}=50$~m  and we also allow the lateral acceleration ${a_{\max}^{\rm lat}=12}$~m/s$^2$. (b) Lateral deviation $e_{\rm C}$ and heading angle error $\theta_{\rm C}$. (c) Steering angles $\gamma_{\rm des}$, $\gamma_{\rm fb}$ and $\gamma_{\rm ff}$. (d) Lateral acceleration at the center of rear axle $a_{\rm R}^{\rm lat}$ and desired longitudinal acceleration $a_{\rm des}$. (e) Desired speed $v_{\rm des}$, longitudinal velocity $\sigma_{1}$, and road curvature $\kappa_{\rm C}$. (f) The ratio $\iota$. (g) Acceleration terms $a_{1}$ and $a_{2}$. (h) Lateral force-to-weight ratios $\mu_{\rm R}$ and $\mu_{\rm F}$. \label{fig:curv_path_err_trq_steer_long_ctrl_aggr}}
\end{figure}

We simulate the vehicle model \eqref{eqn:EOM_010_rear_closed_loop2} with the path-following controller (\ref{eqn:steer_controller},\ref{eqn:steer_controller_ff},\ref{eqn:steer_controller_fb}) and the longitudinal controller (\ref{eqn:ades},\ref{eqn:tgt_speed}) using the parameters in Table~\ref{tab:veh_param}.  Fig.~\ref{fig:curv_path_err_trq_steer_long_ctrl}(a-d) show the path-following performance, which is similar to that in Fig.~\ref{fig:curv_path_err}, except that in panel (d) the peak lateral acceleration is smaller as the longitudinal controller reduces the speed at the corners. This panel also shows that the desired longitudinal acceleration is bounded by ${a_{\max}^{\rm long}=6}$~m/s$^2$. Panel (e) shows the time profile of the desired speed tracked by the actual speed as well as the changes of road curvature. Notice that the speed decreases once the curvature increases according to our design. Panels (f) and (g) show the coefficients defined in \eqref{eqn:ratios} and one may observe ${\iota \ll 1}$ and ${a_1,a_2 \ll a_{\rm des}}$. That is, for the driving scenario considered in Fig.~\ref{fig:curv_path_err_trq_steer_long_ctrl}, one may omit these in \eqref{eqn:fbck_linearization_rewrite} and use $F_{\rm R} \approx m_1 a_{\rm des}$ when commanding the driving force. Finally, panel (h) depicts the lateral force-to-weight ratios given in \eqref{eqn:lat_force_norm}. These suggest that, in  normal driving conditions (i.e., dry asphalt), there is sufficient friction to maintain the motion of the automobile. Also while in the transient phase the coefficients differ a little, we have ${\mu_{\rm R} \approx \mu_{\rm F}}$ once the vehicle closely follows the path. This implies that the two wheels would reach the sliding limit simultaneously if the friction becomes smaller.

In order to investigate a more aggressive driving scenario we reduce the parameter $s_{\rm T}$ from 250 meters to 50 meters which results in the minimum turning radius ${1/\kappa_{\rm max} \approx 16}$~m. Correspondingly, we increase the lateral acceleration limit to $a_{\max}^{\rm lat}=12$~m/s$^{2}$. Recall that this parameter influences both the largest allowed steering angle feedback \eqref{eqn:max_allow_steer_ang} as well as the desired speed  \eqref{eqn:tgt_speed}. The simulation results are shown in Fig.~\ref{fig:curv_path_err_trq_steer_long_ctrl_aggr} where panels (a-c) show qualitatively similar behavior as seen in Fig.~\ref{fig:curv_path_err_trq_steer_long_ctrl}(a-c), but the steering angle reaches much larger values since the path has much sharper corners. Panel (d) shows that the lateral acceleration also gets much higher compared to Fig.~\ref{fig:curv_path_err_trq_steer_long_ctrl}(d), while the longitudinal acceleration approaches zero. The latter is explained by the speed profiles in panel (e), where the longitudinal velocity approaches the (constant) desired speed. Panels (f) and (g) show that ${\iota \ll 1}$ and ${a_2\ll a_{\rm des}}$ still hold, but $a_1$ becomes comparable with $a_{\rm des}$. This coefficient is expected to grow further for maneuvers where rapid change of the steering angle is needed (i.e., $\dot\gamma$ becomes large) like sudden lane changes. Finally, the lateral force-to-weight ratios in panel (h) show qualitatively similar behavior to those in Fig.~\ref{fig:curv_path_err_trq_steer_long_ctrl}(d), but they reach much higher values, which can make it challenging for the automobile to stay on track. Once the kinematic constraints are violated the vehicle model needs to be changed to accommodate the sliding, but developing those models is beyond the scope of this paper.

\section{Conclusion}\label{sec:conclusion}

The Appellian approach was utilized to derive single track models that can describe the versatile maneuvering capabilities of automated vehicles. The models were categorized based on the modeling assumptions regarding the wheel-ground contact, the longitudinal dynamics, and the steering dynamics. It was shown that when the vehicle was driven by force/torque, the Lagrangian approach led to singularities in the equations of motion, while using the Appellian approach, we were able to obtain non-singular equations. The Lagrangian approach was used to derive nonholonomic constraining forces that ensure that the vehicle stays on track.

By re-writing the equations of motion using path coordinates, low-complexity nonlinear controllers were constructed that enable automated vehicles to execute a large variety of complex maneuvers. The corresponding motion planning and control algorithms are of low complexity and can be evaluated in a fast manner. This allows one to minimize time delays in the control loops, which is particularly important in safety critical scenarios. Such property is becoming more important as vehicles are moving towards higher levels of automation, requiring more and more complex perception algorithms with larger and larger computational needs.

There are many technological, economical and legal challenges to overcome if one wants to make highly automated vehicles deployable on public roads. Here we highlight four challenges related to vehicle dynamics and control.

The first challenge concerns modeling. The Appellian models we presented in this paper are built to capture the backbone dynamics of automobiles. They assume rigid skates and wheels with point contact to the ground. Nevertheless, the Appellian approach can also be utilized to build higher fidelity models which take into account that flexibility of pneumatic tires \cite{BerAveHeTakOro2022}. Such models may be used to test the performance of the low-complexity nonlinear controllers
developed through the backbone models and to evaluate the performance limits of automated vehicles.

The second challenge is related to the performance, adaptability and robustness of controllers.
Automated vehicles are expected to  perform as good as human drivers in a versatile set of conditions in terms of weather, road surface and behavior of neighboring vehicles. Parameterizing controllers so that they can automatically adapt to changing environments is a challenging task. Rather than hand tuning the low-complexity controllers developed in this paper, one may complement them with controllers learned from the behavior of human drivers \cite{Ave2021, Wubing_TVT_2022}. Maintaining safety under varying conditions also requires robustness to disturbances. This may be achieved by extending the theory of control barrier functions and synthesizing robust safety critical controllers \cite{Alan2022}.

The third challenge lies in having a gap between control theory and its practical applications. During the research phase, attention is mostly attracted to the study on stability, robustness and adaptability in order to ensure the eventual settlement to desired steady states. In contrast, in dynamic traffic environments, automated vehicles frequently deal with transient events, such as cut-ins, cut-outs, lane changes, take-offs, stopping at traffic lights, etc. Controllers neglecting transient response may generate ``overreaction", jerky behaviors (sway and surge motions), and oscillations in such scenarios, which can make human occupants very uncomfortable. Controllers that can handle transient responses well without using large computational resources are urgently needed in the automotive industry.

Finally, a significant challenge is related to how to integrate automated vehicles to transportation systems so that they do not only benefit their passengers but also positively influence the safety and efficiency of the overall transportation network. One way to achieve this is to utilize wireless vehicle-to-everything (V2X) communication which can enable vehicles to collect high-quality motion information about the traffic environment they are embedded in. Integrating such information into vehicle controllers may lead to large benefits even for lean penetration of automation and connectivity \cite{AveBanOro2022}.

\begin{acknowledgements}
D\'{e}nes Tak\'{a}cs would like to thank the Rosztoczy Foundation for their generous support.
\end{acknowledgements}

\section*{Funding}
This research was partially supported by the National Research, Development, and Innovation Office of Hungary under grant no.~NKFI-128422.

\section*{Conflict of interest}
The authors declare having no conflict of interest.

\section*{Availability of data and material}
The data generated will be made available  online.


\bibliographystyle{spmpsci}      			

\appendix       

\newpage
\section{Derivations \label{app:deriv}}

The first derivatives of (\ref{eqn:skate pos}) with respect to time are given by
\begin{equation}\label{eqn:skate_dxdy_rf}
\begin{split}
\dot{x}_{\rm R}  &= \dot{x}_{\rm G}+d\,\dot{\psi}\sin\psi\ ,
\\
\dot{y}_{\rm R}   &= \dot{y}_{\rm G}-d\,\dot{\psi}\cos\psi\ ,
\\
\dot{x}_{\rm F}  &= \dot{x}_{\rm G}-(l-d)\,\dot{\psi}\sin\psi\ ,
\\
\dot{y}_{\rm F}  &= \dot{y}_{\rm G}+(l-d)\,\dot{\psi}\cos\psi\ ,
\end{split}
\end{equation}
while the second derivatives of \eqref{eqn:skate pos} read
\begin{equation}\label{eqn:skate_ddxddy_rf}
\begin{split}
\ddot{x}_{\rm R}  &= \ddot{x}_{\rm G}+d\,\ddot{\psi}\sin\psi+d\,\dot{\psi}^{2}\cos\psi\ ,\\
\ddot{y}_{\rm R}   &= \ddot{y}_{\rm G}-d\,\ddot{\psi}\cos\psi+d\,\dot{\psi}^{2}\sin\psi\ ,\\
\ddot{x}_{\rm F}  &= \ddot{x}_{\rm G}-(l-d)\,\ddot{\psi}\sin\psi-(l-d)\,\dot{\psi}^{2}\cos\psi\ ,\\
\ddot{y}_{\rm F}  &= \ddot{y}_{\rm G}+(l-d)\,\ddot{\psi}\cos\psi-(l-d)\,\dot{\psi}^{2}\sin\psi\ .
\end{split}
\end{equation}

The first derivatives of \eqref{eqn:010_dxdy} are
\begin{equation}\label{eqn:010:ddxddyddpsi}
\begin{split}
\ddot{x}_{\rm G} &= \dot{\sigma}_{1} \Big(\cos\psi-\dfrac{d}{l}\sin\psi\tan\gamma \Big)
-\dfrac{d}{l}\sigma_{1}\dot{\gamma}\dfrac{\sin\psi}{\cos^{2}\gamma}
\\
&-\dfrac{\sigma_{1}^{2}}{l}\tan\gamma\Big(\sin\psi+\dfrac{d}{l}\cos\psi\tan\gamma\Big)\ ,
\\
\ddot{y}_{\rm G} &= \dot{\sigma}_{1} \Big(\sin\psi+\dfrac{d}{l}\cos\psi\tan\gamma\Big)
+\dfrac{d}{l}\sigma_{1}\dot{\gamma}\dfrac{\cos\psi}{\cos^{2}\gamma}
\\
&+\dfrac{\sigma_{1}^{2}}{l}\tan\gamma\Big(\cos\psi-\dfrac{d}{l}\sin\psi\tan\gamma\Big)\ ,
\\
\ddot{\psi} &= \dfrac{\dot{\sigma}_{1}}{l}\tan\gamma+\dfrac{\sigma_{1} \dot{\gamma}}{l\cos^{2}\gamma}\ .
\end{split}
\end{equation}

The first derivatives of \eqref{eqn:001_dxdy} are
\begin{equation}\label{eqn:001:ddxddyddpsiddgamma}
\begin{split}
\ddot{x}_{\rm G} &= -\frac{V^{2}}{l}\tan\gamma\Big(\sin\psi+\dfrac{d}{l}\cos\psi\tan\gamma\Big)
-\frac{d}{l}V\sigma_{2}\frac{\sin\psi}{\cos^{2}\gamma}\ ,
\\
\ddot{y}_{\rm G} &= \frac{V^{2}}{l}\tan\gamma\Big(\cos\psi-\dfrac{d}{l}\sin\psi\tan\gamma\Big)
+\frac{d}{l} V\sigma_{2}\frac{\cos\psi}{\cos^{2}\gamma}\ ,
\\
\ddot{\psi} &= \frac{V\sigma_{2}}{l\cos^{2}\gamma}\ ,
\\
\ddot{\gamma} &= \dot{\sigma}_{2}\, .
\end{split}
\end{equation}
Taking the first derivative of \eqref{eqn:011_dxdy}, one can see that $\ddot{x}_{\rm G}$, $\ddot{y}_{\rm G}$ and $\ddot{\psi}$ are the same as those given in \eqref{eqn:010:ddxddyddpsi}, and $\ddot{\gamma}$ is the same as that given in \eqref{eqn:001:ddxddyddpsiddgamma}.

Taking the first derivative of \eqref{eqn:110:appell:dxdydpsidphi1dphi2dgamma}, one can see that $\ddot{x}_{\rm G}$, $\ddot{y}_{\rm G}$ and $\ddot{\psi}$ are the same as those given in \eqref{eqn:010:ddxddyddpsi}, and
\begin{equation}\label{eqn:110:appell:ddxddyddthetaddphi1ddphi2}
\begin{split}
\ddot{\varphi}_{\rm R} &= \frac{\dot{\sigma}_{1}}{r}\ ,
\\
\ddot{\varphi}_{\rm F} &= \frac{\dot{\sigma}_{1}}{r\cos\gamma}+\frac{\sin\gamma}{r\cos^{2}\gamma}\sigma_{1}\dot{\gamma}\ .
\end{split}
\end{equation}

Taking the first derivative of \eqref{eqn:101:dxdydpsidphi1dphi2dgamma}, one can see that $\ddot{x}_{\rm G}$, $\ddot{y}_{\rm G}$ and $\ddot{\psi}$ are the same as those given in \eqref{eqn:001:ddxddyddpsiddgamma}, and
\begin{equation}\label{eqn:101:appell:ddxddyddthetaddphi1ddphi2}
\begin{split}
\ddot{\varphi}_{\rm R} &= 0\ ,
\\
\ddot{\varphi}_{\rm F} &= \frac{\sin\gamma}{r\cos^2\gamma}V \dot{\gamma}\ .
\end{split}
\end{equation}

Taking the first derivative of \eqref{eqn:111:dxdydpsidphi1dphi2dgamma}, one can see that $\ddot{x}_{\rm G}$, $\ddot{y}_{\rm G}$ and $\ddot{\psi}$ are the same as given in
\eqref{eqn:010:ddxddyddpsi}, $\ddot{\gamma}$ is the same as given in \eqref{eqn:001:ddxddyddpsiddgamma}, and $\ddot{\varphi}_{\rm R}$ and $\ddot{\varphi}_{\rm F}$ are the same as given in \eqref{eqn:110:appell:ddxddyddthetaddphi1ddphi2}.

\section{Lagrangian Approach \label{app:Lagrange}}

Here we apply the Lagrangian approach for the model presented in Section~\ref{sec:model_skate_dynamic_assign} as an example. The generalized coordinates are chosen as ${x}_{\rm G}$, ${y}_{\rm G}$, and $\psi$, and the corresponding Lagrange equation of second kind \eqref{eq:RouthVoss} become
\begin{equation}\label{eqn:Lag_eqn_org}
\begin{split}
\frac{{\rm d}}{{\rm d}\,t}\frac{\partial\, T}{\partial\, \dot{x}_{\rm G}}-\frac{\partial\, T}{\partial\, {x}_{\rm G}} &= Q_{1}+\lambda_{1}\, A_{11}+\lambda_{2}\, A_{21}\ ,
\\
\frac{{\rm d}}{{\rm d}\,t}\frac{\partial\, T}{\partial\, \dot{y}_{\rm G}}-\frac{\partial\, T}{\partial\, {y}_{\rm G}} &= Q_{2}+\lambda_{1}\, A_{12}+\lambda_{2}\, A_{22}\ ,
\\
\frac{{\rm d}}{{\rm d}\,t}\frac{\partial\, T}{\partial\, \dot{\psi}}-\frac{\partial\, T}{\partial\, \psi} &= Q_{3}+\lambda_{1}\, A_{13}+\lambda_{2}\, A_{23}\ ,
\end{split}
\end{equation}
where $T$ is the kinetic energy of the system, $Q_{1}$, $Q_{2}$ and $Q_{3}$ are the generalized forces corresponding to generalized coordinates ${x}_{\rm G}$, ${y}_{\rm G}$ and $\psi$, respectively. The Lagrange multipliers $\lambda_{1}$ and $\lambda_{2}$ are related to the two kinematic constraints in \eqref{eqn:skate_consraint_no_slip}. Moreover, $A_{11}$, $A_{12}$ and $A_{13}$ are the coefficients of $\dot{x}_{\rm G}$, $\dot{y}_{\rm G}$ and $\dot{\psi}$ in the first equation in \eqref{eqn:skate_consraint_no_slip}, while $A_{21}$, $A_{22}$ and $A_{23}$ are the coefficients of $\dot{x}_{\rm G}$, $\dot{y}_{\rm G}$ and $\dot{\psi}$ in the second equation in \eqref{eqn:skate_consraint_no_slip}. Those coefficients are
\begin{align}
A_{11} &=\sin\psi\ ,&A_{21} &=\sin(\psi+\gamma)\ , \nonumber
\\
A_{12} &=-\cos\psi\ ,& A_{22}&=-\cos(\psi+\gamma)\ , \label{eqn:lag_coeff_Aij}
\\
A_{13} &=d\ ,& A_{23}&=-(l-d)\cos\gamma\ . \nonumber
\end{align}

The kinetic energy is given by
\begin{equation}\label{eqn:kineticEnergy_org}
\begin{split}
T &= \frac{1}{2}m\,(\dot{x}_{\rm G}^{2}+\dot{y}_{\rm G}^{2})+\frac{1}{2}J_{\rm G} \dot{\psi}^{2}
\\
&+\frac{1}{2}m_{\rm R} \,(\dot{x}_{\rm R} ^{2}+\dot{y}_{\rm R}  ^{2})+\frac{1}{2}J_{\rm R}  \,\dot{\psi}^{2}
\\
&+\frac{1}{2}m_{\rm F} \,(\dot{x}_{\rm F} ^{2}+\dot{y}_{\rm F} ^{2})+\frac{1}{2}J_{\rm F} \,(\dot{\psi}+\dot{\gamma})^{2}\ .
\end{split}
\end{equation}
By substituting the derivative of \eqref{eqn:skate pos}, one can get
\begin{equation}
\begin{split}
T &= \frac{1}{2}m\,(\dot{x}_{\rm G}^{2}+\dot{y}_{\rm G}^{2})+\frac{1}{2}J_{\rm G} \,\dot{\psi}^{2}+\frac{1}{2}J_{\rm R}  \,\dot{\psi}^{2}+\frac{1}{2}J_{\rm F} \,(\dot{\psi}+\dot{\gamma})^{2} \\
&+\frac{1}{2}m_{\rm R} \big((\dot{y}_{\rm G}-d\,\dot{\psi}\,\cos\psi)^{2}+(\dot{x}_{\rm G}+d\,\dot{\psi}\,\sin\psi)^{2}\big)\label{eqn:kineticEnergy}\\
&+\frac{1}{2}m_{\rm F} \Big(\big(\dot{y}_{\rm G}+(l-d)\dot{\psi}\cos\psi\big)^{2}+(\dot{x}_{\rm G}-(l-d)\dot{\psi}\sin\psi)^{2}\Big)\ .
\end{split}
\end{equation}

To obtain the generalized forces, we calculate the virtual power of the active forces
\begin{equation}\label{eqn:010:virtualpower2}
\begin{split}
\delta P &=
\begin{bmatrix}
F_{\rm R} \cos\psi & F_{\rm R} \sin\psi & 0
\end{bmatrix}_{\mathcal{F}}
\begin{bmatrix}
\delta\dot{x}_{\rm R}  \\ \delta\dot{y}_{\rm R}  \\0
\end{bmatrix}_{\mathcal{F}} \\
&+
\begin{bmatrix}
F_{\rm F} \cos(\psi+\gamma) & F_{\rm F} \sin(\psi+\gamma) & 0
\end{bmatrix}_{\mathcal{F}}
\begin{bmatrix}
\delta\dot{x}_{\rm F}  \\ \delta\dot{y}_{\rm F}  \\ 0
\end{bmatrix}_{\mathcal{F}} \\
&= \big(F_{\rm R} \cos\psi+F_{\rm F} \cos(\psi+\gamma)\big)\,\delta\dot{x}_{\rm G}  \\
&+\big(F_{\rm R} \sin\psi+F_{\rm F} \sin(\psi+\gamma)\big)\,\delta\dot{y}_{\rm G} \\
&+F_{\rm F} (l-d)\sin\gamma\;\delta\dot{\psi}\ ,
\end{split}
\end{equation}
implying that
\begin{equation}\label{eqn:gen_forces}
\begin{split}
Q_{1}&=F_{\rm R} \cos\psi+F_{\rm F} \cos(\psi+\gamma)\ ,
\\
Q_{2}&=F_{\rm R} \sin\psi+F_{\rm F} \sin(\psi+\gamma)\ ,
\\
Q_{3}&=F_{\rm F} (l-d)\sin\gamma\ .
\end{split}
\end{equation}

Substituting (\ref{eqn:lag_coeff_Aij},\ref{eqn:kineticEnergy},\ref{eqn:gen_forces}) into \eqref{eqn:Lag_eqn_org}, the Lagrangian equations become
\begin{equation}\label{eqn:010}
\begin{split}
&\big((m_{\rm R} +m_{\rm F} )d-m_{\rm F} l\big)\,\dot{\psi}^{2}\cos\psi+(m+m_{\rm R} +m_{\rm F} )\ddot{x}_{\rm G}
\\
&\quad+\big((m_{\rm R} +m_{\rm F} )d-m_{\rm F} l\big)\,\ddot{\psi}\sin\psi 
\\
&\quad= F_{\rm R} \cos\psi +F_{\rm F} \cos(\psi+\gamma)+\lambda_{1}\sin\psi+\lambda_{2}\sin(\psi+\gamma)\   ,
\\
&\big((m_{\rm R} +m_{\rm F} )d-m_{\rm F} l\big)\,\dot{\psi}^{2}\sin\psi+(m+m_{\rm R} +m_{\rm F} )\ddot{y}_{\rm G}
\\
&\quad-\big((m_{\rm R} +m_{\rm F} )d-m_{\rm F} l\big)\,\ddot{\psi}\cos\psi 
\\
&\quad= F_{\rm R} \sin\psi+F_{\rm F} \sin(\psi+\gamma)-\lambda_{1}\cos\psi-\lambda_{2}\cos(\psi+\gamma)\   ,
\\
&\big((m_{\rm R} +m_{\rm F} )d-m_{\rm F} l\big)\,\ddot{x}_{\rm G}\,\sin\psi
\\
&-\big((m_{\rm R} +m_{\rm F} )d-m_{\rm F} l\big)\,\ddot{y}_{\rm G}\,\cos\psi
\\
&\quad+\big(J_{\rm R}  +J_{\rm F} +J_{\rm G} +m_{\rm R} \,d^{2}+m_{\rm F} (l-d)^{2}\big)\ddot{\psi}+J_{\rm F} \,\ddot{\gamma}
\\
&\quad=F_{\rm F} (l-d)\sin\gamma+\lambda_{1}\,d-\lambda_{2}(l-d)\cos\gamma\ .
\end{split}
\end{equation}
The first two equations result in the Lagrange multipliers
\begin{equation}\label{eqn:lagrange_nu}
\begin{split}
\lambda_{1} & = \dfrac{1}{\sin\gamma}\Big(F_{\rm R} \cos\gamma +F_{\rm F} +m_{3}d \ddot{\psi}\sin\gamma
-m_{3}d \dot{\psi}^{2}\cos\gamma\,
\\
&\qquad\quad\,\,\,\,\,-m_{1} \big(\ddot{x}_{\rm G}\cos(\psi+\gamma)+\ddot{y}_{\rm G}\sin(\psi+\gamma)\big)\Big)\ ,
\\
\lambda_{2} &=\frac{1}{\sin\gamma}\Big(-F_{\rm R} -F_{\rm F} \cos\gamma+m_{3}d \dot{\psi}^{2}\,
\\
&\qquad\quad\,\,\,\,\,+m_{1} \big(\ddot{x}_{\rm G}\cos\psi+\ddot{y}_{\rm G}\sin\psi\big)\Big)\ ,
\end{split}
\end{equation}
where $m_{1}$ and $m_{2}$ are given in (\ref{eqn:const_m1_m2}), and
\begin{equation}\label{eqn:const_m3}
  m_{3} = m_{\rm R} -\dfrac{l-d}{d}m_{\rm F} \ .
\end{equation}

Substituting \eqref{eqn:lagrange_nu} into the third equation in \eqref{eqn:010} yields
\begin{equation}\label{eqn:skate_6th_dynamics}
\begin{split}
  &\Big(-(m_{1}-m_{3}) d \sin\psi\sin\gamma +m_{1} l \cos\psi\cos\gamma\Big)\ddot{x}_{\rm G}
  \\
  &+\Big((m_{1}-m_{3}) d \cos\psi\sin\gamma +m_{1} l \sin\psi\cos\gamma\Big)\ddot{y}_{\rm G}
  \\
  &+\Big(J_{\rm G} +J_{\rm R}  +J_{\rm F} +m_{\rm F}  l (l-d)\Big)\ddot{\psi}\sin\gamma +J_{\rm F}  \ddot{\gamma}\sin\gamma
  \\
  &+m_{3} d l \dot{\psi}^{2}\cos\gamma-F_{\rm R}  l \cos\gamma - F_{\rm F}  l =0\ .
\end{split}
\end{equation}
Combining this equation with the first derivatives of (\ref{eqn:skate_consraint_no_slip}), one can obtain three linear equations on $\ddot{x}_{\rm G}$, $\ddot{y}_{\rm G}$, and $\ddot{\psi}$. Note that the solutions for $\ddot{x}_{\rm G}$, $\ddot{y}_{\rm G}$, and $\ddot{\psi}$ are not independent, one can choose the solution for any of $\ddot{x}_{\rm G}$, $\ddot{y}_{\rm G}$, and $\ddot{\psi}$. Here, we choose to solve for
\begin{align}
\label{eqn:010:ddpsi}
\ddot{\psi} &= \frac{\Big(F_{\rm R} +\frac{F_{\rm F} }{\cos\gamma}\Big)\frac{\tan\gamma}{l}+m_{1}\frac{1}{\sin\gamma\cos\gamma}\dot{\gamma}\,\dot{\psi}-\frac{J_{\rm F} }{l^{2}}\,\ddot{\gamma}\,\tan^{2}\gamma}{m_{1}+m_{2}\tan^{2}\gamma}\ ,
\end{align}
where $m_{1}$ and $m_{2}$ are given in (\ref{eqn:const_m1_m2}).
By solving $\dot{x}_{\rm G}$ and $\dot{y}_{\rm G}$ from \eqref{eqn:skate_consraint_no_slip} in terms of $\dot{\psi}$, we obtain the governing equations
\begin{equation}\label{eqn:EOM_010_singular}
\begin{split}
\dot{x}_{\rm G} &=(l\cos\psi\cot\gamma-d\sin\psi)\overline{\sigma}_{1}\ ,
\\
\dot{y}_{\rm G} &=(l\sin\psi\cot\gamma+d\cos\psi)\overline{\sigma}_{1}\ ,
\\
\dot{\psi} &=\overline{\sigma}_{1}\ ,
\\
\dot{\overline{\sigma}}_{1} &= \frac{\Big(F_{\rm R} +\frac{F_{\rm F} }{\cos\gamma}\Big)\frac{\tan\gamma}{l}+m_{1}\frac{\dot{\gamma}\,\overline{\sigma}_{1}}{\sin\gamma\cos\gamma}-\frac{J_{\rm F} }{l^{2}}\,\ddot{\gamma}\,\tan^{2}\gamma}{m_{1}+m_{2}\tan^{2}\gamma}\ .
\end{split}
\end{equation}

These equations are singular at ${\gamma=0}$, that is, they cannot describe the rectilinear motion. This singularity can be solved by using Appellian approach that results in \eqref{eqn:EOM_010}. As mentioned in Section~\ref{sec:constrainforces}, one can
substitute the first derivative of \eqref{eqn:010_dxdy} (cf.~\eqref{eqn:010:ddxddyddpsi}) into \eqref{eqn:lagrange_nu} to
eliminate $\ddot{x}_{\rm G}$, $\ddot{y}_{\rm G}$, and $\ddot{\psi}$.

This results in formulas that depend on the velocities $\dot{x}_{\rm G}$, $\dot{y}_{\rm G}$, $\dot{\psi}$ and $\dot{\sigma}_{1}$ that are given by \eqref{eqn:EOM_010} and lead to
\begin{equation}\label{eqn:010:nu1nu2}
\begin{split}
\lambda_1 &=\dfrac{(m_{2}-m_{4}) \tan\gamma}{m_{1}+m_{2} \tan^{2} \gamma}\,\bigg(F_{\rm R} +\dfrac{F_{\rm F} }{\cos\gamma}\bigg)
-(m_{1} - m_{4})\dfrac{\sigma_{1}^{2}}{l}\tan\gamma
 \\
&- \dfrac{m_{4} \sigma_{1}\dot{\gamma}}{\cos^{2} \gamma} +\dfrac{m_{1}+m_{4}\tan^{2} \gamma}{m_{1}+m_{2} \tan^{2} \gamma}\bigg(\dfrac{m_{2} \sigma_{1}\dot{\gamma}}{\cos^{2} \gamma}+\dfrac{J_{\rm F} }{l} \ddot{\gamma}\bigg) \ ,
\\
\lambda_2 &= -\dfrac{1}{m_{1}+m_{2} \tan^{2} \gamma}\bigg(m_{2} F_{\rm R}  \dfrac{\tan\gamma}{\cos\gamma}+(m_{2}-m_{1}) F_{\rm F}  \tan\gamma
 \\
&+m_{1} \dfrac{m_{2} \sigma_{1}\dot{\gamma} }{\cos^{3} \gamma} +m_{1} \dfrac{J_{\rm F} }{l}\dfrac{\ddot{\gamma}}{\cos\gamma} \bigg)
-m_{4} \dfrac{\sigma_{1}^{2}}{l}\dfrac{\tan\gamma}{\cos\gamma}\ ,
\end{split}
\end{equation}
which are singular at ${|\gamma|=\pi/2}$. Below we show that these multipliers are indeed related to the lateral constraining forces that keep the skates on track.

\section{Using Newton's Law \label{app:model_newton}}

\begin{figure}[!t]
  \centering
  \includegraphics[scale = 1]{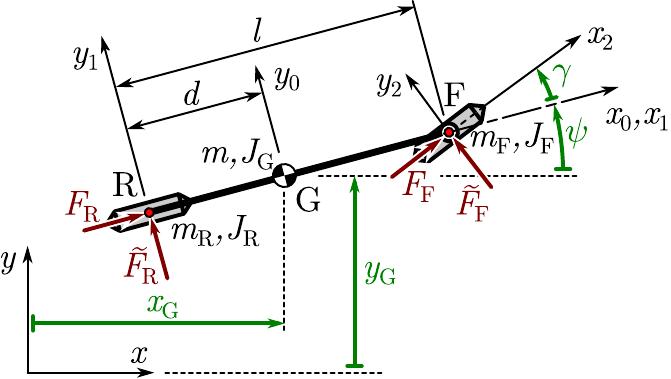}\\
  \caption{The constraining forces acting in the lateral directions of the rear and front wheels. \label{fig:mechmodel_001_Newton}}
\end{figure}

Here we derive the governing equations for the mechanical model studied in Section~\ref{sec:model_skate_dynamic_assign} using the Newtonian approach. The mechanical model with the lateral constraining forces $\tilde{F}_{\rm R}$ and $\tilde{F}_{\rm F}$ acting on the skates are shown in Fig.~\ref{fig:mechmodel_001_Newton}. We relate these to Lagrange multipliers  \eqref{eqn:010:nu1nu2}.

\begin{figure}[t]
  \centering
  \includegraphics[scale = 1]{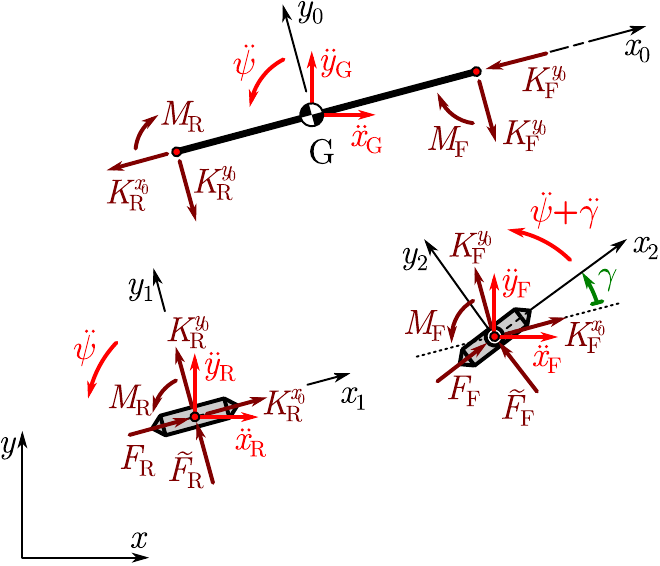}\\
  \caption{Free-body-diagrams of the vehicle body and the skates. \label{fig:mechmodel_001_FBD}}
\end{figure}

In order to derive the Newton equation we separate the three bodies that constitute the system, namely, the vehicle body, the skate at the rear, and the skate at the front. The corresponding free body diagrams are illustrated in Fig~\ref{fig:mechmodel_001_FBD}. The components of the internal forces between the skates and the vehicle body are denoted by $K^{x_0}_{{\rm R}}$, $K^{y_0}_{{\rm R}}$,  $K^{x_0}_{{\rm F}}$ and $K^{y_0}_{{\rm F}}$. The torques acting between the skates and the vehicle body are referred to as $M_{\rm R}$ and $M_{\rm F}$. For the sake of simplicity, the same notations are used for the counter forces, but their directions are opposite in the figures according to Newton's third law.
Thus, one can apply Newton's second law for the  skates, and for the vehicle body. The resulting equations are given in the $\mathcal{F}_0$ frame:
\begin{equation}\label{eqn:Newton_3body}
\begin{split}
     & m_{\rm R}  (\ddot{x}_{\rm R}  \cos\psi +\ddot{y}_{\rm R}   \sin\psi)= K^{x_0}_{\rm R}+F_{\rm R} \ ,
     \\
    & m_{\rm F}  (\ddot{x}_{\rm F}  \cos\psi +\ddot{y}_{\rm F}  \sin\psi)= K^{x_0}_{\rm F}+F_{\rm F} \cos\gamma - \tilde{F}_{\rm F}\sin\gamma\ ,
    \\
     & m (\ddot{x}_{\rm G} \cos\psi +\ddot{y}_{\rm G} \sin\psi)= -K^{x_0}_{\rm R}-K^{x_0}_{\rm F}\ ,
     \\
     & m_{\rm R}  (-\ddot{x}_{\rm R}  \sin\psi +\ddot{y}_{\rm R}   \cos\psi)= K^{y_0}_{\rm R}+\tilde{F}_{\rm R}\ ,
     \\
     & m_{\rm F}  (-\ddot{x}_{\rm F}  \sin\psi +\ddot{y}_{\rm F}  \cos\psi)= K^{y_0}_{\rm F}+F_{\rm F} \sin\gamma + \tilde{F}_{\rm F}\cos\gamma\ ,
     \\
     & m (-\ddot{x}_{\rm G} \sin\psi +\ddot{y}_{\rm G} \cos\psi) = -K^{y_0}_{\rm R}-K^{y_0}_{\rm F}\ ,
     \\
     & J_{\rm R}  \ddot{\psi}= M_{\rm R} \ ,
     \\
     & J_{\rm F} (\ddot{\psi}+\ddot{\gamma})= M_{\rm F} \ ,
     \\
     & J_{\rm G} \ddot{\psi}=K^{y_0}_{\rm R} d-K^{y_0}_{\rm F} (l-d) -M_{\rm R} -M_{\rm F} \ .
\end{split}
\end{equation}

Based on the first eight equations in \eqref{eqn:Newton_3body}, one can determine all constraining forces and torques. We are interested in the lateral constraining forces acting on the skates, which read
\begin{equation}\label{eqn:skate_F_lat_gen}
\begin{split}
\tilde{F}_{\rm R} & = \dfrac{1}{\sin\gamma}\Big(-F_{\rm R} \cos\gamma-F_{\rm F} -m_{3}d \ddot{\psi}\sin\gamma+m_{3}d \dot{\psi}^{2}\cos\gamma \\
&\qquad\quad\,\,\,\,\,+m_{1} \big(\ddot{x}_{\rm G}\cos(\psi+\gamma)+\ddot{y}_{\rm G}\sin(\psi+\gamma)\big)\Big)\ ,\\
\tilde{F}_{\rm F} &=\dfrac{1}{\sin\gamma}\Big(F_{\rm R} +F_{\rm F} \cos\gamma-m_{3}d \dot{\psi}^{2}   \\
&\qquad\quad\,\,\,\,\,-m_{1} (\ddot{x}_{\rm G}\cos\psi+\ddot{y}_{\rm G}\sin\psi)\Big)\ ,
\end{split}
\end{equation}
where we used the derivatives of \eqref{eqn:skate pos} (cf.~\eqref{eqn:skate_ddxddy_rf}),  $m_{1}$ and $m_{2}$ are given in (\ref{eqn:const_m1_m2}), and $m_{3}$ is given in \eqref{eqn:const_m3}.

After substituting all the constraining forces and torques into the last equation of \eqref{eqn:Newton_3body}, one can obtain
\begin{equation}\label{eqn:skate_6th_dynamics1}
\begin{split}
  &\Big(-(m_{1}-m_{3}) d \sin\psi\sin\gamma +m_{1} l \cos\psi\cos\gamma\Big)\ddot{x}_{\rm G}
  \\
  &+\Big((m_{1}-m_{3}) d \cos\psi\sin\gamma +m_{1} l \sin\psi\cos\gamma\Big)\ddot{y}_{\rm G}
  \\
  &+\Big(J_{\rm G} +J_{\rm R}  +J_{\rm F} +m_{\rm F}  l (l-d)\Big)\ddot{\psi}\sin\gamma +J_{\rm F}  \ddot{\gamma}\sin\gamma \\
  &+m_{3} d l \dot{\psi}^{2}\cos\gamma-F_{\rm R}  l \cos\gamma - F_{\rm F}  l =0\ ,
\end{split}
\end{equation}
which is the same as \eqref{eqn:skate_6th_dynamics}.
Then, following the same steps as in Section~\ref{app:Lagrange}, one can obtain the dynamics as given in \eqref{eqn:EOM_010_singular}. We point out again the singularity at ${\gamma=0}$. Moreover, comparing \eqref{eqn:skate_F_lat_gen} to \eqref{eqn:lagrange_nu} one may notice that
\begin{equation}\label{eqn:nu_F_lat_relation}
  \tilde{F}_{\rm R} = -\lambda_{1}\ ,
  \qquad
  \tilde{F}_{\rm F} = -\lambda_{2}\ ,
\end{equation}
that is, the lateral forces $\tilde{F}_{\rm R}$ and $\tilde{F}_{\rm F}$ are identical to the Lagrange multipliers $\lambda_1$ and $\lambda_2$ (except the negative signs). These forces prevent the skates from side slips given sufficiently large friction coefficients and normal forces. Again following the same steps as in Section~\ref{sec:Lagrange}, one can obtain the form \eqref{eqn:010:F1F2} that is identical with \eqref{eqn:010:nu1nu2} (except the negative signs).

\section{Coordinate Transformation \label{append:coord_trans}}

\begin{figure}[!b]
\begin{center}
\includegraphics[scale=1]{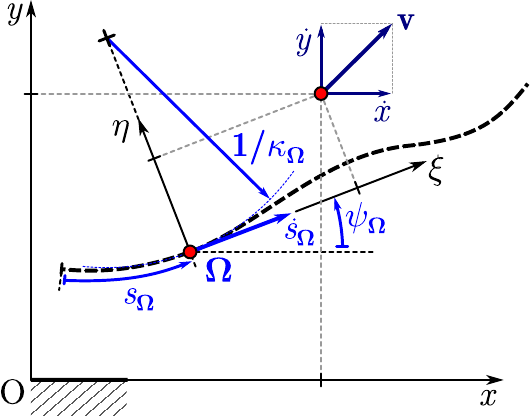}
\end{center}
\caption{Coordinate transformation between the Earth-fixed frame and a frame traveling along a given path. \label{fig:coord_transf}}
\end{figure}

In this part, we discuss the transformation between the Earth-fixed frame ${(x,y)}$ and the path-reference frame ${(\xi,\eta)}$; see  Fig.~\ref{fig:coord_transf}. We make the following assumptions
\begin{enumerate}
  \item ${(x,y,z)}$ is the Earth-fixed frame (denoted as $\mathcal{F}$) already used in Section~\ref{sec:model_skate} and \ref{sec:model_wheels}.
  \item ${(\xi,\eta,\zeta)}$ is the path-reference frame (denoted as $\mathcal{F}_{\rm \Omega}$) with the origin located at $\mathrm{\Omega}$, where $\xi$ and $\eta$ are along the tangential and normal directions of the path at point $\mathrm{\Omega}$, respectively. Note that the frame $\mathcal{F}_{\rm \Omega}$ is translating and rotating as $\mathrm{\Omega}$ moves along the path.
  \item The position of $\mathrm{\Omega}$ is referred to ${(x_{\rm \Omega}, y_{\rm \Omega})}$ when expressed in frame $\mathcal{F}$, and the heading angle and curvature at point $\mathrm{\Omega}$ are $\psi_{\rm \Omega}$ and $\kappa_{\rm \Omega}$, respectively. Note that $x_{\rm \Omega}$, $y_{\rm \Omega}$, $\psi_{\rm \Omega}$ and $\kappa_{\rm \Omega}$ are all changing in time.
\end{enumerate}

Let us consider a arbitrary point whose position are given by ${(x,y)}$ and ${(\xi, \eta)}$ in $\mathcal{F}$ and $\mathcal{F}_{\rm \Omega}$, respectively. Based on coordinates transformation (cf. Fig.~\ref{fig:coord_transf}), we have
\begin{equation}\label{eqn:xy_s_eta_transf}
\begin{split}
 \xi &= (x-x_{\rm \Omega})\cos\psi_{\rm \Omega}+(y-y_{\rm \Omega})\sin \psi_{\rm \Omega}\ ,\\
 \eta & = -(x-x_{\rm \Omega})\sin\psi_{\rm \Omega}+(y-y_{\rm \Omega})\cos \psi_{\rm \Omega}\ ,
\end{split}
\end{equation}
from which the derivatives read
\begin{equation}\label{eqn:transf_s_eta_deriv}
\begin{split}
 \dot{\xi} & =(\dot{x}-\dot{x}_{\rm \Omega})\cos\psi_{\rm \Omega}+(\dot{y}-\dot{y}_{\rm \Omega})\sin \psi_{\rm \Omega}+\eta \dot{\psi}_{\rm \Omega}\ ,
 \\
 \dot{\eta} & =-(\dot{x}-\dot{x}_{\rm \Omega})\sin\psi_{\rm \Omega}+(\dot{y}-\dot{y}_{\rm \Omega})\cos \psi_{\rm \Omega}-\xi \dot{\psi}_{\rm \Omega}\ .
\end{split}
\end{equation}
Using the arclength $s_{\rm \Omega}(t)$ along the path of the point $\rm \Omega$, we obtain
\begin{equation}\label{eqn:EOM_ref_path_dt}
\begin{split}
    \dot{x}_{\rm \Omega} &= \dfrac{\textrm{d} x_{\rm \Omega}}{\textrm{d} t} =\dfrac{\textrm{d} x_{\rm \Omega}}{\textrm{d} s_{\rm \Omega}}\dfrac{\textrm{d} s_{\rm \Omega}}{\textrm{d} t} = \cos\psi_{\rm \Omega}\dot{s}_{\rm \Omega}\ ,
    \\
    \dot{y}_{\rm \Omega} &= \dfrac{\textrm{d} y_{\rm \Omega}}{\textrm{d} t} =\dfrac{\textrm{d} y_{\rm \Omega}}{\textrm{d} s_{\rm \Omega}}\dfrac{\textrm{d} s_{\rm \Omega}}{\textrm{d} t} = \sin\psi_{\rm \Omega}\dot{s}_{\rm \Omega}\ ,
    \\
    \dot{\psi}_{\rm \Omega} &=\dfrac{\textrm{d} \psi_{\rm \Omega}}{\textrm{d} t} =\dfrac{\textrm{d} \psi_{\rm \Omega}}{\textrm{d} s_{\rm \Omega}}\dfrac{\textrm{d} s_{\rm \Omega}}{\textrm{d} t} =\kappa_{\rm \Omega}\dot{s}_{\rm \Omega}\ .
\end{split}
\end{equation}
Substituting these formulas into \eqref{eqn:transf_s_eta_deriv} yields
\begin{equation}\label{eqn:transf_s_eta_deriv_2}
\begin{split}
 \dot{\xi} & =\dot{x}\cos\psi_{\rm \Omega} + \dot{y}\sin \psi_{\rm \Omega} -(1-\kappa_{\rm \Omega}\eta )\dot{s}_{\rm \Omega}\ ,
 \\
 \dot{\eta} & =-\dot{x}\sin\psi_{\rm \Omega}+\dot{y}\cos \psi_{\rm \Omega}-\xi \kappa_{\rm \Omega}\dot{s}_{\rm \Omega}\ .
\end{split}
\end{equation}

\section{Nonlinear Wrapper Functions \label{app:wrapper}}

In Section~\ref{sec:control}, we used the wrapper function \eqref{eqn:satfunction} to improve the performance of the controller. One may notice that this belongs to a larger class of wrapper functions defined by
\begin{equation}\label{eqn:sat_func_set}
  G =\left\{g_{n}(x)\,\Big|\, g'_{n}(x)=\dfrac{1}{\big(1+(c\, x)^{2}\big)^\frac{n}{2}},\, n=2,\, 3,\, \ldots\right\}\ ,
\end{equation}
where $c$ is a constant.
Utilizing the requirements that $g_{n}(x)$ is a bounded odd function, one can solve \eqref{eqn:sat_func_set} and obtain
\begin{equation}
\begin{split}
  g_{2}(x) & = \dfrac{1}{c}\arctan (c x)\,,\qquad c=\dfrac{\pi}{2 g_{\rm sat}}\ ,
  \\
  g_{3}(x) & = \dfrac{x}{\sqrt{1+(c\,x)^{2}}}\,, \qquad c = \dfrac{1}{g_{\rm sat}}\ ,
  \\
  g_{n}(x) & = \dfrac{n-3}{n-2} g_{n-2}(x)+\dfrac{1}{n-2}\dfrac{x}{\big(1+(c\,x)^{2}\big)^{\frac{n}{2}-1}}\,,  \\
  & c=
    \begin{cases}
      \dfrac{(n-3)(n-5)\cdots 1}{(n-2)(n-4)\cdots 2}\dfrac{\pi}{2 g_{\rm sat}}, & n = 4,\, 6,\, 8,\,\ldots \ , \vspace{4pt}
      \\
      \dfrac{(n-3)(n-5)\cdots 2}{(n-2)(n-4)\cdots 3}\dfrac{1}{g_{\rm sat}}, & n = 5,\, 7,\, 9,\, \ldots \ .
    \end{cases}
\end{split}
\end{equation}
Indeed the wrapper function \eqref{eqn:satfunction} is the second element of the series while $n\to \infty$ yields
\begin{equation}\label{eqn:satfunction_linear}
    g_{\infty}(x) =\min\big\{\max\{x, -g_{\rm sat}\}, \,g_{\rm sat}\big\}\ .
\end{equation}

\begin{figure}[!h]
\begin{center}
\setlength{\unitlength}{0.012500in}%
\includegraphics[scale=1]{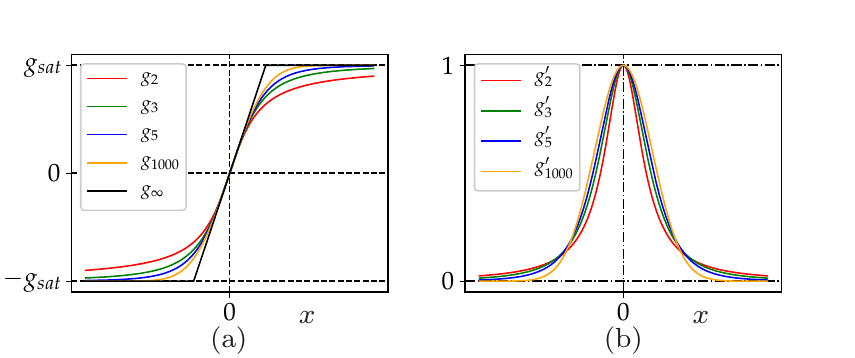}
\end{center}
\caption{(a) Wrapper function $g_{n}(x)$. (b) Downscale factor $g'_{n}(x)$. \label{fig:sat_func_deriv}}
\end{figure}

Fig.~\ref{fig:sat_func_deriv}(a) shows the wrapper functions $g_{n}(x)$ while Fig.~\ref{fig:sat_func_deriv}(b) depicts derivatives $g'_{n}(x)$ for $n=2,\, 3,\, 5,\, 1000$. The later illustrates how much the gains downscale as $|x|$ increases. Such downscaling allows the usage of larger linear gains yielding better tracking performance for small errors and less overshoot for larger errors. When $n$ is larger the downscaling occurs faster.

\newpage
\section{Derivatives of Steering Angle \label{append:deriv_steer_angle}}

Using (\ref{eqn:steer_controller_ff},\ref{eqn:steer_controller_fb},\ref{eqn:satfunction},\ref{eqn:steer_nonlinear_fb}) the steering controller \eqref{eqn:steer_controller} reads as
\begin{equation}
  \gamma = \gamma_{\rm ff}+\gamma_{\rm fb}
= \arctan (\kappa_{\rm C}\,l) + \dfrac{2\,\gamma_{\rm sat}}{\pi}\arctan \Big(\dfrac{\pi}{2\,\gamma_{\rm sat}} \gamma_{\rm fb}^{1}\Big)\ ,
\end{equation}
Taking the time derivative yields
\begin{equation}\label{eqn:gamma_dot}
  \dot{\gamma} = \dot{\gamma}_{\rm ff}+\dot{\gamma}_{\rm fb}
  =\dfrac{l \dot{\kappa}_{\rm C}}{1+l^{2}\kappa_{\rm C}^{2}} +
  \dfrac{\dot{\gamma}_{\rm fb}^{1}}{1+\left(\frac{\pi}{2\gamma_{\rm sat} } \gamma_{\rm fb}^{1}\right)^{2}}
  \ .
\end{equation}

The derivatives $\dot{\kappa}_{\rm C}$ and $\dot{\gamma}_{\rm fb}^{1}$ can be calculated by differentiating \eqref{eqn:curv_s_func} and \eqref{eqn:steer_nonlinear_fb} with respect to time:
\begin{equation}\label{eqn:kappa_dot}
\begin{split}
  \dot{\kappa}_{\rm C} &= \dfrac{\pi\dot{s}_{\rm C}}{s_{\rm T}} \kappa_{\max}\sin\left(\dfrac{2\pi}{s_{\rm T}}s_{\rm C} \right)\ ,
  \\
  \dot{\gamma}_{\rm fb}^{1} &= k_{1}\left(\dot{\theta}_{\rm C}+\dfrac{k_{2}\,\dot{e}_{\rm C}}{1+k_{2}^{2}\,e_{\rm C}^{2}} \right)\ ,
\end{split}
\end{equation}
where $\dot{s}_{\rm C}$, $\dot{e}_{\rm C}$, and $\dot{\theta}_{\rm C}$ are given by \eqref{eqn:EOM_s_err}.

The derivative of \eqref{eqn:gamma_dot} becomes
\begin{equation}\label{eqn:gamma_ddot}
\begin{split}
  \ddot{\gamma} = \ddot{\gamma}_{\rm ff}+\ddot{\gamma}_{\rm fb}
   &= \dfrac{l \ddot{\kappa}_{\rm C}(1+l^{2}\kappa_{\rm C}^{2})-2 l^{3} \kappa_{\rm C} \dot{\kappa}_{\rm C}^{2}}{(1+l^{2}\kappa_{\rm C}^{2})^{2}} \ ,
  \\
  &+ \dfrac{\ddot{\gamma}_{\rm fb}^{1} \left(1+\bigr(\frac{\pi}{2\gamma_{\rm sat} } \gamma_{\rm fb}^{1}\bigr)^{2}\right) -\frac{\pi}{\gamma_{\rm sat} }\gamma_{\rm fb}^{1} (\dot{\gamma}_{\rm fb}^{1})^{2}}{\left(1+\bigr(\frac{\pi}{2\gamma_{\rm sat} } \gamma_{\rm fb}^{1}\bigr)^{2}\right)^{2}} \ ,
\end{split}
\end{equation}
which contain the derivatives of \eqref{eqn:kappa_dot}:
\begin{equation}\label{eqn:kappa_ddot}
\begin{split}
  \ddot{\kappa}_{\rm C} &= \dfrac{\pi\ddot{s}_{\rm C}}{s_{\rm T}} \kappa_{\max}\sin\left(\dfrac{2\pi}{s_{\rm T}}s_{\rm C} \right) +
  2\left(\dfrac{\pi\dot{s}_{\rm C}}{s_{\rm T}}\right)^{2} \kappa_{\max}\cos\left(\dfrac{2\pi}{s_{\rm T}}s_{\rm C} \right)\ ,
  \\
  \ddot{\gamma}_{\rm fb}^{1} &= k_{1}\left(\ddot{\theta}_{\rm C}+\dfrac{k_{2}\,\ddot{e}_{\rm C}(1+k_{2}^{2}\,e_{\rm C}^{2}) -2 k_{2}^{3}\,e_{\rm C}\dot{e}_{\rm C}^{2}}{(1+k_{2}^{2}\,e_{\rm C}^{2})^{2}} \right) \ ,
\end{split}
\end{equation}
and
\begin{equation}\label{eqn:dEOM_s_err}
\begin{split}
 \ddot{s}_{\rm C}&=\dfrac{\dot{\sigma}_{1}\cos\theta_{\rm C}-\sigma_{1} \dot{\theta}_{\rm C}\sin\theta_{\rm C}}{1-\kappa_{\rm C}e_{\rm C}}
  +\dfrac{\sigma_{1} \cos\theta_{\rm C} (\dot{e}_{\rm C}\kappa_{\rm C} +e_{\rm C}\dot{\kappa}_{\rm C} ) }{(1-\kappa_{\rm C}e_{\rm C})^{2}}\ ,
 \\
 \ddot{e}_{\rm C} & =\dot{\sigma}_{1}\sin\theta_{\rm C}+\sigma_{1} \dot{\theta}_{\rm C}\cos\theta_{\rm C}\ ,
 \\
 \ddot{\theta}_{\rm C} &=
 \dfrac{\dot{\sigma}_{1}\tan\gamma }{l}
 +\dfrac{\sigma_{1}\dot{\gamma}}{l \cos^{2}\gamma}
 -\dfrac{\sigma_{1} \kappa_{\rm C}\cos\theta_{\rm C}(\dot{e}_{\rm C}\kappa_{\rm C} +e_{\rm C}\dot{\kappa}_{\rm C}) }{(1-\kappa_{\rm C}e_{\rm C})^{2}}\ ,
 \\
 &-\dfrac{\dot{\sigma}_{1}\kappa_{\rm C}\cos\theta_{\rm C}+\sigma_{1}\dot{\kappa}_{\rm C}\cos\theta_{\rm C} -\sigma_{1}\kappa_{\rm C}\dot{\theta}_{\rm C}\sin\theta_{\rm C} }{1-\kappa_{\rm C}e_{\rm C}}\ ,
\end{split}
\end{equation}
that are the derivatives of \eqref{eqn:EOM_s_err}.

\end{document}